\documentclass[opre,nonblindrev]{informs3} % current default for manuscript submission

\DoubleSpacedXI % Made default 4/4/2014 at request
\usepackage{enumitem}
    \usepackage{epsfig}
    \usepackage{amsmath}
    \usepackage{amsfonts}
    \usepackage{amssymb}
\usepackage{graphicx}
\usepackage{algpseudocode}
\usepackage{algorithm}
\usepackage{bm}
\usepackage{float}
    \usepackage{eucal}
    \usepackage{subfigure}
\usepackage{capt-of}
\usepackage{caption}
\usepackage[T1]{fontenc}

\usepackage{booktabs}

\newcommand{\RR}{\mathbb{R}}

\newcommand{\cA}{\mathcal{A}}
\newcommand{\cB}{\mathcal{B}}

\newcommand{\cV}{\mathcal{V}}
\newcommand{\cF}{\mathcal{F}}
\newcommand{\cU}{\mathcal{U}}
\newcommand{\cS}{\mathcal{S}}
\newcommand{\cW}{\mathcal{W}}
\newcommand{\cP}{\mathcal{P	}}
\newcommand{\bcV}{up(\mathcal{V})}

\newcommand{\bcU}{up(\mathcal{U})}
\newcommand{\bcA}{up(\mathcal{A})}
\newcommand{\bcB}{up(\mathcal{B})}

\newcommand{\bone}{\mathbf{1}}

\newcommand{\br}{\mathbf{r}}
\newcommand{\bR}{\mathbf{R}}
\newcommand{\bQ}{\mathbf{Q}}
\newcommand{\bF}{\mathbf{F}}
\newcommand{\bx}{\mathbf{x}}
\newcommand{\by}{\mathbf{y}}
\newcommand{\bu}{\mathbf{u}}
\newcommand{\bp}{\mathbf{p}}
\newcommand{\bq}{\mathbf{q}}
\newcommand{\bv}{\mathbf{v}}
\newcommand{\bz}{\mathbf{z}}
\newcommand{\ch}{\mathbf{ch}}

\usepackage[dvipsnames]{xcolor}
\usepackage[colorlinks=true,breaklinks=true,bookmarks=true,urlcolor=blue,
     citecolor=blue,linkcolor=blue,bookmarksopen=false,draft=false]{hyperref}

\def\EMAIL#1{\href{mailto:#1}{#1}}% When hyperref is used, otherwise outcomment 
\TheoremsNumberedThrough     % Preferred (Theorem 1, Lemma 1, Theorem 2)
\EquationsNumberedThrough    % Default: (1), (2), ...
\begin{document}
%%%%%%%%%%%%%%%%

% Outcomment only when entries are known. Otherwise leave as is and 
%   default values will be used.
%\setcounter{page}{1}
%\VOLUME{00}%
%\NO{0}%
%\MONTH{Xxxxx}% (month or a similar seasonal id)
%\YEAR{0000}% e.g., 2005
%\FIRSTPAGE{000}%
%\LASTPAGE{000}%
%\SHORTYEAR{00}% shortened year (two-digit)
%\ISSUE{0000} %
%\LONGFIRSTPAGE{0001} %
%\DOI{10.1287/xxxx.0000.0000}%

% Author's names for the running heads
% Sample depending on the number of authors;
% \RUNAUTHOR{Jones}
% \RUNAUTHOR{Jones and Wilson}
 \RUNAUTHOR{Kamble, Loiseau, and Walrand}
% \RUNAUTHOR{Jones et al.} % for four or more authors
% Enter authors following the given pattern:
%\RUNAUTHOR{}

% Title or shortened title suitable for running heads. Sample:
% \RUNTITLE{Bundling Information Goods of Decreasing Value}
% Enter the (shortened) title:
\RUNTITLE{An ADP Approach to Repeated Games with Vector Losses}

% Full title. Sample:
% \TITLE{Bundling Information Goods of Decreasing Value}
% Enter the full title:
\TITLE{An Approximate Dynamic Programming Approach to Adversarial Online Learning}
%   \EMAIL field can be repeated if more than one author
\ARTICLEAUTHORS{%
\AUTHOR{Vijay Kamble}
\AFF{Department of Information and Decision Sciences,\\ University of Illinois at Chicago\\ \EMAIL{kamble@uic.edu}}
\AUTHOR{Patrick Loiseau}
\AFF{Univ. Grenoble Alpes, INRIA, CNRS, Grenoble INP, LIG, France\\ Max-Planck Institute for Software Systems (MPI-SWS), Germany\\
 \EMAIL{patrick.loiseau@inria.fr}}
\AUTHOR{Jean Walrand}
\AFF{Dept. of Electrical Engineering and Computer Sciences,\\ University of California, Berkeley\\ \EMAIL{walrand@berkeley.edu}}
% Enter all authors
} % end of the block
% Block of authors and their affiliations starts here:
% NOTE: Authors with same affiliation, if the order of authors allows, 
%   should be entered in ONE field, separated by a comma. 

\ABSTRACT{%
%Motivated by the problem of regret minimization in repeated decision-making in adversarial environments, 
We describe an approximate dynamic programming (ADP) approach to compute approximations of the optimal strategies and of the minimal losses that can be guaranteed in discounted repeated games with vector-valued losses. 
Such games prominently arise in the analysis of regret in repeated decision-making in adversarial environments, also known as adversarial online learning.
At the core of our approach is a characterization of the lower Pareto frontier of the set of expected losses that a player can guarantee in these games as the unique fixed point of a set-valued dynamic programming operator. When applied to the problem of regret minimization with discounted losses, our approach yields algorithms that achieve markedly improved performance bounds compared to off-the-shelf online learning algorithms like Hedge. These results thus suggest the significant potential of ADP-based approaches in adversarial online learning.}

%This fixed point can be approximated by an iterative application of this DP operator compounded by a polytopic set approximation, beginning with a single point. Each iteration can be computed by solving a set of linear programs corresponding to the vertices of the polytope. We derive rigorous bounds on the error of the resulting approximation and the performance of the corresponding approximately optimal strategies.
%We discuss an application to regret minimization in repeated decision-making in adversarial environments, where we show that this approach can be used to compute approximately optimal policies and approximations of the minimax optimal regret when the action sets are finite. We illustrate this approach by computing provably approximately optimal algorithms for the problem of prediction using expert advice under discounted $\{0,1\}-$losses. 
% Sample
%\KEYWORDS{deterministic inventory theory; infinite linear programming duality; 
%  existence of optimal strategies; semi-Markov decision process; cyclic schedule}
%\MSCCLASS{Primary: 90B05; secondary: 90C40, 90C90}
%\ORMSCLASS{Primary: Inventory/production: deterministic multi-item;
%  secondary: dynamic programming/optimal control: deterministic 
%  semi-Markov; programming: infinite dimensional}
%\HISTORY{Received November 20, 2003; revised March 8, 2004, and March 26, 2004.}

% Fill in data. If unknown, outcomment the field
%\MSCCLASS{}
%\ORMSCLASS{Primary: ; secondary: }
%\HISTORY{}

\maketitle
%%%%%%%%%%%%%%%%%%%%%%%%%%%%%%%%%%%%%%%%%%%%%%%%%%%%%%%%%%%%%%%%%%%%%%

% Samples of sectioning (and labeling) in MOOR.
% NOTE: (1) all section levels end with a period,
%       (2) capitalization is as shown (sentence style, not title style)
%
%\section{Introduction.}\label{intro} %%1.
%\subsection{Duality and the classical EOQ problem.}\label{class-EOQ} %% 1.1.
%\subsection{Outline.}\label{outline1} %% 1.2.
%\subsubsection{Cyclic schedules for the general deterministic SMDP.}
%  \label{cyclic-schedules} %% 1.2.1
%\section{Problem description.}\label{problemdescription} %% 2.

\section{Introduction.}
In several decision-making scenarios in uncertain and potentially adversarial environments, a decision-maker cares about multiple objectives at the same time. For example, in defense operations, an agent might be interested in simultaneously defending multiple targets against an enemy. In repeated decision-making in an unknown environment, an agent may want to perform as well in hindsight as every policy in a given class of policies. In asymmetric information games where a player lacks some information that other players have, a natural goal for the player is to choose a strategy that gives appropriate worst-case guarantees simultaneously across the different underlying possibilities (e.g., countries lacking knowledge about the arsenal and defense technologies possessed by other countries). One can model many such scenarios as a vector-valued sequential game between the agent and an adversary. 

In this paper, we analyze a simple class of such sequential games: two-player repeated games with vector-valued losses. These are two-player games in which a single-stage, simultaneous-move game with actions that result in vector-valued losses to one player is repeated many times. The player that incurs these losses wants to minimize them while the adversary wants to maximize them; however, since the losses are multi-dimensional, the relative importance of the different components matters in the choice of a good strategy for the minimizing player. 
%Focusing on the case where the losses are discounted over time, we describe an approximate dynamic programming (ADP) approach for approximating the lower Pareto frontier of the set of upper bounds on expected losses that the loss-minimizing player can simultaneously guarantee across the different dimensions. In other words, 

Focusing on the case where the losses are discounted over time, we describe an approximate dynamic programming (ADP) approach to calculate the best bounds on the losses that a payer can guarantee {\it simultaneously} in all the components. Formally, our approach approximates the lower Pareto frontier of the set of all points $\mathbf{b}\in\mathbb{R}^K$ (where $K$ is the dimension of the loss vector), such that the loss-minimizing player can guarantee that the expected losses in the game are contained in \emph{the lower corner set} $\{\bx\in \mathbb{R}^K: \bx\leq \mathbf{b}\}$.\footnote{An extension to general convex polytopes of the form $\{\bx\in \mathbb{R}^K: A\bx\leq \mathbf{b}\}$ follows by considering games with appropriate linear transformations of the vector losses.  The corresponding results in the single-dimensional case, i.e., where $A\bx\in \mathbb{R}$ (i.e., $K=1$), are well-known and we review them in Section~\ref{apx:scalar} in the Appendix, where we also discuss that it is not possible to characterize achievable simultaneous guarantees in vector repeated games by a reduction to the single-dimensional case.} We characterize this optimal Pareto frontier as the unique fixed point of a set-valued dynamic programming (DP) operator. %, which also simultaneously characterizes the strategies that achieve its different points. 
Using this characterization, we propose two computational schemes to derive good policies. The first scheme computes simple finite-state controllers that approximately achieve different points on the optimal frontier, but becomes prohibitive for larger values of $K$.  The second scheme addresses such higher-dimensional settings at the cost of loss in optimality. 
%and it relies on numerical solutions to a certain non-convex optimization problem of finding the best policy with a fixed budget of states.  

%\footnote{In this section, we also discuss that it is not possible to characterize achievable simultaneous guarantees in vector repeated games by a reduction to the single-dimensional case.}

%The Pareto frontier of the set of such bounds is shown to be the fixed point of set-valued dynamic programming equations that we use to construct policies that approximate that frontier.

The main application that motivates our study is the problem of regret minimization in repeated decision-making in adversarial environments, also known as {\it adversarial online learning} \cite{Cesa-Bianchi06a, hazan2016introduction, shalev2007online}.\footnote{Adversarial online learning is one extreme of the online learning framework in which the environment is assumed to be fully adversarial; there are several other models in this framework, which make assumptions that restrict the degree of adversarial behavior of the environment, resulting in less conservative and at the same time less robust decision-making algorithms.} In this problem, one wishes to find an adaptive policy that minimizes the incremental loss relative to the loss of the best fixed action that could have been chosen in hindsight, also known as the {\it regret}.  Algorithms exist that achieve, with high probability, an average regret that is $\textup{O}(1/\sqrt{T})$ over a fixed horizon $T$. In the infinite-horizon case with losses discounted over time by a factor $\beta \in (0,1)$, similar algorithms achieve an expected discounted average regret of $\textup{O}(\sqrt{1 - \beta})$ as $\beta \to 1$. Such algorithms are called {\it no-regret} algorithms, since the average regret vanishes as $T\rightarrow \infty$ or as $\beta\rightarrow 1$. However, the performance of such algorithms can be far from optimal for a fixed $T$ or $\beta$.  Our new approach addresses this shortcoming by constructing policies that are approximately optimal, under the assumption that the action spaces for the decision-maker and the adversary are finite.% in settings where the actions of the adversary lie in a finite set.

As an illustration, we use our approach to compute near-optimal algorithms for the well-known problem of prediction using expert advice with binary, discounted losses and 2 experts [16]. We show that these algorithms achieve lower regret guarantees than those achieved by existing benchmark policies in adversarial online learning, including the well-known Hedge algorithm (also known as the ``exponentially weighted average forecaster'' [16]). To the best of our knowledge, this is the first class of provably near-optimal algorithms for this setting. We also numerically evaluate our scheme for designing good policies for regret minimization in higher-dimensional settings, in which we continue to find that that these policies achieve lower regret guarantees compared to Hedge. Our results thus suggest the significant potential in using ADP approaches to designing effective adversarial online learning algorithms. 

%Finally, we address a practical concern that our approximation scheme becomes computationally prohibitive in higher dimensions. For such cases, we define the optimization problem of finding a regret-optimal finite-state stationary policy with a fixed budget of states. Although this problem is a non-convex optimization problem, we show that this problem can be efficiently solved to local optimality for reasonably-sized instances, yielding policies that, in many cases, achieve better performance guarantees than off-the-shelf online learning algorithms. Our results thus suggest a significant potential in using ADP approaches to designing effective adversarial online learning algorithms. 

\subsection{Organization of the paper}
 The paper is organized as follows. We first discuss related literature in Section~\ref{sec:related}. In Section~\ref{sec:model}, we formally introduce the model of repeated games with vector-values losses and define the objective of characterizing the set of losses that can be guaranteed by a player. We also discuss the connection of the problem to adversarial online learning.  In Section~\ref{sec:dynprog}, we introduce the set-valued dynamic programming approach for characterizing the optimal guarantees and the corresponding optimal strategies. In Section~\ref{sec:approx}, we present an approximation procedure to design near-optimal policies based on our characterization. Section~\ref{sec:heuristic} discusses a procedure to design good policies in higher dimensional settings. 
 %Section~\ref{sec:aumann} presents the application to zero-sum repeated games with incomplete information on one side. 
Section~\ref{sec:regmin} presents a numerical evaluation of the algorithms resulting from our approach in the context of regret minimization.
 %via the problem of prediction with expert advice. 
We discuss potential shortcomings of our approach in Section~\ref{sec:conclusion} and conclude the paper. The proofs of all our results are presented in Section~\ref{apx:proofs} the Appendix.

\section{Related literature}\label{sec:related}
{\bf Repeated games with vector losses.} \label{subsec:related}Blackwell \cite{Blackwell56a} pioneered the study of two-player vector-valued repeated games under the long-run average loss criterion. He described necessary and sufficient conditions for any convex set of loss vectors to be \emph{approachable} by a player, which means that there exists a strategy for the player that ensures that the long-run average loss approaches this set almost surely regardless of the adversary's actions. He also defined an adaptive randomized strategy that ensures this. 
Further, he proved the following remarkable ``minimax'' theorem: any convex set is either approachable by the player or is \emph{excludable} by the adversary, which means that the adversary has a strategy that ensures that the long-run average loss remains \emph{outside} this set almost surely. These initial results established the now well known ``approachability'' framework to analyze these games.

Approachability theory has developed significantly over the years; see \cite{Perchet14a} or \cite{laraki2015advances} for a survey. Necessary and sufficient conditions for approachability of general sets have been established in \cite{spinat2002necessary}. \cite{Vieille92a} considers a weaker notion called ``weak approachability'' and it is shown that every set is either weakly approachable or weakly excludable. There have also been several extensions of this framework beyond the setting of repeated games with finite action spaces, e.g., to stochastic games \cite{milman2006approachable}, to repeated games with payoffs in infinite-dimensional spaces \cite{Lehrer03a}, to repeated games with partial monitoring \cite{perchet2011internal,perchet2011approachability,perchet2014unified}, etc.

Despite these advances in our understanding of vector repeated games, the characterization and computation of loss vectors that can be guaranteed in expectation under the discounted loss criterion and the strategies that achieve these guarantees has remained a significant gap. Discounting of losses over time has natural interpretations in practice, e.g., it may capture a low-risk rate of return on investment. Hence, the closing of this gap in our work has significant practical ramifications. %This paper takes a major step towards bridging this gap using an ADP approach. 
{\bf Set-valued dynamic programming.} A well-known use of set-valued dynamic programs in the context of dynamic games is due to Abreu, Pearce, and Stacchetti \cite{Abreu86a,Abreu90a}. They characterize the set of pure strategy subgame-perfect equilibrium payoffs in non-zero sum repeated games with imperfect monitoring as the fixed point of a set-valued DP operator. Although we have a similar fixed point characterization of the optimal Pareto frontier, there are important differences in the machinery involved in iterative computation of this fixed point. Our ADP approach and related error bounds critically rely on the new metric that we define on the space of Pareto frontiers of convex and compact sets and on the fact that our DP operator is a contraction in this metric. On the other hand, their iterative computation scheme relies on the monotonicity of the DP operator;\footnote{A set-valued operator $\mathcal{B}$ is monotone if $A\subseteq A'$ implies that $\mathcal{B}(A)\subseteq \mathcal{B}(A')$ \cite{Abreu90a}} and because they don't define a metric space, they do not obtain error bounds. Moreover, theirs is an exact scheme akin to `value iteration' in DP \cite{Bertsekas05b} and implementing it in practice would require defining appropriate set-approximations that are finitely parameterized. %This non-trivial issue has not been addressed in these works. 
Our carefully defined polytopic set approximation scheme squarely addresses this issue in our setting.
%The former characterization relies on a monotonicity property of the operator, which obviates the necessity of defining a metric, while our characterization relies on the fact that the operator is a contraction in an appropriate metric.  

For the use of dynamic programming in zero-sum dynamic games, one can refer to the classic paper by Shapley \cite{shapley1953stochastic} on stochastic games. For a general theory of dynamic programming in control problems and Markov decision processes (MDPs), see \cite{Bertsekas05a, Bertsekas05b, puterman2014markov}. %We discuss relevant literature on regret minimization and online learning in Section~\ref{sec:lit-OL}.

{\bf Regret minimization in repeated games.} The first study of regret minimization in repeated games dates back to the pioneering work of Hannan \cite{Hannan57a}, who introduced the notion of regret optimality in repeated games and proposed the earliest known no-regret algorithm. Since then, numerous other such algorithms have been proposed, particularly for the problem of prediction using expert advice, see \cite{Littlestone94a,Vovk90a,Cesa-Bianchi97a,Freund99a}, one particularly well-known class being the Multiplicative Weights update class of algorithms. Regret minimization with discounted losses has been considered before in \cite{Cesa-Bianchi06a,chernov2010prediction,Perchet14a}. Other settings with limited feedback have been considered, most notably the multi-armed bandit setting \cite{Auer02a,Bubeck12a}. Stronger notions of regret such as internal regret, have also been studied \cite{Foster97a,Cesa-Bianchi03a,blum2007external,Stoltz05a}. \cite{Cesa-Bianchi06a} offers a fairly comprehensive survey of the literature.

%While the average regret goes to zero if the weights satisfy a non-summability condition, lower bounds exist (\cite{Cesa-Bianchi06a}, Thm 2.7) that show that the optimal regret in the infinitely repeated game is bounded away from $0$ if the weights are summable, which is the case with discounting. 
%Natural extensions of no-regret algorithms incur a cumulative regret of $\textup{O}(1/\sqrt{1-\beta})$ (and hence an average regret of $\textup{O}(1/\sqrt{1-\beta})(1-\beta)= \textup{O}(\sqrt{1-\beta})$) in this case; for instance see Thm. 2.8 in \cite{Cesa-Bianchi06a} and Prop. 2.22 in \cite{Perchet14a}. %\cite{chernov2010prediction} considers the case of reverse discounting where future losses are given a higher weight that current ones.
%But such a weighting goes against the notion of time value of money.

The results on exact regret minimization are few. In an early work, Cover \cite{cover1966behavior} gave the optimal algorithm for the problem of prediction using expert advice over any finite horizon $T$, for the case of $2$ experts, and where the losses are $\{0,1\}$. Recently, \cite{gravin2014towards} extended the result to the case of $3$ experts for both the finite horizon and geometrically distributed random horizon problems. Although a geometric time horizon model appears to be related to the infinite horizon model with discounted losses, the two problem formulations define regret differently, and thus lead to different optimal regrets. We discuss this in Section \ref{apx:gps2} in the Appendix. \cite{abernethyoptimal} considers a related problem, where a gambler places bets from a finite budget repeatedly on a fixed menu of events, the outcomes of which are adversarially chosen from $\{0,1\}$ (you win or you lose), and characterizes the minimax optimal strategies for the gambler and the adversary. \cite{luo2013towards} considers a similar repeated decision-making problem where an adversary is restricted to pick loss vectors (i.e., a loss for each action of the decision-maker in a stage) from a set of basis vectors, and characterizes the minimax optimal strategy for the decision-maker under both, a fixed and an unknown horizon. 
Most of the approaches in these works are specific to their settings, and exploit the assumptions on the structure of the loss vectors. 
%In particular, many of these works rely on a particular nice property specific to these settings, which is that the optimal strategy of the adversary is a controlled random walk. If the losses are simple, for instance if they are the unit basis vectors, then this random walk can be exactly analyzed to compute the optimal regret. Knowing the optimal regret then simplifies the computation of the optimal strategy of the decision-maker \cite{gravin2014towards,luo2013towards,abernethyoptimal}. 
But if the loss vectors are arbitrary, these approaches are difficult to generalize and it is generally recognized that characterizing the optimal regret and algorithm is difficult \cite{luo2013towards}.

Finally, all of the above examples deal with games with finite action spaces, which is the setting that we are concerned with. But there are many works that consider exact minimax optimality in repeated games with general action sets, with specific types of loss functions; see \cite{koolen2014efficient,bartlett2015minimax,koolen2015minimax} and references therein.

\section{Model.}\label{sec:model}
For the remainder of the paper, $\mathbf{1}$ and $\mathbf{0}$ denote the the vector of ones and zeros, respectively, in $\mathbb{R}^K$.

Consider a two-player vector-valued game $\mathbb{G}$ defined by an action set $A=\{1,\cdots,l\}$ for player 1, who is the decision-maker and whom we will call Alice, and the action set $B=\{1,\cdots,m\}$ for player 2 who is the adversary and whom we will call Bob. For each pair of actions $a\in A$ and $b\in B$, Alice incurs a vector-valued loss $\mathbf{r}(a, b) \in \RR^K$. 
%For simplicity, we restrict the discussion to the case where $K=2$, i.e. $\mathbf{r}(a,b)=\big(r_1(a,b),r_2(a,b)\big)$, although the results hold for any finite $K$.  

The game $\mathbb{G}$ is played repeatedly in stages $t=1,2,3,\cdots, T$. Let $\mathbb{G}^T$ denote this $T$-stage repeated game. In each stage $t$, both Alice and Bob simultaneously pick their actions $a_t$ and $b_t$ respectively, and Alice bears the vector of losses $\mathbf{r}(a_t,b_t)$. Fix a discount factor $\beta\in[0,1)$. Then the vector of total discounted losses is defined as:
\begin{equation}
\sum_{t=1}^{T}\beta^{t-1}\mathbf{r}(a_t,b_t)=\bigg(\sum_{t=1}^{T}\beta^{t-1}r_k(a_t,b_t);\, k=1,\cdots,K\bigg).
\end{equation}

An adaptive randomized strategy $\pi_A$ for Alice specifies for each stage $t$, a mapping from the set of observations till stage $t$, i.e., $H_t=(a_1,b_1,\cdots,a_{t-1},b_{t-1})$, to a probability distribution on the action set $A$, denoted by $\Delta(A)$. Let $\Pi_A$ be the set of all such strategies of Alice. Similarly, let $\Pi_B$ be the set of all adaptive randomized strategies for Bob. For a pair of strategies $\pi_A$ and $\pi_B$, the expected discounted loss on component $k$ in the repeated game is given by:
\begin{equation}
R^T_k(\pi_A,\pi_B)=\textup{E}_{\pi_A,\pi_B}\bigg[\sum_{t=1}^T \beta^{t-1}r_k(a_t,b_t)\bigg],
\end{equation} 
where the expectation is over the randomness in the strategies $\pi_A$ and $\pi_B$. Alice would like to minimize her loss in every component $k$.  However, reducing the loss in one dimension typically implies increasing the loss in another dimension. For instance, consider the situation of protecting two different targets against attacks: devoting more resources to protect one target makes the other more vulnerable. Accordingly, it is important to characterize the set of best possible tradeoffs between the different dimensions of the loss.  

Consider a fixed strategy $\pi_A\in \Pi_A$. If Alice plays this strategy, then irrespective of the strategy chosen by Bob, Alice guarantees that the long term expected vector loss is no larger than
$$\bigg(\max_{\pi^k_B\in \Pi_B}R^T_k(\pi_A,\pi^k_B);\, k=1,\cdots,K \bigg)$$
along each dimension. Let the set of all such \emph{simultaneous upper bounds} that correspond to \emph{all} the strategies $\pi_A\in \Pi_A$ be defined as:
\begin{equation}
\cW^T \triangleq\bigg\{\bigg(\max_{\pi^k_B\in \Pi_B}R^T_k(\pi_A,\pi^k_B);\, k=1,\cdots,K \bigg): \pi_A\in \Pi_A\bigg\}.\label{eq:wt}
\end{equation}
Then characterizing the best possible tradeoffs across the different dimensions amounts to finding the \emph{minimal} points in the set $\cW$, i.e., its \emph{lower Pareto frontier}, which is the set 
\begin{equation}
\label{eq.Ustar}
\cV^T\triangleq\Lambda(\cW^T ) \triangleq\{\bx \in \cW^T : \forall\, \bx'  \in \cW^T\setminus\{\bx\}, \, \exists\, k  \textrm{ s.t. }x_k < x'_k\}, 
\end{equation}
since all other points are strictly sub-optimal. Our goal in this paper is to characterize and approximate the set $\cV^{\infty}$  that can be achieved in the infinite horizon game $\mathbb{G}^\infty$ and compute approximately optimal strategies for Alice in $\Pi_A$ that approximately guarantee different points in it. 

\subsection{Application to adversarial online learning}
As an application of our model and objective, we now describe how they lead to the solution to the problem of regret minimization in repeated games in the framework of adversarial online learning. This conclusion follows from the transformation of the regret minimization problem into a vector-valued repeated game that we describe below. 

We first define the problem of regret minimization in repeated games. Let $G$ be a two player game with $l$ actions  $A =\{1,\ldots,l\}$ for Alice (the decision-maker), who is assumed to be the minimizer, and $m$ actions $B =\{1,\ldots,m\}$ for Bob (the adversary), in keeping with the previous notation. For each pair of actions $a\in A$ and $b\in B$, the corresponding loss for Alice is $L(a,b)\in \mathbb{R}$. 
%The losses for different pairs of actions are known to Alice. 
The game $G$ is played repeatedly for $T$ stages $t=1,2,\cdots, T$. In each stage, both Alice and Bob simultaneously pick their actions $a_t\in A$ and $b_t\in B$ and Alice incurs the corresponding loss $L(a_t,b_t)$. 
%At the end of the stage, the action chosen by Bob becomes known to Alice. 
The loss of the repeated game is defined to be the total discounted loss given by $\sum_{t=1}^{T}\beta^{t-1}L(a_t,b_t)$, where $\beta\in (0,1)$. We define the total discounted regret of Alice as:
\begin{equation}
\sum_{t=1}^T\beta^{t-1}L(a_t,b_t) - \min_{a\in A}\sum_{t=1}^T\beta^{t-1}L(a,b_t),
\end{equation}
which is the difference between her actual discounted loss and the loss corresponding to the single best action that could have been chosen against the sequence of actions chosen by Bob in hindsight. 

Alice chooses an adaptive randomized strategy $\pi_A\in\Pi_A$. 
% specifies for each stage $t$, a mapping from the set of observations till stage $t$, i.e., $H_t=(a_1,b_1,\cdots,a_{t-1},b_{t-1})$, to a probability distribution on the action set $A$, denoted by $\Delta(A)$. Let $\Pi_A$ be the set of all such strategies of Alice. 
Bob is assumed to choose a deterministic oblivious strategy, i.e., his choice is simply a sequence of actions $\mathbf{b}=(b_1,b_2,b_3,\cdots,b_T)\in B^T$ chosen before the start of the game.\footnote{A deterministic, oblivious adversary is a standard assumption in regret-minimization literature~\cite{Auer02a,Bubeck12a,Cesa-Bianchi06a}. This makes sense when ``nature'' is modeled as an adversary in an application, which would be the case for instance in weather forecasting. Also see Chapter 3 in \cite{Bubeck12a} for a discussion of different adversary models and their implications on different definitions of regret.} %Let $\Pi_B$ be the set of all such sequences. %For fixed strategies $\pi^A$ and $\pi^B$ of Alice and Bob, the expected discounted regret is given by:
%\begin{equation}\label{def:regret}
%\mathcal{R}(\Pi_A, \Pi_B)=\textup{E}_{\pi_A,\pi_B}\bigg[\max_{a\in A} \bigg(\sum_{t=1}^\infty\beta^{t-1}(L(a_t,b_t) -L(a,b_t))\bigg)\bigg],
%\end{equation}
%where the expectation is over the randomization in the strategies $\pi_A$ and $\pi_B$. 
Alice's problem of minimizing her worst-case expected discounted regret of Alice is defined as:
\begin{subequations}
\begin{align}
&\min_{\pi_A\in\Pi_A}\max_{\mathbf{b}\in B^T}\textup{E}_{\pi_A}\bigg[\sum_{t=1}^T\beta^{t-1}L(a_t,b_t) -\min_{a\in A}\sum_{t=1}^T\beta^{t-1}L(a,b_t)\bigg]\nonumber\\
%&=\min_{\pi_A\in\Pi_A}\max_{\pi_B\in \Pi_B}\textup{E}_{\pi_A}\bigg[\max_{a\in A}\bigg(\sum_{t=1}^T\beta^{t-1}L(a_t,b_t) -\sum_{t=1}^T\beta^{t-1}L(a,b_t)\bigg)\bigg]\\
&~~=\min_{\pi_A\in\Pi_A}\max_{\mathbf{b}\in B^T}\max_{a\in A}\textup{E}_{\pi_A}\bigg[\sum_{t=1}^T\beta^{t-1}(L(a_t,b_t) -L(a,b_t))\bigg]\label{pseudo}\\
&~~\overset{(a)}{=}\min_{\pi_A\in\Pi_A}\max_{\pi_B\in \Pi_B}\max_{a\in A}\textup{E}_{\pi_A, \pi_B}\bigg[\sum_{t=1}^T\beta^{t-1}(L(a_t,b_t) -L(a,b_t))\bigg]\bigg]\label{pseudo2}.
\end{align}
\end{subequations}
%and the strategy $\pi^*_A$ for Alice that guarantees this value. 
%\footnote{Here one can see that there is no loss of generality in assuming that the adversary is deterministic. Indeed if $\Pi_B$ is allowed to be the set of possible randomizations over T-length sequences of Bob's actions, the optimal strategy of Bob in the problem
%\begin{equation*}
%\max_{\pi_B\in \Pi_B}\textup{E}_{\pi_A,\pi_B}\bigg[\sum_{t=1}^T\beta^{t-1}L(a_t,b_t) -\min_{a\in A}\sum_{t=1}^T\beta^{t-1}L(a,b_t)\bigg]
%\end{equation*}
%s a deterministic sequence (see Proposition $33$ in \cite{audibert2010regret}).} 
%Since Bob's strategy $\pi_B$ is deterministic and oblivious, the expectation in \eqref{eqn:regretdef} is effectively only over the randomness in Alice's strategy, and moreover the second term does not depend on Alice's strategy. Thus we can exchange the expectation and the inner maximization and equivalently write~\eqref{eqn:regretdef} as:
%\begin{equation}
%\min_{\pi_A\in\Pi_A}\max_{\pi_B\in \Pi_B}\max_{a\in A}\textup{E}_{\pi_A, \pi_B}\bigg[\sum_{t=1}^T\beta^{t-1}(L(a_t,b_t) -L(a,b_t))\bigg].
%\end{equation}
%This quantity is called the \emph{pseudo-regret} in regret minimization literature (see Chapter 3 in \cite{Bubeck12a}). 
%The expected regret is equivalent to the pseudo-regret only when the adversary's strategy is deterministic and oblivious; in general if the adversary's strategy is randomized and adaptive, then the pseudo-regret is smaller than the expected regret.
Equality (a) says that in problem~\eqref{pseudo}, there is no loss to Alice if Bob is allowed to choose any adaptive randomized strategy $\pi_B\in\Pi_B$ instead of restricting him to choosing a deterministic oblivious strategy. This is because one can show that Alice's optimal strategy in problem \eqref{pseudo} need not depend on her own past actions; see Lemma~\ref{lma:box} and Theorem~\ref{thm:optstrategy} below. Hence, strategies in $\Pi^B$ are not more powerful than those in $B^T$ against the optimal strategy of Alice. The problem \eqref{pseudo2} is called the problem of minimizing the {\it pseudo-regret} in the online learning literature (see Chapter 3 in \cite{Bubeck12a}). As we saw, it is equivalent to minimizing the expected regret when the adversary is restricted to $B^T$. 

Now the connection to our original vector-valued repeated game is straightforward. Define a vector-valued game $\mathbb{G}$, in which, for a pair of actions $a\in A$ and $b\in B$, the vector of losses is $\br(a,b)$ with $K=l$ components (recall that $|A|=l$), where $r_k(a,b)=L(a,b)-L(k,b)$. $r_k(a,b)$ is the single-stage additional loss that Alice bears by choosing action $a$ instead of action $k$, when Bob chooses $b$: referred to as the ``single-stage regret'' with respect to action $k$.
%\begin{lemma}
%$$\min_{\pi_A\in\Pi_A}\max_{\pi_B,\in \Pi_B}\textup{E}_{\pi_A,\pi_B}\bigg[\max_{a\in A} %(1-\beta)\bigg(\sum_{t=1}^\infty\beta^{t-1}(L(a_t,b_t) -L(a,b_t))\bigg)\bigg]$$
%$$=\min_{\pi_A\in\Pi_A}\max_{\pi_B,\in \Pi_B}\max_{a\in A}\textup{E}_{\pi_A,\pi_B}\bigg[ %(1-\beta)\bigg(\sum_{t=1}^\infty\beta^{t-1}(L(a_t,b_t) -L(a,b_t))\bigg)\bigg]$$
%\end{lemma}
%\begin{proof}
%$$\min_{\pi_A\in\Pi_A}\max_{\pi_B,\in \Pi_B}\textup{E}_{\pi_A,\pi_B}\bigg[\max_{a\in A} %(1-\beta)\bigg(\sum_{t=1}^\infty\beta^{t-1}(L(a_t,b_t) -L(a,b_t))\bigg)\bigg]$$
%$$=\min_{\pi_A\in\Pi_A}\max_{\pi_B,\in \Pi_B}(1-\beta) \bigg(\sum_{t=1}^\infty\beta^{t-1}\textup{E}_{\pi_A,%\pi_B}[L(a_t,b_t)] -\textup{E}_{\pi_B}[\min_{a\in A}\sum_{t=1}^\infty\beta^{t-1}L(a,b_t)]\bigg)$$
%\end{proof}
%For $\pi_A\in \Pi_A$ and $\mathbf{b}\in B^T$, the expected loss on component $k$ in this vector-valued repeated game over horizon $T$ is given by 
%\begin{eqnarray}
%R^T_k(\pi_A,\mathbf{b})&=& \textup{E}_{\pi_A}\bigg[\sum_{t=1}^T \beta^{t-1}r_k(a_t,b_t)\bigg].
%\end{eqnarray}
%where the expectation is over the randomness in Alice's strategy. 
%By playing a fixed strategy $\pi_A\in \Pi_A$, irrespective of the strategy chosen by Bob, Alice guarantees that the total expected loss on component $k$ is no more than $\max_{\mathbf{b}_k\in B^T}R^T_k(\pi_A,\mathbf{b}_k)$. 
%Suppose that we determine the set of all \emph{simultaneous upper bounds} that correspond to \emph{all} the strategies $\pi_A\in \Pi_A$, defined as:
Defining the set $\cW^T$ as in \eqref{eq:wt} for the repeated game  $\mathbb{G}^T$, it is clear that the minimax optimal regret can be written as:
%\begin{equation}
%\cW^T \triangleq\bigg\{\bigg(\max_{\mathbf{b}_k\in B^T}R^T_k(\pi_A,\mathbf{b}_k) \bigg)_{k=1,\cdots,m}: \pi_A\in \Pi_A\bigg\},
%\end{equation}
%which is the set of all \emph{simultaneous upper bounds} that correspond to \emph{all} the strategies $\pi_A\in \Pi_A$.
$$\min_{\pi_A\in\Pi_A}\max_{\pi_B\in \Pi_B}\max_{a\in A}\textup{E}_{\pi_A}\bigg[ \sum_{t=1}^T\beta^{t-1}(L(a_t,b_t) -L(a,b_t))\bigg]=\min_{\bx\in \cW^T}\max_k x_k.$$
Hence, to compute the minmax optimal regret, it suffices to compute $\cW^T$. In fact, it suffices to compute its lower Pareto frontier $\cV^{T}$, and the strategies that achieve this frontier, since all other points are strictly sub-optimal. %In particular, we are interested in $\Lambda(\cW^{\infty})$. 

\section{Set-valued dynamic programming.}\label{sec:dynprog}
%\subsection{Overview of the approach}\label{subsec:overview}
We first present an informal description of our approach. %for characterizing and approximating $\cV^{\infty}$. 
Let $\cV^0=\{\mathbf{0}\}$, i.e., the singleton set containing only the zero vector in $\mathbb{R}^K$. We can show that one can obtain the set $\mathcal{V}^{T+1}$ from the set $\mathcal{V}^{T}$, by decomposing Alice's strategy in $\mathbb{G}^{T+1}$ into a strategy for the 1st stage, and a continuation strategy for the remainder of the game from stage $2$ onwards as a function of the action chosen by both the players in the 1st stage. The inductive argument results from the fact that the minimal guarantees that she can guarantee from stage 2 onwards are exactly the set $\mathcal{V}^T$. Suppose that at the start of $\mathbb{G}^{T+1}$, Alice fixes the following plan for the entire game: she will play a mixed strategy $\bm{\alpha}\in \Delta(A)$ in stage 1. Then depending on her realized action $a$ and Bob's action $b$, from stage 2 onwards she will play a continuation strategy that achieves the upper-bound $\bR(a,b)\in \mathcal{V}^T$ (she will choose one such point $\bR(a,b)$ for every $a\in A$ and $b\in B$). Note that it is strictly sub-optimal for Alice to choose any points outside $\mathcal{V}^T$ from stage $2$ onwards. Now this plan for the entire game $\mathbb{G}^{T+1}$ gives Alice the following simultaneous upper bounds on the expected losses on the $K$ dimensions:
 
$$\bigg(\max_{b\in B}\sum_{a\in A}\alpha_a\big[r_k(a,b)+\beta R_k(a,b)\big];k=1,\cdots, K\bigg).$$
%\footnote{Note that here we are implicitly assuming that Bob's actions from stage 2 onwards can depend on Alice's action in stage 1, which is not possible if Bob is assumed to follow an oblivious strategy. But we will show later that this relaxation does not make any difference: the optimal set of guarantees remains the same.}.$$
By varying the choice of $\bm{\alpha}$ and the map $\bR(a,b)$ we can obtain the set of all the simultaneous upper bounds that Alice can achieve in the $(T+1)$-stage game. The lower Pareto frontier of this set is exactly $\cV^{T+1}$. Thus there is an operator $\Phi$, such that 
$$\mathcal{V}^{T+1}=\Phi(\mathcal{V}^T)$$
for any $T\geq 0$. In what follows, we will show that this operator is a contraction in the space of lower Pareto frontiers of compact and convex sets, with an appropriately defined metric. This space is shown to be complete, and thus the sequence $\mathcal{V}^T$ converges in the metric to a set $\mathcal{V}^*$, which is the unique fixed point of this operator $\Phi$. As one would guess, this $\mathcal{V}^*$ is indeed the set $\cV^{\infty}$ of minimal simultaneous upper bounds that Alice can achieve in the infinitely repeated game $\mathbb{G}^{\infty}$. 

The rest of this section formalizes these arguments. We will begin the formal presentation of our results by first defining the space of Pareto frontiers that we will be working with.
\subsection{A space of Pareto frontiers in $[0,1]^K$}
We work with the following basic definitions.
%Consider the box set in $\mathbf{R}^K$, $\Omega=\{(x,y): x_1\leq x\leq x_2$ and $y_1\leq y\leq y_2\}$. And consider the set of all subsets of $\Omega$. Define the following metric on this set:

%For two subsets $\cA$ and $\cB$ of $\Omega$, we say that $d(\cA,\cB)\leq \epsilon$ if 
%\begin{eqnarray}
%\forall U\in \cA, \, \exists\, V \in \cB& \textrm{ s.t. } &V\preceq U+ \epsilon\mathbf{1} \textrm{ and }\\
%\forall V\in \cB, \, \exists\, U \in \cA& \textrm{ s.t. }&U\preceq V+ \epsilon\mathbf{1}. 
%\end{eqnarray}
%Here $\preceq$ denotes a component-wise inequality and $\mathbf{1}$ is a vector of  $1$s.
%For the remainder of the paper, $\mathbf{1}$ and $\mathbf{0}$ denote the the vector of ones and zeros, respectively, in $\mathbb{R}^K$.
\begin{definition}
(a) Let $\bu, \bv \in \RR^K$.  We say that $\bu \preceq \bv$ if $u_k \leq v_k$ for all $k$.
Also, we say that $\bu \prec \bv$ if $\bu \preceq \bv$ and $\bu \neq \bv$.  For some $\epsilon\geq 0$ if $\bu \preceq \bv + \epsilon \mathbf{1}$, we say that $\bv$ {\em $\epsilon$-dominates} $\bu$. If $\epsilon = 0$, we simply say that $\bv$ dominates $\bu$.

(b) A {\em Pareto frontier} in $[0, 1]^K$ is a subset $\cV$ of $[0, 1]^K$
such that no $\bv \in \cV$ is dominated by another element of $\cV$.

%(c) The {\em Upper Pareto frontier} 
%of $\cS  \subset [0, 1]^K$ is the set of elements of $\cS $ that are
%not dominated by another element of $\cS $
(c) For two Pareto Frontiers $\cU$ and $\cV$, we say that $\cV$ $\epsilon$-dominates $\cU$ if for every point $\bv\in\cV$, there is a point $\bu\in \cU$ that it $\epsilon$-dominates. If $\epsilon = 0$, then we simply say that $\cV$ dominates $\cU$.

(d) The {\em lower Pareto frontier} (or simply  {\em Pareto frontier})
of $\cS  \subset [0, 1]^K$, denoted by $\Lambda(\cS)$, is the set of elements of $\cS $ that do not dominate
any another element of $\cS $.
\end{definition}
%Figure~\ref{fig:lpf} shows the lower Pareto frontiers of some sets in $[0,1]^K$:
%\begin{figure}[htb]
%\begin{center}
%\includegraphics[width=2.5in,angle=0]{LPF.pdf}
%\caption{Lower Pareto frontiers of some sets in  $[0,1]^K$.}
%\label{fig:lpf}
%\end{center}
%\end{figure} 

The Pareto frontier of a set may be empty, as is certainly the case when the set is open.  But one can show that the Pareto frontier of a non-empty compact set is always non-empty. %The proof is presented in Section~\ref{apx:proofs} in the Appendix.

\begin{figure}[h]
\centering
\begin{minipage}[t]{.4\textwidth}
\centering
\includegraphics[width=2.5in,angle=0]{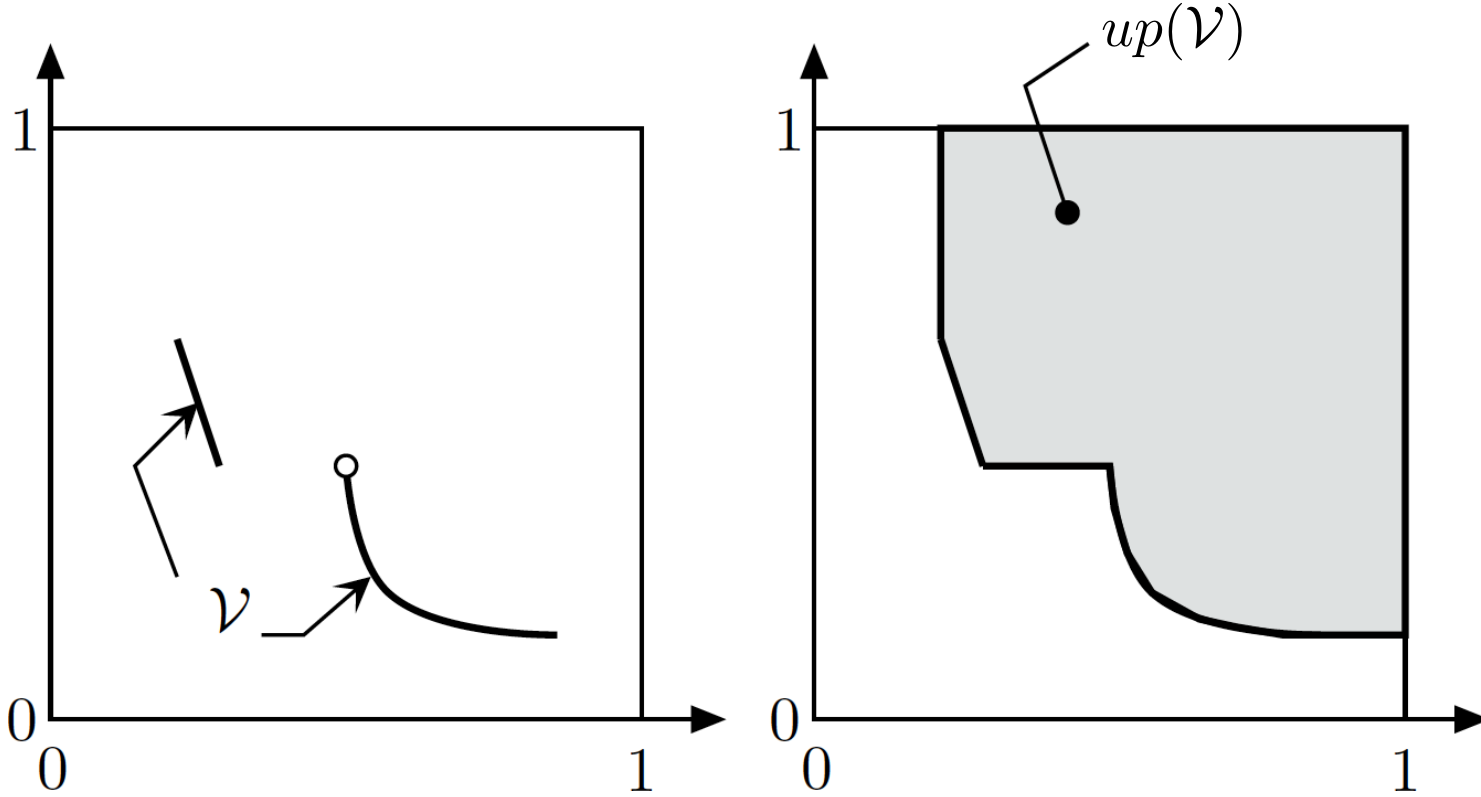}
\caption{A Pareto frontier $\cV$ and its upset $\bcV$ in $[0,1]^2$.}
\label{fig:upset}
\end{minipage}\hfill
\begin{minipage}[t]{.5\textwidth}
\centering
\includegraphics[width=1.8in]{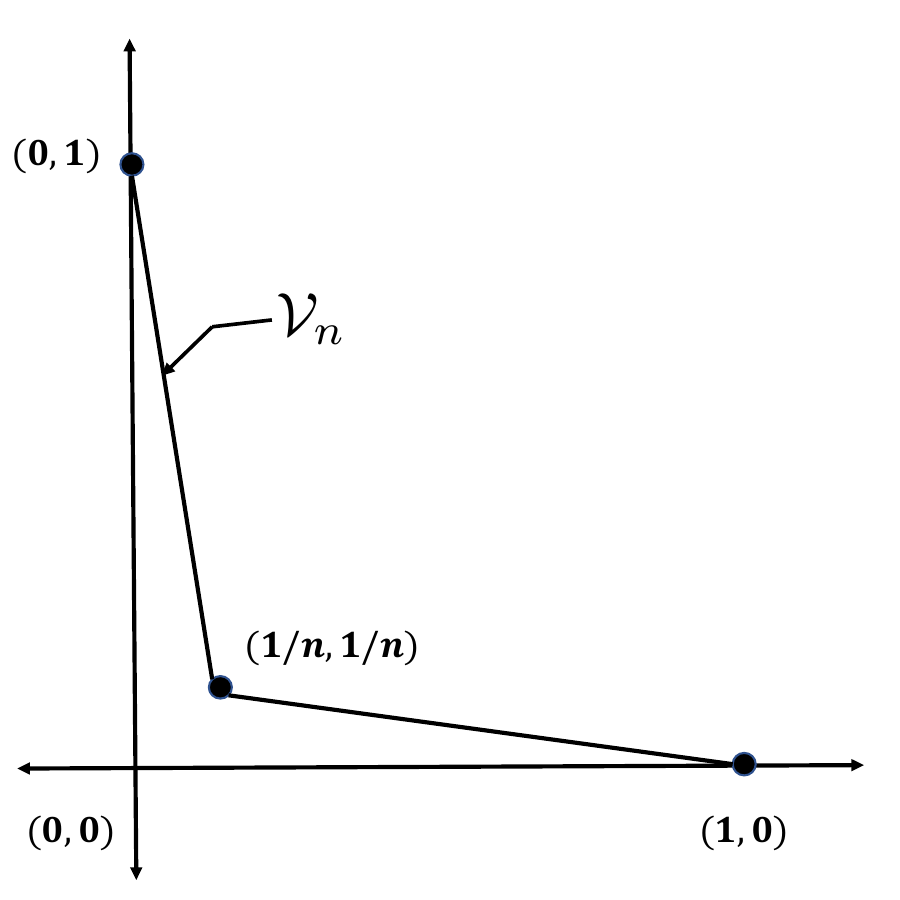}
\caption{Approximations of $(1-\beta)\mathcal{V}^*$ for different $\beta$ values with corresponding errors.}\label{fig:metric}
\end{minipage}
\end{figure}

\begin{lemma}\label{lma:comp}
Suppose that $\mathcal{S}$ is a non-empty compact subset of $\mathbb{R}^K$. Then $\Lambda(\mathcal{S})$ is non-empty.
\end{lemma}
Since compactness is equivalent to a set being closed and bounded in Euclidean spaces, any closed subset of $[0,1]^K$ has a non-empty Pareto frontier.  We next define the {\em upset} of a set, illustrated in Figure \ref{fig:upset}.

\begin{definition}
Let $\cA$ be a subset of $\cB\subseteq\mathbb{R}^K$.  The {\em upset} of $\cA$ in $\cB$ is defined as $\bcA = \{\bx \in \cB \mid x_k \geq y_k \mbox{ for all } k, \mbox{ for some } \by \in \cA\}$, i.e., $\bcA$ is the set of all points in $\cB$ that dominate some point in $\cA$. Equivalently, $\bcA = \{\bx \in \cB \mid \bx = \by + \bv, \mbox{ for some } \by \in \cA \mbox{ and } \bv \succeq \mathbf{0}\}$. 
\end{definition}

For a subset of $[0,1]^K$, we will refer to its upset in $[0,1]^K$ as simply its upset. It is immediate that the upset of a closed and convex subset of $[0,1]^K$ is closed and convex. 
%The following result is useful, and we state is here without a proof.
%\begin{lemma}\label{lma:comp}
%Suppose that $S$ is a compact subset of $\mathbb{R}^K$. Then $\Lambda(S)$ is non-empty.
%\end{lemma}
%Note that a set in $\mathbb{R}^K$ is compact iff it is closed and bounded. 
We define the following space of Pareto frontiers.

\begin{definition}
 $\cF$ is the space of Pareto frontiers in $[0,1]^K$ whose upset is closed and convex. 
 \end{definition}
 %It is immediate that $d$ is a metric on $\cF$. 
%We first show the equivalence of the two metrics:
It is easy to show that $\cF$ can be equivalently defined as the space of lower Pareto frontiers of closed and convex subsets of $[0,1]^K$.\footnote{One direction is clear since a Pareto frontier in $\cF$ is the lower Pareto frontier of its upset, which is closed and convex. The other direction follows from the observation that the upset of the lower Pareto frontier of a set is the upset of the set itself and the upset of a closed and convex set is closed and convex.} We will now define a metric on this space. We first recall the definition of {\em Hausdorff} distance induced by the $\mathcal{L}^\infty$ norm.

\begin{definition}
Let $\cA$ and $\cB$ be two subsets of $\RR^K$.  The Hausdorff distance
$h(\cA, \cB)$ between the two sets is defined as
\[
h(\cA, \cB) = \max\{\sup_{\bx \in \cA} \inf_{\by \in \cB} ||\bx - \by||_{\infty}, \,\sup_{\by \in \cB} \inf_{\bx \in \cA} ||\bx - \by||_{\infty} \}.
\]

\end{definition}
The Hausdorff distance defines a metric on the space of non-empty closed subsets of $[0,1]^K$, and further, this space is compact and hence complete in this metric \cite{henrikson1999completeness}. On the other hand, it only defines a pseudometric on space of all non-empty subsets of $[0,1]^K$.
Now, a possible straightforward metric on the space $\mathcal{F}$ could be the one defined by the Hausdorff distance. But, as we discuss in Section~\ref{sec:convclo} in the Appendix, if $K>2$, then a Pareto frontier in $\cF$ may not be closed, and hence the Hausdorff distance at best defines a pseudometric on $\cF$. Moreover, even this pseudometric is not appropriate for our purposes as demonstrated by the following example.\\
{\bf Example:} Consider a sequence of Pareto frontiers $(\mathcal{V}_n)_{n\in\mathbb{N}}$ where $\mathcal{V}_n$ is the union of the line segment joining $(0,1)$ and $(1/n,1/n)$, and the segment joining $(1/n,1/n)$ and $(1,0)$, as depicted in Figure~\ref{fig:metric}. Then we would like this sequence of frontiers to converge to the Pareto frontier defined by the singleton set $\{(0,0)\}$, but the Hausdorff distance between $\mathcal{V}_n$ and $\{(0,0)\}$ doesn't vanish as $n\rightarrow\infty$. Under the Hausdorff metric, the sequence $(\mathcal{V}_n)_{n\in\mathbb{N}}$ converges to the union of the line segment joining $(0,1)$ and $(0,0)$, and the segment joining $(0,0)$ and $(1,0)$, which is not in $\mathcal{F}$.
%thus $\mathcal{F}$ is not closed in this metric.

It is thus clear that we need to define a different metric on $\mathcal{F}$. We now proceed to define one with the desired properties. We define the distance between two Pareto frontiers in $\cF$ as the Hausdorff distance between their upsets.\footnote{Since the upsets of the sets in $\cF$ are compact, the $\sup$ and the $\inf$ in the definition of the Hausdorff distance can be replaced by $\min$ and $\max$ respectively.\label{fn:hausdorff}}
\begin{definition}\label{def:metric1}
For two Pareto frontiers $\cU$ and $\cV$ in $\cF$, we define the distance $d(\cU,\cV)$ between them as
$d(\cU,\cV)\triangleq h(\bcU,\bcV)$.
\end{definition}
We can then show that $d$ is a metric on $\cF$, and $\cF$ is compact in the metric $d$. The latter essentially follows from the compactness of the space of closed subsets of $[0,1]^K$ in the Hausdorff metric. It also immediately follows that $\cF$ is complete. 
%The proof of the following set of claims is presented in Section~\ref{apx:proofs} in the Appendix.

\begin{proposition}\label{lma:complete}
a) $d$ is a metric on $\cF$.\\
b) Let $\big(\cV_n\big)_{n\in\mathbb{N}}$ be a sequence in $\cF$. Then there is a subsequence $\big(\cV_{n_k}\big)_{k\in\mathbb{N}}$ and a $\cV\in \cF$ such that $d(\cV_{n_k},\cV)\rightarrow 0$.
%c) Let $\big(\cV_n\big)_{n\in\mathbb{N}}$ be a sequence in $\cF$. Suppose that $\sup_{m,k>n}d(\cV_m,\cV_k)\rightarrow 0$. Then there exists a unique $\cV  \in \cF$  such that $d(\cV_n,\cV)\rightarrow 0.$
\end{proposition} 
In the proof, it becomes clear that $d$ induces these properties not just on $\cF$, but also on the more general space of Pareto frontiers in $[0,1]^K$ whose upset is closed (though not necessarily convex). 
Finally, we end this section by presenting another way of defining the same metric $d$ on $\cF$.  
\begin{definition}
For two Pareto frontiers $\cV$ and $\cU$ in $\cF$, define
\begin{equation}
e(\cU,\cV)\triangleq\inf\{\epsilon\geq 0: \forall \,\,\bu\in \mathcal{U}, \, \exists\, \bv \in \mathcal{V} \textrm{ s.t. } \bv\preceq \bu+ \epsilon\mathbf{1}\}.
\end{equation}
\end{definition}
In other words, $e(\cU,\cV)$ is the smallest $\epsilon\geq 0$ such that $\cU$ $\epsilon$-dominates $\cV$ (note that $e$ is not a symmetric distance).\footnote{The $\inf$ in the definition can be replaced by a $\min$ since $e(\cU,\cV)$ can be equivalently defined as $\inf\{\epsilon\geq 0: \forall \,\,\bu\in up(\cU), \, \exists\, \bv \in up(\cV) \textrm{ s.t. } \bv\preceq \bu+ \epsilon\mathbf{1}\}$, and $up(\cU)$ and $up(\cV)$ are compact sets for any $\cU,\cV\in\cF$.\label{fn:e}} We can then show the following.
\begin{proposition}\label{prop:hausdorffequi}
For any two Pareto frontiers $\cV$ and $\cU$ in $\cF$,
\[
d(\cU,\cV) =  \max(e(\cU,\cV),e(\cV,\cU))
\]
\end{proposition}
This means that the distance $d$ between two frontiers $\cV$ and $\cU$ is less than or equal to $\epsilon$ if both $\cU$ and $\cV$ $\epsilon$-dominate each other. This way of defining $d$ is attractive since it does not require defining upsets of the Pareto frontiers as we do in Definition~\ref{def:metric1}. %The proof of this equivalence is presented Section~\ref{apx:proofs} in the Appendix.

\subsection{Dynamic programming operator and the existence of a fixed point.}
By scaling and shifting the losses, we assume without loss of generality that $r_k(a, b) \in [0, 1-\beta]$
for all $(a, b, k)$.  Accordingly, the total discounted rewards of the game take values in $[0, 1]$ irrespective of the time horizon. Now, for any set $\mathcal{S}\subseteq [0,1]^K$, define the following operator $\Psi$ that maps $\mathcal{S}$ to a subset of $\mathbb{R}^K$:

\begin{equation}
\Psi(\mathcal{S})=\bigg\{\bigg(\max_{b\in B}\sum_{a\in A}\alpha_a\big[r_k(a,b)+\beta R_k(a,b)\big];\, k=1,\cdots,K\bigg): \bm{\alpha}\in\Delta(A),\,\,\bR(a,b)\in \mathcal{S}\,\,\forall\, a\in A,\,b\in B\bigg\}.
\end{equation}
This operator can be interpreted as follows. Assuming that $\mathcal{S}$ is the set of vectors of simultaneous upper bounds on expected losses that Alice can ensure in $\mathbb{G}^T$, $\Psi(\mathcal{S})$ is the set of vectors of simultaneous upper bounds on expected losses that she can ensure in $\mathbb{G}^{T+1}$. If $\mathcal{S}$ is convex then $\Psi(\mathcal{S})$ is not necessarily convex as we demonstrate in the following example. 

\begin{figure}[h]
\centering
\begin{minipage}[t]{.4\textwidth}
\centering
\includegraphics[width=1.5in]{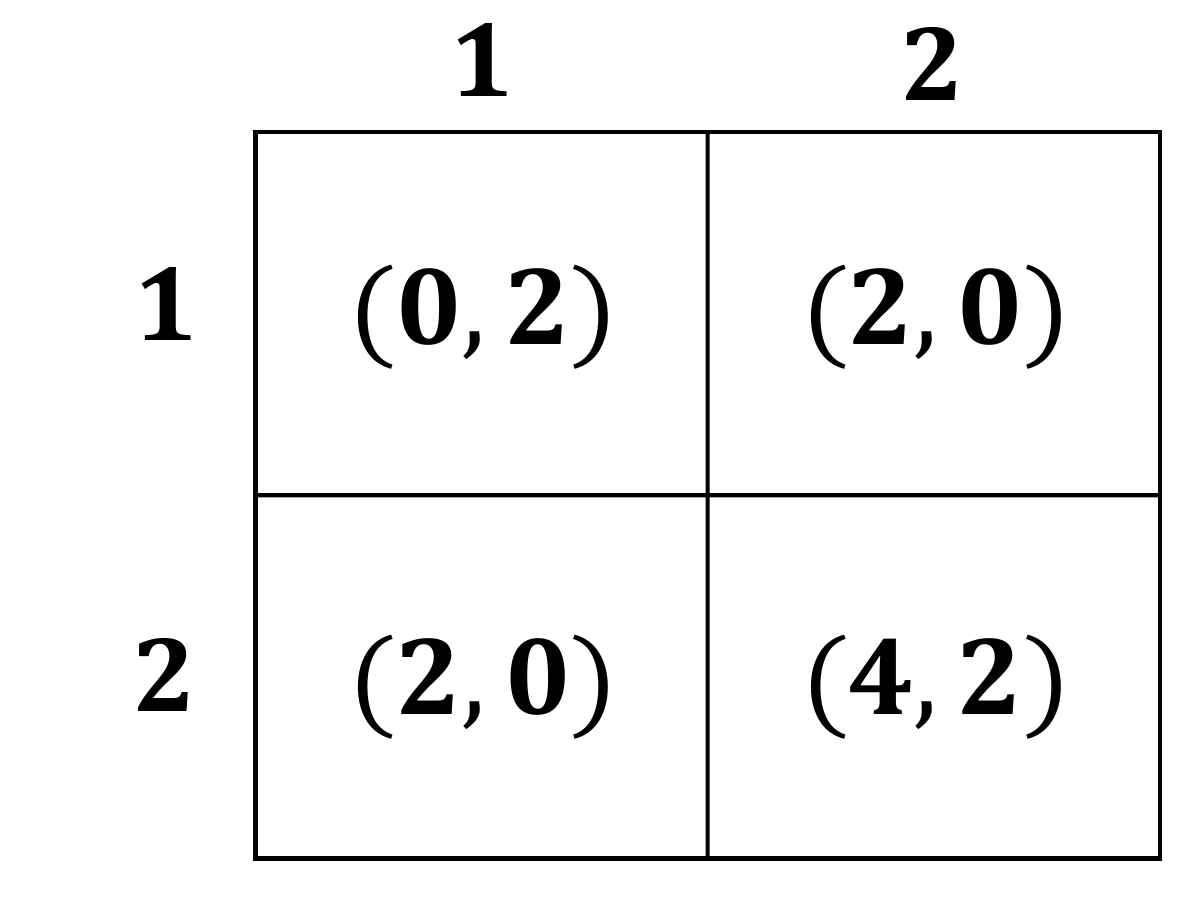}
\caption{A game with vector losses.}\label{fig:conv-tab}
\end{minipage}\hfill
\begin{minipage}[t]{.5\textwidth}
\centering
\includegraphics[width=2.5in,angle=0]{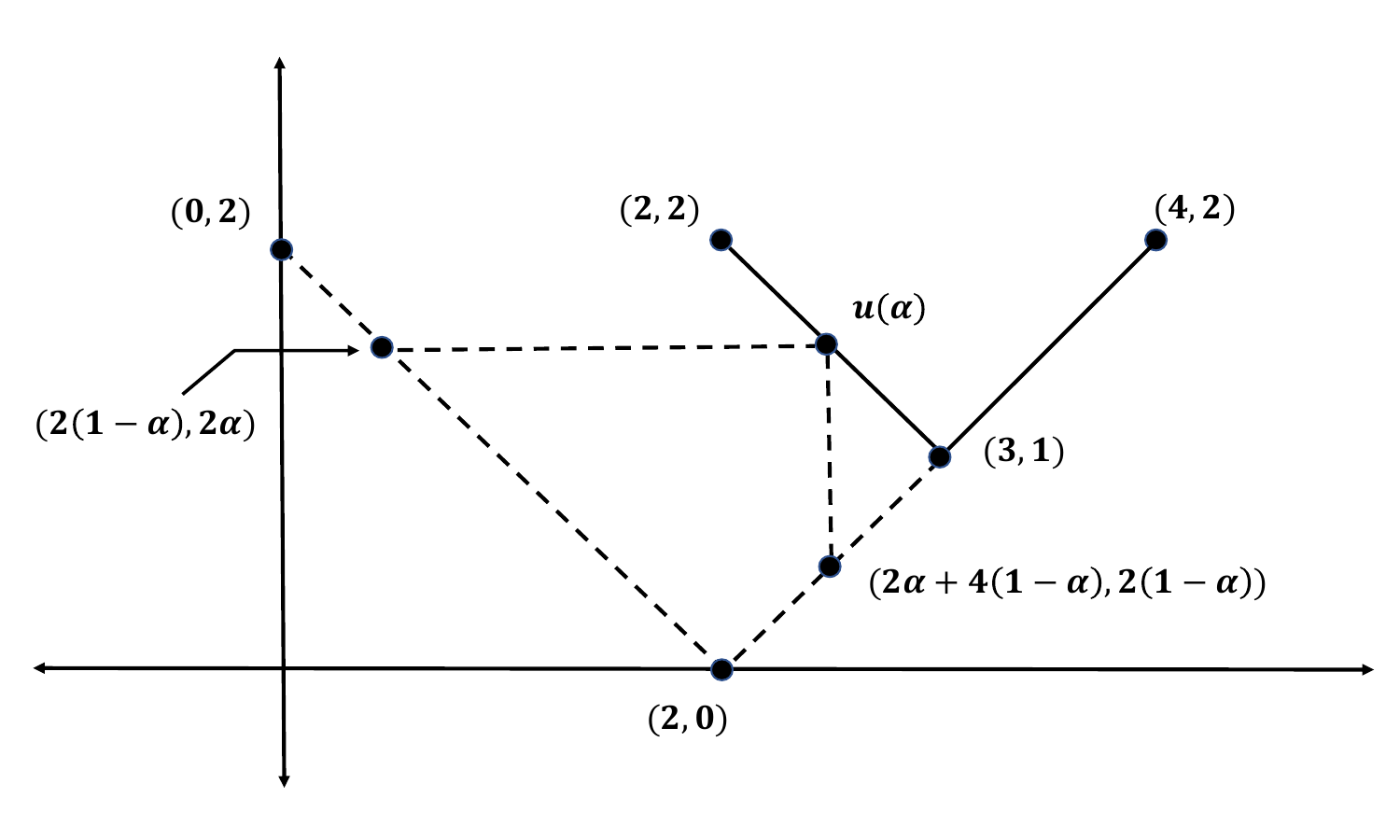}
\caption{Construction of the set $\Psi(\mathcal{S})$ where $\mathcal{S}=\{(0,0)\}$, for the game shown in Figure~\ref{fig:conv-tab}.}
\label{fig:preconv}
\end{minipage}
\end{figure}
{\bf Example:} Consider the game depicted in Figure~\ref{fig:conv-tab}. Suppose that $\mathcal{S}=\{(0,0)\}$, which is convex. Then for any discount factor $\beta$ and any $(\alpha,1-\alpha)$ where $\alpha\in[0,1]$, one obtains the guarantee:
$$u(\alpha)=(\max(2(1-\alpha), 2\alpha + 4(1-\alpha)), \max(2\alpha,2(1-\alpha))).$$
This is depicted in Figure~\ref{fig:preconv}. Thus, by varying $\alpha$, we find that the set $\Psi(\mathcal{S})$ is the union of the line segment joining points $(2,2)$ and $(3,1)$ and the segment joining the points $(3,1)$ and $(4,2)$. Clearly, this set is not convex.

Despite the fact that the operator $\Psi$ does not preserve convexity, we can nevertheless show that if $\cV$ is a Pareto frontier in $\cF$ (which means that it is the Pareto frontier of a convex and closed set), then the Pareto frontier of $\Psi(\cV)$ is also in $\cF$ (observe that in the example above, the Pareto frontier of $\Psi(\{(0,0)\})$ is the line segment joining points $(2,2)$ and $(3,1)$). Further, we can also show that if $\cV \in \cF$ is the set of vectors of simultaneous upper bounds on expected losses that Alice can ensure in $\mathbb{G}^T$, then in any optimal plan for Alice in $\mathbb{G}^{T+1}$, the continuation strategy from stage $2$ onwards need not depend on her own action in stage $1$. 

\begin{lemma}\label{lma:box}
Let $\cV\in\cF$. Then:
\begin{enumerate} 
\item $\Lambda(\Psi(\cV)) \in \mathcal{F}$.
\item Any point  $\bu$ in $\Lambda(\Psi(\cV))$ is of the form: 
$$\bu=\bigg(\max_{b\in B}\bigg [\sum_{a\in A}\alpha_ar_k(a,b)+\beta Q_k(b)\bigg];\, k=1,\cdots,K\bigg).$$ where $\bQ(b) \in \cV$ for all $b\in B$.
\end{enumerate}
\end{lemma}
%The proof of this result is presented in Appendix~\ref{apx:proofs}. 
We next define the following dynamic programming operator $\Phi$ on $\cF$.
\begin{definition}\label{operator1}
(Dynamic programming operator) For $\cV \in \cF$, we define $\Phi(\cV)=\Lambda(\Psi(\cV))$.
\end{definition}
From Lemma \ref{lma:box}, we know that $\Phi(\cV) \in \cF$ whenever $\cV \in \cF$.
Next, we claim that $\Phi$ is a contraction in the metric $d$.
\begin{lemma}\label{lma:contraction} $e(\Phi(\mathcal{U}),\Phi(\mathcal{V}))\leq \beta e(\mathcal{U},\mathcal{V})$, and hence $d(\Phi(\mathcal{U}),\Phi(\mathcal{V}))\leq \beta d(\mathcal{U},\mathcal{V}).$
\end{lemma}

Finally, the completeness of $\cF$ and the fact that $\Phi$ is a contraction in $d$ directly implies the following result as a consequence of the Banach fixed point theorem \cite{munkres1975topology}. 
%we show that the dynamic programming operator has a unique fixed point and starting from a Pareto frontier in $\cF$, the sequence of frontiers obtained by a repeated application of this operator converges to this point.

\begin{theorem}\label{thm:main1}
For any $\cV\in\cF$, the sequence $(\cA_n=\Phi^n(\cV))_{n\in\mathbb{N}}$
converges in the metric $d$ to the Pareto frontier $\cV^* \in \cF$, which is the unique fixed point of the operator $\Phi$, i.e., the unique solution of $\Phi(\cV)=\cV$.
\end{theorem}
We can then show that $\cV^*$ is indeed the optimal set $\cV^{\infty}$ that we are looking for.

\begin{theorem}\label{thm:main2}
 $\cV^{\infty}=\cV^*$.
 \end{theorem}

\subsection{Optimal strategies: existence and structure.}\label{subsec:optstrategy}
%In this section, we look at another way of characterizing the Pareto frontier. 
%First we can show the following:
%\begin{theorem}
%$F^{\cV}$ is a continuous function of $\cV$, i.e. for any sequence $\{\cV_n\}$ such that 
%$$\lim_{n\rightarrow\infty}d(\cV_n,\cV^*)=0,$$
%we have that 
%$$\lim_{n\rightarrow\infty}\|F_{\cV_n}-F_\mathcal{V^*}\|=0$$
%\end{theorem}

\begin{figure}[H]
\begin{center}
\includegraphics[width=1.8in,angle=0]{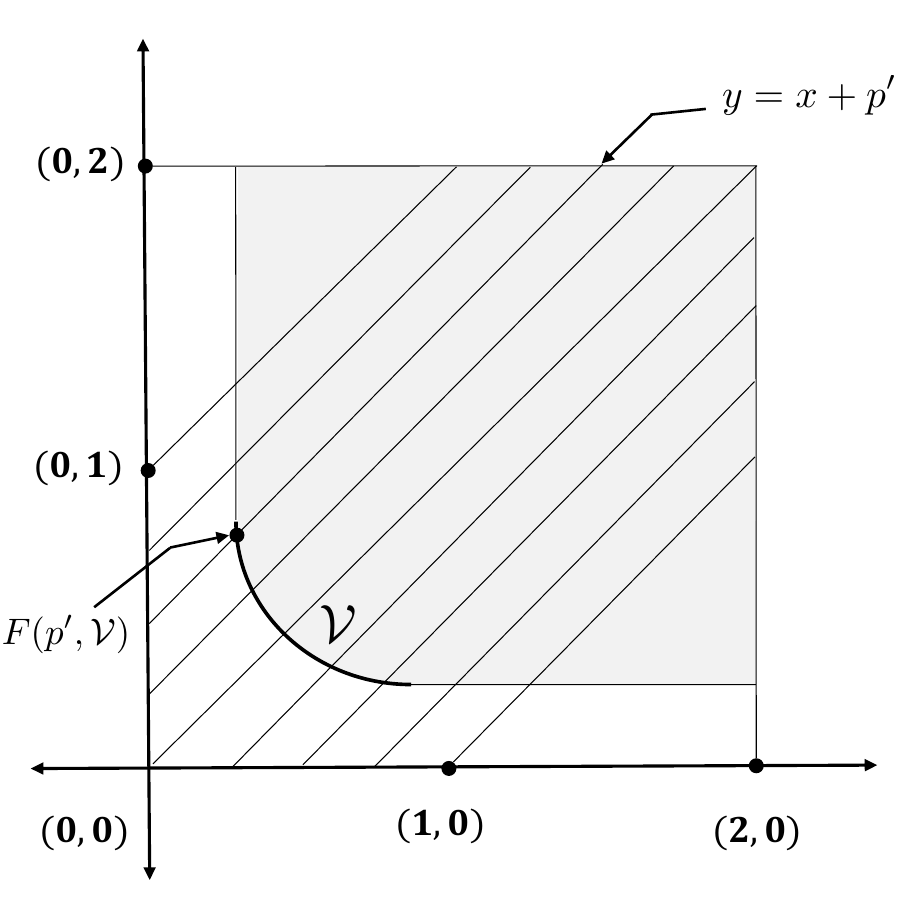}
\caption{Parameterization of $\cV$.}
\label{fig:quant}
\end{center}
\end{figure}

For a Pareto frontier $\cV \in \cF$, one can define a one-to-one function from some compact parameter set $\mathcal{P}$ to  $\cV$. Such a function parameterizes the Pareto frontier.
%For instance, let $\mathcal{P}$ be the $K-1$ dimensional simplex, i.e., the set $\Delta^K=\{\mathbf{p}\in \mathbb{R}^K:\sum_{k=1}^Kp_k =1; \,\,p_k\geq 0\,\forall\, k\}$, and consider the function $\bF: \mathcal{P} \times \cF\to \mathbb{R}^K$ defined as
%\begin{equation} \label{e.FV}
%F(\mathbf{p}, \cV) =\argmin_{x\in \cV} \{\sum_{k=1}^Kp_kx^2_k\}.
%\end{equation}
%One can show that this minimization problem has a unique optimum for each $\mathbf{p}$ and further, for each point in $\cV$, there is a $\mathbf{p}\in \mathcal{P}$ that maps to this point.
 %consider the set of lines 
%\begin{equation}
%\{y=x + p:p\in [-1,1]\}.
%\end{equation}
%Then one can define $F^{\cV}:[-1,1]\rightarrow \cV$ to be the point of intersection of the upset of $\cV$ in $[0,2]^2$ with the line $y=x+p$. %This function is clearly one-to-one since any point on $\cV$ intersects exactly one of the lines. 
We present one such parameterization that will be used later in our approximation procedure. Define the set 
$$\mathcal{P} \triangleq \cup_{k=1}^{K} \{(p_1,\cdots,p_{k-1},0,p_k,\cdots,p_{K-1});\,\, p_r\in [0,1]\textrm{ for all } r=1,\cdots,K-1\}.$$ $\mathcal{P}$ is thus the union of $K$, $K-1$ dimensional faces of the hypercube $[0,1]^K$, where each face is obtained by pinning the value along one dimension to 0. For instance for $K=2$, we have $\mathcal{P} = [0,1]\times \{0\} \cup \{0\}\times [0,1]$, i.e., the union of the line segment joining $(0,0)$ and $(0,1)$, and the segment joining $(0,0)$ and $(1,0)$. Now consider the function $\bF: \mathcal{P} \times \cF\to \mathbb{R}^K $, where we define,
\begin{equation} \label{param}
\bF(\mathbf{p},\cV)=\argmin_{\mathbf{x}} t
\end{equation}
$$\textrm{s.t. } \mathbf{x}= t\mathbf{1}+\mathbf{p},\,\,t\in\mathbb{R},$$
$$\bx\succeq \bu,\,\,\bu\in \cV. \nonumber$$
$\bF(\mathbf{p},\cV)$ is essentially the component-wise smallest point of intersection of the line $\mathbf{x}= t\mathbf{1}+\mathbf{p}$ (for a fixed $\mathbf{p}$) with the upset of $\cV$ in $[0,2]^K$. In $\mathbb{R}^2$, this is simply the family of lines $y=x+p'$ where $p' = p_2 - p_1$, for $p' \in [-1,1]$ (see Figure \ref{fig:quant}).
Then for a given $\cV$, the function $\bF(.,\cV):\cP\to up(\cV)$ defines a map such that for every point $\bu$ on $\cV$, there is a unique $\mathbf{p}\in\cP$ that maps \emph{only} to that point. This is the $\bp$ such that the line $\mathbf{x}= t\mathbf{1}+\mathbf{p}$ intersects $\cV$ at $\bu$ (if the line intersects $\cV$ at two or more points, then one of those points is dominated by the other(s), which is a contradiction). Note that for some values of $\bp$, the line $\mathbf{x}= t\mathbf{1}+\mathbf{p}$ may not intersect $\cV$, but it will definitely intersect the upset of $\cV$ in $[0,2]^K$, which is why in~\eqref{param}, we optimize over $\mathbf{x}$ that dominate $\bu \in\cV$, rather than directly optimizing over $\bu \in\cV$.

We can now express the DP operator in the form of such a parametrization. Assume that $\cV^*$ is such that $\cV^* = \Phi(\cV^*)$.
Then for $\mathbf{p} \in \mathcal{P}$,
one can choose $\bm{\alpha}(\bp) \in \Delta(A)$  and $\bq(b,\bp) \in \mathcal{P}$ for each $b\in B$ such that for $k\in\{1,\cdots,K\}$,
\begin{equation}\label{eqn:strategymap}
F_k(\mathbf{p},{\cV^*}) =  \max_{b\in B}\{\sum_{a \in A} \alpha_a(\bp)r_k(a, b) + \beta F_k(\bq(b,\bp),{\cV^*})\}.
\end{equation}

Then we have the following result.

\begin{theorem}\label{thm:optstrategy}
For any $\bp_1 \in \mathcal{P}$, the upper bound $\bF(\bp_1,\cV^*)\in \cV^*$ on losses is guaranteed by Alice in the infinite horizon game by first choosing action $a_1 \in A$
with probability $\alpha_{a_1}(\bp_1)$. Then if Bob chooses an action $b_1 \in B$,  the optimal guarantees to choose
from the second step onwards are then $\beta \bF(\bp_2,{\cV^*})$ in $\beta \cV^*$, where $\bp_2 = \bq(b_1,\bp_1)$, which can be guaranteed by Alice by choosing action $a_2 \in A$ with probability $\alpha_{a_2}(\bp_2)$, and so on.   
\end{theorem}
This implies that $\mathcal{P}$ can be thought of as a compact \emph{state space} for the strategy. In the remainder of the paper, however, we will refer to these states as {\it modes} to distinguish them from the classical notion of an exogenously defined state in, e.g., Markov decision processes. Each mode is associated with an immediate optimal randomized action and a transition rule that depends on the observed action of Bob. In order to attain a point in $\cV^*$, Alice starts with the corresponding mode, plays the associated randomized action, transitions into another mode depending on Bob's observed action as dictated by the rule, plays the randomized action associated with the new mode and so on. In particular, the strategy does not depend on the past actions of Alice and it depends on the past actions of Bob only through this information state, i.e., the mode, that Alice keeps track of. It is interesting to note that unlike in stochastic games or Markov decision processes (MDPs) \cite{puterman2014markov}, the state transitions are not exogenously defined but they are endogenously specified by the dynamic programming operator.

\section{Approximation.}\label{sec:approx}
In general, except for simple examples (such an example is presented in Section~\ref{sec:exact} in the Appendix), it is difficult to analytically compute $\cV^*$ and the optimal strategies $\{(\bm{\alpha}(\bp),q(b,\bp)):\bp\in \mathcal{P}\}$ that satisfy (\ref{eqn:strategymap}) by simply using the fixed point relation. Hence, we now propose an approximate dynamic programming procedure to approximate the optimal Pareto frontier and devise approximately optimal strategies. In order to do so, we first define an appropriate finitely parameterized approximation of any Pareto frontier where one gets an increasingly finer approximation as the size of the parameter space increases. 
%This carefully chosen approximation scheme is instrumental in driving our results on the approximation guarantees. 

Consider the following approximation scheme for a Pareto frontier $\mathcal{V}\in \cF$. 
%For an integer $N$, choose $2N+1$ lines defined as:
%$\{y=x\pm \frac{k}{N}: k=1,2,\cdots, N\}.$
For a fixed positive integer $N$, define the set 
\begin{equation}\label{set:pn}
\mathcal{P}_N = \cup_{k=1}^{K} \{(p_1,\cdots,p_{k-1},0,p_k,\cdots,p_{K-1});\,\, p_r\in \{0,\frac{1}{N},\frac{2}{N},\cdots,\frac{N-1}{N},1\}\,\,\forall\,\,r=1,\cdots,K-1\}.
\end{equation}
In words, $\cP_N$ is obtained by approximating each of the $K$, $K-1$ dimensional faces in $\cP$ by a uniformly distributed grid of $(N+1)^{K-1}$ points. The number of distinct points in this set is 
\begin{equation}\label{eq:hkn}
H(K,N)\triangleq (N+1)^{K} - N^K.
\end{equation}
Next, define the approximation operator to be 
%vector-values function $\bF^N(p,\cV):\{0,\pm\frac{1}{N},\pm\frac{2}{N},\cdots,\pm\frac{N-1}{N}, \pm 1\}\times \cF\rightarrow \mathbb{R}^2$, where for each $p$,
%\begin{equation}\label{eqn:approx}
%\bF^N(p,\cV)=\argmin_{(x,y)} x+y 
%\end{equation}
%$$\textrm{s.t. } x\geq u_1,\,\,y\geq u_2,\,\,\bu\in \cV,\,\,y=x+p\nonumber.$$
%$\bF^N(p,\cV)$ is hence just the function $\bF(p,\cV)$ defined in \eqref{param} restricted to $p\in \{0,\pm\frac{1}{N},\pm\frac{2}{N},\cdots,\pm\frac{N-1}{N}, \pm 1\}$.
 \begin{equation}\label{eq:defapprox}
 \Gamma_{N}(\mathcal{V})= \Lambda\bigg(\ch\bigg(\big\{\bF(\bp,\cV): \bp\in \mathcal{P}_N \big\}\bigg)\bigg),
 \end{equation}
where $\bF(\bp,\cV)$ was defined in \eqref{param}. Here $\ch$ denotes the closed convex hull of a set. Thus $\Gamma_{N}(\mathcal{V}) \in\cF$ is the Lower Pareto frontier of a convex polytope, and it has at most $H(K,N)$ vertices, where each vertex is the point of intersection of the line $\mathbf{x}=t\mathbf{1}+\bp$ with the upset of $\cV$ for some $\bp\in\mathcal{P}_N$. 
The following approximation guarantee is instrumental in driving our results.

\begin{proposition}\label{lma:approx}
Consider a $\cV\in \cF$. Then
$$e(\Gamma_{N}(\cV),\cV)=0 \textrm{ and } e(\cV, \Gamma_{N}(\cV))\leq \frac{1}{N},$$
and hence 
$$d(\cV, \Gamma_{N}(\cV))\leq \frac{1}{N}.$$
\end{proposition}
%Now suppose that $\mathcal{V}$ is in $\cF$. Then we know that $\Phi(\mathcal{V})$ is in $\cF$ and it is the Pareto frontier of a convex and compact set (its upset, for instance); thus 

Next, we can express the compound operator $\Gamma_{N}\circ \Phi$ via a set of explicit optimization problems as in~\eqref{param}, that only take $\cV$ as input:
\begin{equation}
\bF(\bp,\Phi(\cV))=\argmin_{\mathbf{x}} t\label{operator}
\end{equation}
$$\textrm{s.t. }\mathbf{x}=t\mathbf{1}+\bp,\,\,t\in\mathbb{R},$$
$$\bx \succeq \sum_{a\in A}\alpha_a \br(a,b)+\beta \bQ(b)\,\,\forall\,\,b\in B,$$
$$\bm{\alpha}\in \Delta(A),\,\,\bQ(b)\in\cV\,\,\forall\,\, b\in B.$$

If $\cV\in\cF$ is the lower Pareto frontier of a convex polytope, then this is a linear program, and further $\Gamma_{N}\circ \Phi(\mathcal{V})$ is also the lower Pareto frontier of a convex polytope. We then we have the following result.
\begin{proposition}\label{prop:approxfrontier}Let $\mathcal{G}_0=\{\mathbf{0}\}$ and let $\mathcal{G}_n=(\Gamma_{N}\circ \Phi)^n(\mathcal{G}_0)$. Then
\begin{equation}
e(\mathcal{G}_n,\cV^*) \leq \beta^n \textrm{ and } e(\cV^*,\mathcal{G}_n) \leq \frac{1}{N}\bigg(\frac{1-\beta^n}{1-\beta}\bigg)+\beta^n.
\end{equation}
And thus $$d(\mathcal{V}^*,\mathcal{G}_n)\leq \frac{1}{N}\bigg(\frac{1-\beta^n}{1-\beta}\bigg)+\beta^n.$$
%Thus $$\limsup_{n\rightarrow\infty}d(\mathcal{V}^*,\mathcal{G}_n)=\frac{1}{N}\textup{O}\bigg(\frac{1}{(1-\beta)}\bigg).$$
\end{proposition}
Hence for any $\epsilon$, there is a pair $(N,n)$ such that $d(\cV^*,\mathcal{G}_n)\leq \epsilon$. This result implies an iterative procedure for approximating $\cV^*$ by successively applying the compound operator $\Gamma_{N}\circ \Phi$ to $\mathcal{G}_0$, by solving the linear program in \eqref{operator} for each $\bp\in\mathcal{P}_N$ at each step. Since $\mathcal{G}_n$ is a lower Pareto frontier of a convex polytope, with at most $H(K,N)$ vertices for each $n$, the size of these linear programs remain the same throughout. More details on solving these programs can be found in Section~\ref{apx:linprog} in the Appendix.

The fact that $e(\mathcal{G}_n,\cV^*) \leq \beta^n$ implies that $\mathcal{G}_n$ $\beta^n$-dominates $\cV^*$ for all $n$, and thus the optimal upper bounds in $\cV^*$ cannot be larger than in $\mathcal{G}_n +\beta^n\mathbf{1}$. Thus as $n$ gets larger, the set $\mathcal{G}_n +\beta^n\mathbf{1}$ approaches $\cV^*$ ``from above,'' and in the limit, ends up within a $1/(N(1-\beta))$ distance of $\cV^*$.

\subsection{Extracting an approximately optimal strategy.}\label{subsec:approxpol}
From $\mathcal{G}_n$, one can also extract an approximately optimal strategy $\pi_n$ in the infinite horizon game. 
%Consider the following linear program that computes $\mathcal{G}^{n+1}$ from $\mathcal{G}^{n}$.
%\begin{equation}
%\bF(p,\Phi(\mathcal{G}^n))=\argmin_{(x,y)} x + y\label{operator}
%\end{equation}
%$$\textrm{s.t. }x \geq \sum_{a\in A}\alpha_a r_1(a,b)+\beta Q_1(b)\,\,\forall\,\,b\in B,$$
%$$y \geq\sum_{a\in A}\alpha_a r_2(a,b)+\beta Q_2(b)\,\,\forall\,\,b\in B,$$
%$$y=x+p,\,\,\bm{\alpha}(p)\in \Delta(A),\\
%\,\,\bQ(b)\in\mathcal{G}^n\,\,\forall\,\, b\in B$$
Suppose $\bm{\alpha}^*(\bp)$ and $\bQ^*(b,\bp)$ for $b\in B$ are the optimal values that solve the program~\eqref{operator} to compute $\bF(\bp,\Phi(\mathcal{G}_n))$ for different $\bp\in \mathcal{P}_N$. Then these define an approximately optimal strategy in the following class:  
\begin{definition}
A $H(K,N)-$\emph{mode stationary strategy} $\pi$ is a mapping from each $\bp\in \mathcal{P}_N$ to the pair 
\begin{enumerate}
\item $\bm{\alpha}(\bp) \in \Delta(A)$, and
\item $\bigg(\bq_1(b,\bp),\cdots,\bq_K(b,\bp),\,\bz(b,\bp)\bigg)$, where for all $b\in B$, $\bq_k(b,\bp)\in \mathcal{P}_N$ for all $k = 1,\cdots,K$ and $\bz(b,\bp)\in \Delta^{K}$.
\end{enumerate}
\end{definition}
Here, $\Delta^{K}$ is the unit simplex in $\mathbb{R}^K$. The interpretation is as follows. One starts with some initial \emph{mode}, i.e., a value of $\bp \in\mathcal{P}_N$. Then at any step, if the current mode is $\bp$, then Alice first chooses action $a\in A$ with probability $\alpha_a(\bp)$. Then if Bob plays action $b\in B$, Alice samples the new mode to be $\bq_k(b,\bp)$ with probability $z_k(b,\bp)$ for each $k$, and after having sampled a new mode, plays accordingly thereafter. 

Now, $\bm{\alpha}^*(\bp)$ defines $\bm{\alpha}(\bp)$ in $\pi_n$, and $\bigg(\bq_1(b,\bp),\cdots,\bq_K(b,\bp),\,\bz(b,\bp)\bigg)$ are defined such that they satisfy
\begin{equation}\label{approx:face}
\bQ^*(b,p)=\sum_{k'=1}^Kz_{k'}(b,\bp)\bF( \bq_{k'}(b,\bp),\mathcal{G}_n).
\end{equation}
%Now consider the optimization problem (\ref{operator}) that corresponds to $i=n$, i.e., the problem the procedure would have solved if it was allowed to continue for one more iteration. Now one can extract a $(2N+1)$-mode strategy $\pi_n$ from the solution of this optimization problem as follows. Defining $\bm{\alpha}(p)$ is immediate. Now note that the optimal $\bQ(b,p)$ is such that either $\bQ(b,p)=F^n(1)$ or $\bQ(b,p)=F^n(-1)$, or $\bQ(b,p)=\bz(b,p)F^n(q)+(1-\bz(b,p))F^n(q')$ for some $\bz(b,p)\in[0,1]$ and some $q,q'$ such that $|q-q'|=\frac{1}{N}$. These define $\bz(b,p)$, $q(b,p)$ and $q'(b,p)$ in our strategy. If $\bQ(b,p)=F^n(1)$, then $\bz(b,p)=1$ and $q(b,p)=1$, whereas if $\bQ(b,p)=F^n(-1)$ then $\bz(b,p)=0$ and $q'(b,p)=-1$.

These $\bigg(\bq_1(b,\bp),\cdots,\bq_K(b,\bp),\,\bz(b,\bp)\bigg)$ are directly obtained as the output of the linear program; see Section~\ref{apx:linprog} in the Appendix. The interpretation is as follows. If $\cV$ is the lower Pareto frontier of a convex polytope with each vertex lying on the line $\mathbf{x}=t\mathbf{1}+\bp$ for some $\bp\in\mathcal{P}_N$, $\bQ^*(b,\bp)$ for each $b\in B$ that results from solving~\eqref{operator} will lie on one of the faces of this Pareto frontier. Thus $\bQ^*(b,\bp)$ can be expressed as a convex combination of (at most $K$) extreme points of the face as expressed in~\eqref{approx:face}. 

Let $\mathcal{V}^{\pi_n}$ be the corresponding Pareto frontier that is attained by the strategy $\pi_n$ (each point on this frontier is guaranteed by choosing different possible initial randomizations over the $H(K,N)$ modes). In Section~\ref{apx:pol-eval} in the Appendix, we discuss how this ``policy evaluation'' can be performed by solving a linear program. Simply from the definition of $\cV^*$ as the optimal frontier, we know that $\mathcal{V}^{\pi_n}$ dominates $\cV^*$, i.e.,  $e(\mathcal{V}^{\pi_n},\cV^*)= 0$. But we can further show the following.
\begin{proposition}\label{prop:approx2}
\begin{equation}
d(\mathcal{V}^{\pi_n},\mathcal{V}^*)\leq \frac{1}{N}\bigg(\frac{1-\beta^n}{1-\beta}\bigg)+2\beta^n + \frac{1}{N}\bigg(\frac{2-\beta^n-\beta^{n+1}}{(1-\beta)^2}\bigg).
\end{equation}
%And thus $$\limsup_{n\rightarrow\infty}d(\mathcal{V}^{\pi_n},\mathcal{V}^*)=\frac{1}{N}\textup{O}\bigg(\frac{1}{(1-\beta)^2}\bigg).$$
\end{proposition}
Thus, an approximately optimal strategy can be obtained by choosing an appropriate $(N,n)$.

\begin{remark}
For a fixed $(N,n)$, in order to approximate the optimal frontier, the procedure needs to solve $nH(K,N)$ linear programs to give the corresponding error bound in Proposition~\ref{prop:approxfrontier}. In our implementation described in Section~\ref{apx:linprog} in the Appendix, each linear program is composed of $mH(K,N) + l+1$ variables and $Km + K+1$ constraints. %(recall that $H(K,N)$ is polynomial in $N$ for a fixed $K$). 
One can focus on two terms in the approximation error separately: the first term is the quantization error, which is bounded by $\frac{1}{N(1-\beta)}$, and second is the iteration error, which is bounded by $\beta^n$. The second term is benign since it decays exponentially in $n$. The first term is dominant and requires $N=\frac{1}{(1-\beta)\epsilon}$ to achieve an error of $\epsilon$. To find an $\epsilon$-optimal strategy, we require $N\approx \frac{1}{(1-\beta)^2\epsilon}$. Thus, for fixed values of $\beta$ not too close to $1$, the $N$ required to obtain a good approximation isn't too large. The main concern, however, is that $H(K,N)$ grows exponentially in the dimension $K$, and hence the computation is expected to be prohibitive when $K$ is large. In Section~\ref{sec:heuristic} below, we propose and evaluate a heuristic approach to get around this difficulty at the cost of loss in optimality. 
%Moreover, this computation can be done offline once and for all. The resulting approximately optimal strategy requires a bounded memory of $\textup{O}(\log(H(K,\frac{1}{\epsilon(1-\beta)^2}))$ bits and is simple to implement via a look-up table.}
\end{remark}
%{\bf Remark:} 
%The procedure to approximate the frontier and extract an approximately optimal strategy illustrates that our characterization of the minimax optimal strategy via the fixed point of a dynamic programming operator opens up the possibility of using several approximate dynamic programming procedures; for instance, see Chapter 6 in \cite{Bertsekas05b}. Here we have not tried to determine a computation procedure that achieves the optimal error-complexity tradeoff. 
\section{Optimal finite-mode policies for larger $K$.}\label{sec:heuristic}
%Although our approximation scheme yields near-optimal policies, as we discuss in Remark~\a practical drawback is that the number of modes grows exponentially with the dimension $K$, and hence the computation can become prohibitive when $K$ is large. %However, our experimental results in the previous section suggest that we can obtain good performance guarantees with significantly smaller number of modes than what is suggested by the approximation bounds for our scheme. Based on this observation, 
In this section, we propose a different approach to designing good policies with a small number of modes in settings where $K$ is large, when the computation of near-optimal policies discussed in Section~\ref{sec:approx} becomes prohibtive. 
%As we discuss in Section~\ref{apx:pol-eval} in the Appendix, we can efficiently compute performance guarantees for these policies, which, as we numerically show in the context of regret minimization, can be significantly better than the guarantees under existing algorithms like Hedge. 

%In order to design a good policy with a small number of modes, the modes must be carefully chosen. In a finite-mode stationary policy, modes are simply associated with an immediate randomized strategy and a probabilistic transition rule as a function of adversary's actions. However, when we talk about the choice of modes, we mean the directions along which we would like to minimize our losses, with the solutions of the resulting optimization problems yielding the strategies and transition rules associated with these modes (much in the same way as how the approximately optimal policy was obtained in Section~\ref{subsec:approxpol}). These directions are rays of the form $t\mathbf{1}+\bp$, $t\in\mathbb{R}$, for a fixed point $\bp$ associated with each mode. Henceforth, as we have done before, we refer to these points $\bp$ as modes. %with the implicit understanding that we are talking about the rays implied by these points. 

In order to design a good policy with a small number of modes, the modes must be carefully chosen. 
%In a finite-mode stationary policy, modes are simply associated with an immediate randomized strategy and a probabilistic transition rule as a function of adversary's actions. However, when we talk about the choice of modes, we mean rays of the form $t\mathbf{1}+\bp$, $t\in\mathbb{R}$, for a fixed point $\bp$ associated with each mode; these are the directions along which we would like to minimize our losses, with the solutions of the resulting optimization problems yielding the strategies and transition rules associated with these modes (much in the same way as how the approximately optimal policy was obtained in Section~\ref{subsec:approxpol}). 
%These directions are rays of the form $t\mathbf{1}+\bp$, $t\in\mathbb{R}$, for a fixed point $\bp$ associated with each mode. 
%Henceforth, as we have done before, we refer to these points $\{\bp\}$ as modes. %with the implicit understanding that we are talking about the rays implied by these points. 
%For instance, for the $2$ experts setting, Figure~\ref{fig:conv} suggests that it is relatively more important to have a larger density of modes closer to the origin in order to approximate $\cV^*$ well. 
The following natural question guides our approach in this section: given an instance of a vector repeated game with discounted losses and a finite budget of modes, how do we design the ``best'' stationary policy within this budget? 
The first step is to specify what we mean by ``best.'' Motivated by our goal of regret minimization in repeated decision making, we aim to minimize the losses along the ray $\{\mathbf{x}= t\mathbf{1}; t\in\mathbb{R}\}$, i.e., we wish to minimize $t$ such that the losses on all dimensions are guaranteed to be at most $t$, irrespective of the actions of the adversary. 

Let such a policy start from mode $0$, and let $\bv_0\in \{\mathbf{x}= t\mathbf{1}; t\in\mathbb{R}\}$ denote vector of losses guaranteed by this policy starting from mode $0$, where we wish to minimize $\bv_0$. Let there be $M$ additional modes allowed by our budget, denoted by $i = 1,\cdots, M$, making a total of $M+1$ modes. Let $\mathcal{M}$ denote the set of all modes. Let $\bv_i$ be the vector of losses guaranteed by the policy starting from mode $i$. Associated with each mode $i =0,\cdots, M$, let $\bm{\alpha}_i\in\Delta(A)$ denote the probability distribution over immediate actions and let $(\mathbf{z}_i(b)\in\Delta(\mathcal{M}); \, b\in B)$ denote the randomized transition rule to other modes as a function of the adversary's action. Then the problem of minimizing $\bv_0$ can written as the following optimization problem.
% corresponding to the $\mathbf{0}$ mode; i.e., we want to design a stationary policy with a finite budget of modes, such that the resulting Pareto frontier of guarantees on losses, $\cV$, results in the smallest value of the following optimization problem (see \eqref{param}).
%\begin{equation}
%\bF(\mathbf{0},\cV)=\argmin_{\mathbf{x}} t
%\end{equation}
%$$\textrm{s.t. } \mathbf{x}= t\mathbf{1}+\mathbf{0},\,\,t\in\mathbb{R},$$
%$$\bx\succeq \bu,\,\,\bu\in \cV. \nonumber$$
\begin{subequations}
\begin{align}
 \min_{\bm\alpha, \mathbf{z}, \bv, t} t
 \end{align}
\begin{equation}\textrm{s.t. } \bv_0 = t\mathbf{1},\,\,t\in\mathbb{R},\label{ray}\end{equation}
\begin{equation}\bv_i\succeq \sum_{a\in A}\alpha_{i,a}\br(a,b) + \beta\sum_{j\in \mathcal{M}}z_{i,j}(b) \bv_j,\textrm{ for all }b\in B\textrm{ and } i\in\mathcal{M}.\label{feasible}\end{equation}
\begin{equation}\bm{\alpha}_i\in\Delta(A),\,\, \mathbf{z}_i(b)\in\Delta(\mathcal{M}); \textrm{ for all }b\in B\textrm{ and } i\in\mathcal{M}.\end{equation}
\end{subequations}
Here, the objective and \eqref{ray} express the fact that we are minimizing losses along the said ray. Equation \eqref{feasible} captures the Bellman one-step optimality conditions, which express the fact that the vector guarantees $(\bv_i)$ are feasible under the stationary policy that associates $\bm{\alpha}$ and $\mathbf{z}$ with the different modes (see also Section~\ref{apx:pol-eval} in the Appendix). 

The optimization problem defined above is a quadratically constrained linear program (QCLP) due to the bilinear constraints in \eqref{feasible}, which is known to be non-convex. The size of the problem grows as $\textup{O}(KMm)$, which has a significantly milder dependence on $K$ compared to the approximation approach of Section~\ref{sec:approx}. %However, this problem is a non-convex. 
Although this problem is non-convex, a wide range of optimization algorithms have been developed over the years that efficiently solve even large-scale instances of such programs to local optimality. For example, in a similar spirit as in our case, such QCLPs arise in the context of finding optimal finite-state controllers in partially observable Markov decision processes (POMDPs). Numerical solutions to these QCLPs have been shown to yield significantly better policies than those obtained via other heuristic approaches \cite{amato2006solving,amato2010optimizing}. In our case, we analogously find that in our numerical evaluations in the context of regret minimization discussed in Section~\ref{sec:numev}, the solutions to our QCLP define finite-mode online learning algorithms that provide guarantees that are significantly better than those provided by Hedge. 
\section{Numerical experiments.} \label{sec:regmin}
In this section, we present a numerical evaluation of our approaches discussed in Sections~\ref{sec:approx} and \ref{sec:heuristic} in the context of adversarial online learning.

\subsection{Designing near-optimal strategies for expert selection with binary losses.}\label{sec:example}
First, we illustrate our approximation scheme discussed in Section~\ref{sec:approx} by applying it to the well-known problem of regret minimization in expert selection with binary losses. We design, to the best of our knowledge, the first-known provably near-optimal algorithms for the case of $K=2$ experts and discounted losses, and show that these algorithms guarantee significantly smaller upper bounds on the regret than existing algorithms in adversarial online learning.%, a fact that, to the best of our knowledge, was unknown before. 

\begin{figure}[h]
\centering
\begin{minipage}[t]{.45\textwidth}
\centering
\includegraphics[width=1.2in]{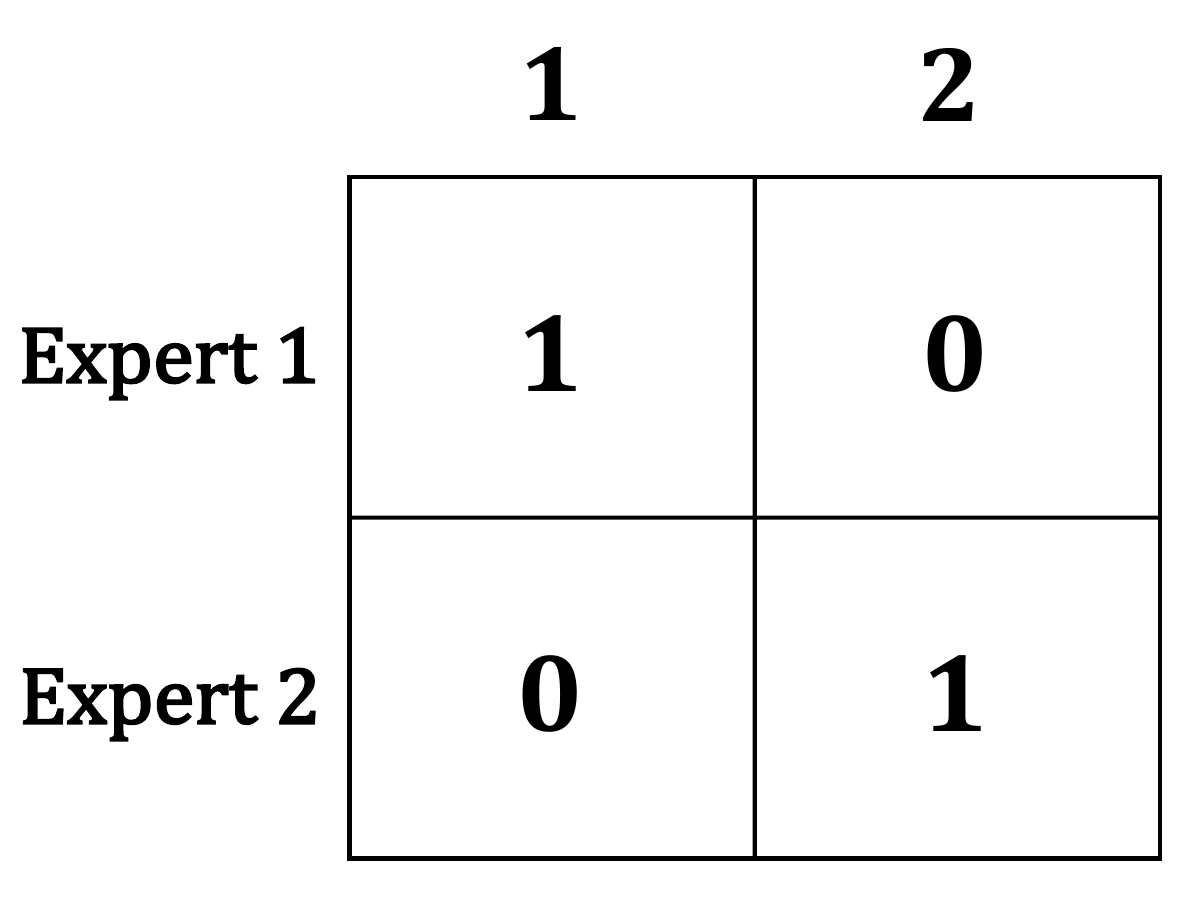}
\caption{Possible loss scenarios with $K=2$ experts.}\label{fig:experts}
\end{minipage}
\hspace{1cm}
\begin{minipage}[t]{.45\textwidth}
\centering
\includegraphics[width=1.2in]{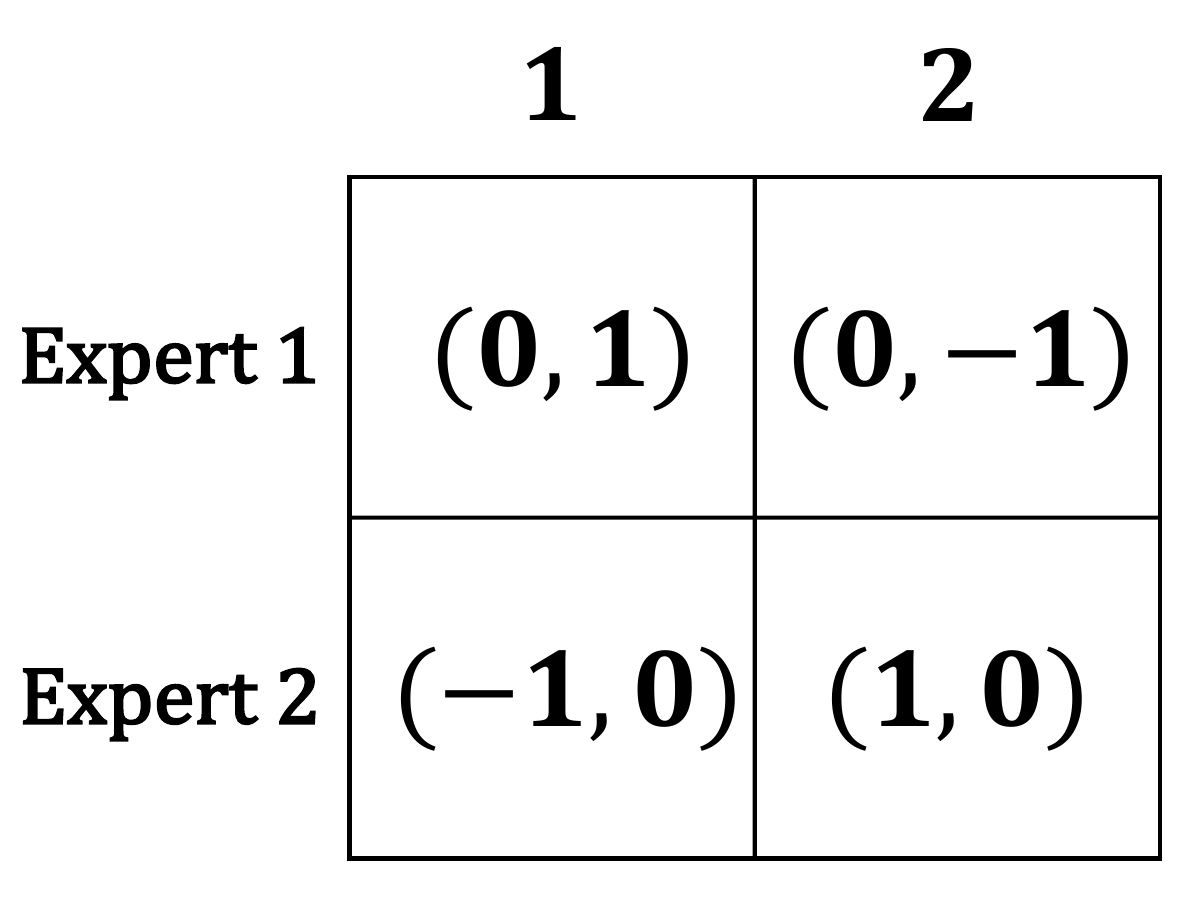}
\caption{Single-stage regret w.r.t. Expert 1 \& 2.}
\label{fig:ssregrets}
\end{minipage}
\end{figure}

The problem of expert selection with binary losses is described as follows. There are $K$ experts who give Alice recommendation for a decision-making task: say predicting which route will be the quickest to commute to work the next day. On each day, Alice decides to act on the recommendation made by one of the experts. The experts' recommendations may be correct or wrong, and if Alice acts on an incorrect recommendation, she bears a loss of $1$; otherwise she does not incur any loss. Each day, any set of experts may be correct while others are wrong. We omit the possibilities that all experts are correct or all experts are wrong, since it is wasteful for the adversary to choose these options. For $K=2$, this model can be represented by the matrix shown in Figure \ref{fig:experts}. The rows correspond to the choice made by Alice and the columns correspond to the different possibilities for the outcomes on each day. The matrix of single-stage regrets in this case is shown in Figure \ref{fig:ssregrets}.

In this case, since $K$ is small, the optimal frontier of regrets can be efficiently computed with high accuracy.
%Proposition~\ref{prop:approxfrontier} implies that these are, in fact, upper bounds on the optimal regret.
%Note that these frontiers appear to converge to the optimal frontier for the average case, i.e., the single point $(0,0)$, as expected. 
%In Table \ref{table:approxstrategy}, for illustration purposes we compute an approximately optimal $11-mode$ strategy $(N=5)$ for $\beta=0.8$. The second column contains the probability of choosing Expert 1 in each of the different modes and columns 3 and 4 give the transition rules to the different modes if Expert 1 incurs a loss and if Expert 2 incurs a loss respectively. If both the experts incur a loss or incur no loss then one stays in the same mode as before (although its is never optimal for the adversary to choose that option). 
%\subsection{The $K=2$ case.} We first illustrate our approximation scheme for the $K=2$ experts setting. Since $K$ is small, in this case
\begin{figure}[h]
\centering
\begin{minipage}[t]{.45\textwidth}
\centering
\includegraphics[width=3.2in]{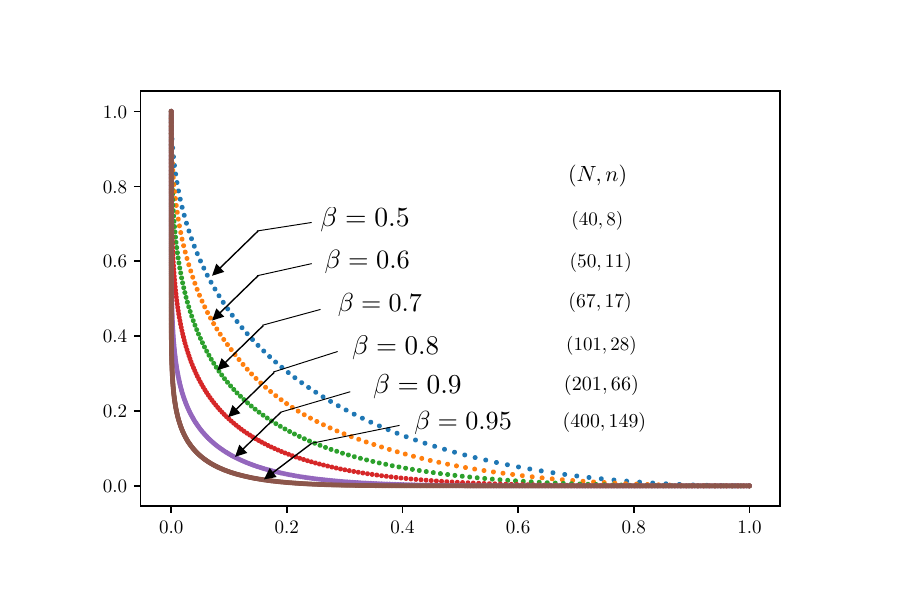}
\caption{Approximations of the optimal frontier $(1-\beta)\mathcal{V}^*(\beta)$ for different $\beta$ values and the associated $(N,n)$.}\label{fig:conv} %$(N,n)$ values in each case are chosen so that the approximation error (Proposition~\ref{prop:approxfrontier}) is less than 0.06. $n$ is chosen so that $\beta^n/(1-\beta)\leq 0.01$.}
\end{minipage}
\hspace{1cm}
\begin{minipage}[t]{.45\textwidth}
\centering
\includegraphics[width=3in]{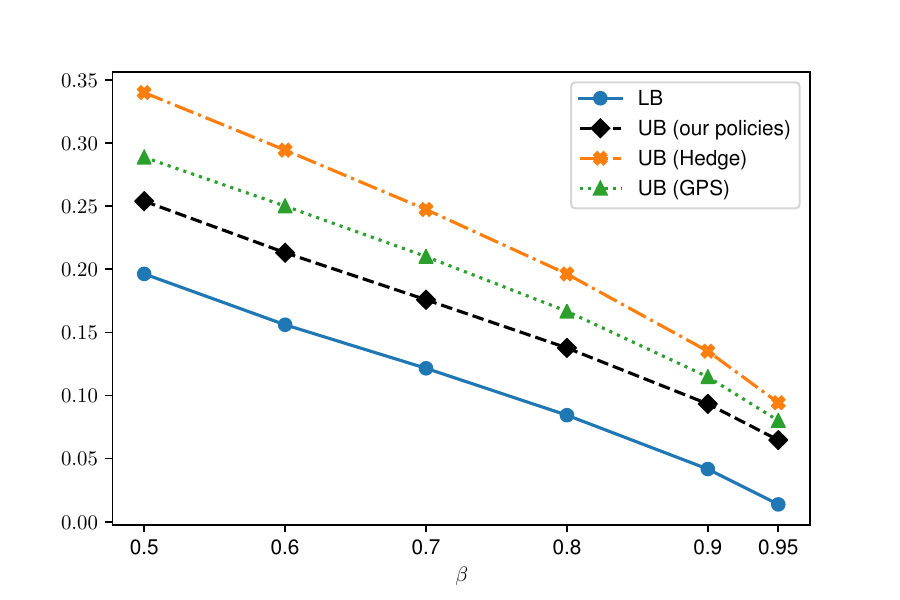}
\caption{Upper bounds on the optimal average discounted regret achieved by the different policies plotted as a function of the discount factor $\beta$. Also plotted is the theoretical lower bound on regret.}\label{fig:cumreg}
\end{minipage}
\end{figure}
%Consider the corresponding vector-valued game that captures the single-stage regret with respect to each expert for the different pairs of choices:
%\begin{table}[h]
%\begin{tabular}{lllll}
%Expert 1 & (0,1) & (0,-1) & (0,0) & (0,0) \\
%Expert 2 & (-1,0) & (1,0) & (0,0) & (0,0) 
%\end{tabular}
%\end{table}
Figure \ref{fig:conv} shows the computed approximately optimal Pareto frontiers of regret for a range of values of $\beta$. $(N,n)$ is chosen in each case so that the error in the approximation of $(1-\beta)\cV^*(\beta)$ (i.e., the optimal Pareto frontier of {\it average} discounted regrets) is at most $0.06$ (Proposition~\ref{prop:approxfrontier}). %We thus obtain lower bounds on the optimal regret in each of these cases. %Further, we can exactly compute the upper bounds on the regret guaranteed by the corresponding finite mode stationary policies that result from our approximation; see Section~\ref{apx:pol-eval} in the Appendix. 
%Further, for these values of $n$, we obtain that the error in the approximation of $(1-\beta)\cV^*(\beta)$ is at most $\beta^n$ from above, resulting in upper bounds on the optimal regret.
%Further, $n$ is chosen in each case so that $\beta^n/(1-\beta)\leq 0.01$, and hence, Proposition~\ref{prop:approxfrontier} guarantees that 0.01 added to the approximations of optimal regret for each $\beta$ result in upper bounds on the optimal regret. 
The lower and upper bounds on the discounted average optimal regret are plotted in Figure~\ref{fig:cumreg}; the lower bounds result from our theoretical guarantees in Proposition~\ref{prop:approxfrontier}, and the upper bounds result from the evaluation of our policies as shown in Section~\ref{apx:pol-eval} in the Appendix. 

In the figure, we also plot the theoretical upper bound on regret guaranteed by two other policies: (a) the well-known exponentially weighted average forecaster, also known as ``Hedge,'' and (b) the optimal algorithm given by Gravin, Peres and Sivan \cite{gravin2014towards} (which we will refer to as GPS) for the 2-experts problem with a geometrically distributed time horizon.\footnote{This model and its relation to our model of discounted losses is discussed in Section~\ref{apx:gps2} in the Appendix.} Hedge guarantees an upper bound of $\sqrt{\log{K}(1-\beta)/(2(1+\beta))}$ on the expected average discounted regret for the $K$ experts problem, which is the best known bound for this problem \cite{Cesa-Bianchi06a}. Hedge is defined in Section~\ref{apx:hedge} in the Appendix. For the 2-experts problem, GPS guarantees an upper bound of $(1/2)\times\sqrt{(1-\beta)/(1+\beta)}$ on the expected average discounted regret (see Footnote~\ref{fn:gps} in the Appendix). The GPS algorithm is presented in Section~\ref{apx:gps} in the Appendix. Both these algorithms achieve significantly higher upper bounds on the regret than those guaranteed by our policies. Additionally, in Section~\ref{apx:badadv} in the Appendix, for $\beta = 0.8$, we design an adversary that induces both Hedge and GPS to exceed the upper bound on the regret guaranteed by our policies, thereby demonstrating their sub-optimality.

\subsection{Numerical evaluation of optimal finite-mode policies}\label{sec:numev}
In this section, we test the approach discussed in Section~\ref{sec:heuristic} for designing optimal finite-mode policies in higher dimensional settings. We consider a set of randomly generated repeated decision-making instances with $l=10$ actions for the decision-maker and $m=10$ actions for the adversary. Each instance is generated by drawing losses corresponding to each pair of actions uniformly in the set $[0,1]$.  We generate 100 such instances. In each instance, the corresponding vector-valued game of single-stage regrets has losses with $K=l=10$ dimensions for each pair of actions. Each dimension tracks the additional regret relative to playing each of the $l=10$ actions. We choose $M=10$, resulting in a budget of $M+1=11$ modes. We consider $\beta \in \{0.8,0.9\}$. We use the open-source nonlinear optimization software APOPT available via Gekko, a Python package and server for optimization \cite{beal2018gekko}, to solve our QCLP.  
%Across the 100 instances, the average time taken by APOPT to yield a solution to our QCLP was $173.14$ seconds, with a standard deviation of $41.69$ seconds.

\begin{figure}[]
\begin{minipage}[H]{.45\textwidth}
\vspace{0pt} 
\centering
\includegraphics[width=3in]{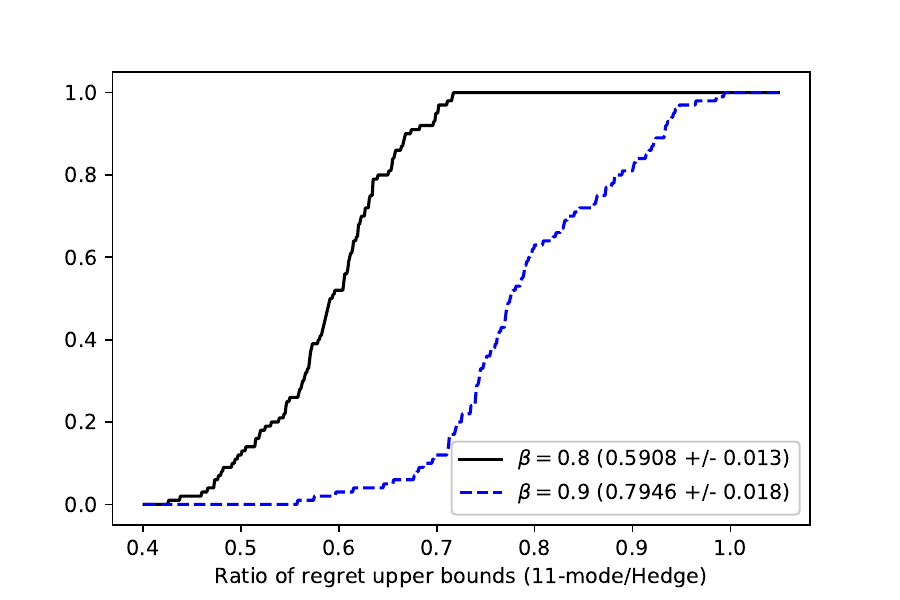}
\captionof{figure}{The empirical c.d.f. of the ratio of the upper bounds guaranteed by our 11-mode policy and by Hedge across 100 instances.}\label{fig:guarat}
%$(N,n)$ values in each case are chosen so that the approximation error (Proposition~\ref{prop:approxfrontier}) is less than 0.06. $n$ is chosen so that $\beta^n/(1-\beta)\leq 0.01$.}
\end{minipage}
\hspace{1cm}
\begin{minipage}[H]{.45\textwidth}
\vspace{0pt} 
\centering
\includegraphics[width=3in]{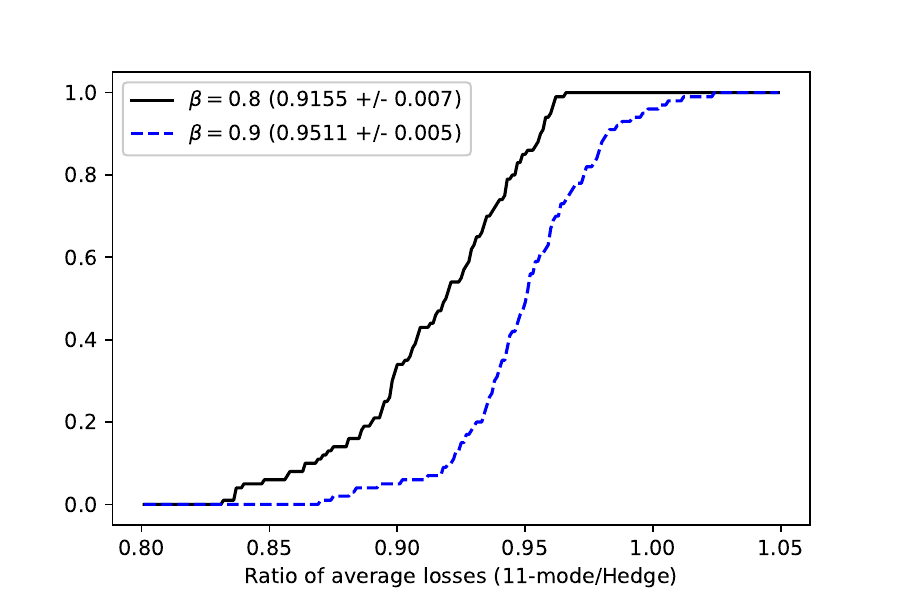}
\captionof{figure}{The empirical c.d.f. of the ratio of the average losses incurred by our 11-mode policy and by Hedge across 100 instances, against 1000 randomly generated adversary action sequences.}\label{fig:avgloss}
\end{minipage}
\end{figure}

In Figure~\ref{fig:guarat}, we plot the empirical cumulative distribution function (c.d.f.) of the ratio of the upper bound guaranteed by our 11-mode stationary policy resulting from the solution of the QCLP and that guaranteed by Hedge, across the 100 instances, for $\beta \in \{0.8,0.9\}$. The mean ratio and its standard error are presented in the legend. When losses are in $[0,a]$, the optimally tuned Hedge algorithm guarantees an average discounted regret of  $a\times\sqrt{\log{K}(1-\beta)/(2(1+\beta))}$; thus the regret upper bound guaranteed by Hedge in each of our instances depends on the maximal loss that the decision-maker can incur in that instance. We find that in all instances and both the settings, the upper bound on the regret guaranteed by our 11-mode stationary policy is smaller than that guaranteed by Hedge. 

Additionally, for each of the $100$ instances, we compare the empirical performance of our 11-mode stationary policy and of Hedge against a set of $1000$ sequences of adversary's actions chosen uniformly at random (for $T = 50$). The empirical c.d.f. of the ratio of average losses incurred by our 11-mode policy and by Hedge across the 100 instances is presented in Figure~\ref{fig:avgloss}. The mean ratio and its standard error are presented in the legend. We observe that our algorithms yield better empirical performance than Hedge across almost all instances in the two settings.

%Against each sequence, we estimate the expected regret incurred by our policy and by Hedge. We finally compare the highest regret incurred by our policy vs. the highest regret incurred by Hedge across the $100$ adversary sequences. The results for the 10 instances with associated error bars are plotted in Figure~\ref{fig:largeKexp2}. We observe that our algorithms yield better empirical performance than Hedge across almost all instances. 
We thus conclude that even though our finite-mode policy is expected to only crudely leverage detailed information about the instance, it can potentially lead to algorithms with significantly better performance than off-the-shelf algorithms in adversarial online learning.

%For $\beta=0.8$, we compare the performance of this algorithm to a $77-mode$ approximately optimal strategy that we derive for the case of two experts over a horizon of length $T=100$. The experiment is performed as follows. In each trial of length $T=100$, the losses are generated randomly and the average discounted regrets of the two algorithms at time $T$ are computed. Each mini-run is of 100 trials (each of length $T=100$), at the end of which we compute the maximum regret that the two algorithms incur w.r.t. Expert 1 and Expert 2 over these trials. Each run is of 100 mini-runs, at the end of which we take the average of the worst-case regrets computed in the mini-runs. Figure~\ref{fig:runs} shows the average worst-case regrets incurred by our algorithm and Hedge, w.r.t. Experts 1 and 2 for 20 runs. It is important to note that this average worst-case regret is expected to be higher than the optimal minimax regret defined in~(\ref{eqn:regretdef}) (since we take the average across mini-runs of the maximum within each mini-run, instead of the other way around, which also means that this is a stronger criterion). As we can see, our algorithm gives a smaller average worst-case regret compared to Hedge on every run. 

%For $\beta\leq 0.5$ our theory leads us to argue that the following trivial strategy is regret-optimal: choose either of the experts with equal probability in the first stage, and from the next stage onwards always choose the expert that incurred no loss in the first stage (repeat if both experts incurred the same loss). A detailed proof is omitted here.
\section{Discussion and conclusion.}\label{sec:conclusion}
We presented a novel approximate dynamic programming approach to approximate the set of minimal guarantees that a player can achieve in a discounted repeated game with vector losses and finite action sets. We showed that this optimal set is the fixed point of a contractive dynamic programming operator and it is the Pareto frontier of a convex and closed set. We also established the structure of the optimal strategies that achieve the different points on this set. We then proposed an iterative procedure to approximately compute this set and also find approximately optimal strategies. The main motivating application of this machinery is to the problem of regret minimization in repeated games within the framework of adversarial online learning.
We illustrated our approach by designing provably approximately optimal strategies for prediction using expert advice with binary losses, for $K=2$ experts. In the process, we demonstrated the suboptimality of well known off-the-shelf adversarial online learning algorithms. Although our approximation approach can become computationally intensive in higher dimensions, we proposed and tested an approach to design well-performing finite-mode policies in such cases based on a QCLP formulation of the problem of finding the optimal stationary strategy with a finite budget of modes. We showed that such policies can result in better performance guarantees compared to existing adversarial online learning algorithms.
%and to approximate the optimal strategy of the uninformed player in Aumann and Maschler's well known model of zero-sum repeated games with incomplete information on one side

It is important to note that adversarial online learning algorithms like Hedge are able to deal with more general adversaries that inflict arbitrary losses lying in a bounded, continuous set, e.g., in $[0,1]$. This power comes at the cost of relatively conservative guarantees on the regret. A drawback of our approach is that we cannot explicitly handle continuous action spaces for the adversary; this is indeed an interesting and important direction for future research. The application of our current approach in such cases would require clustering the vectors of losses inflicted by the adversary based on past data. On the one hand, this approximation may lead to performance loss and one may prefer algorithms like Hedge in such situations. On the other hand, our approach allows one to exploit the structure in adversary's choice of losses, e.g., based on past data, one may conclude that the losses can indeed be effectively clustered, although the choice of the cluster is best modeled as adversarial. In these cases, algorithms like Hedge may be too cautious, and one may wish to incorporate such information in the decision-making process while still seeking robust performance guarantees. Our approach provides a way of doing so by appropriately modifying the set of actions available to the adversary. As we saw from our numerical evaluations in Section~\ref{sec:regmin}, our approach may lead to better performance guarantees than off-the-shelf algorithms like Hedge in these cases.

Finally, we mention that the extension of this approach to the case of long-run average losses in infinitely repeated games appears to be less straightforward, despite the fact that average cost dynamic programming for standard dynamic optimization problems like MDPs is quite well understood. Such an extension, along with the extension to continuous losses, would fill a significant part of the remaining gap in viewing the approximate dynamic programming paradigm as a methodical approach to designing adversarial online learning algorithms. 

\bibliographystyle{plain}
\bibliography{references}
\newpage

\begin{APPENDIX}{}
\section{Proof of all results.}\label{apx:proofs}
\proof{Proof of Lemma \ref{lma:comp}.}
Consider the minimization problem:
$$\min_{\mathbf{x}\in \mathcal{S}} f(\mathbf{x})= \sum_{k=1}^Kx_k.$$
Since $f(\mathbf{x})$ is a continuous function defined on a compact set, it achieves this minimum value at some point $\mathbf{x}^*\in \mathcal{S}$. Hence there cannot be any point $\mathbf{x}'\in\cS$ such that $\mathbf{x}'\prec \mathbf{x}^*$, which means that $\mathbf{x}^*$ is on the Pareto frontier of $\cS$. \hfill\halmos\\
\endproof

\proof{Proof of Proposition \ref{lma:complete}.}

In order to prove the result, we need the following set of results about the Hausdorff distance:
\begin{lemma} \label{lma:hm}

a) $h$ is a metric on the space of closed subsets of $\RR^K$.

%b) Assume that $(\cA_n)_{n\in\mathbb{N}}$ is a Cauchy sequence of closed subsets of $[0, 1]^K$.
%Then there is a unique closed subset $\cA$ of $[0, 1]^K$ such that $h(\cA_n, \cA) \to 0$.
%This set $\cA$ is defined as follows:
%\[
%\cA = \{\bx \in [0, 1]^k :\, \exists \bx_n \in \cA_n\,\forall\,n\, \mbox{ s.t. } \bx_n \to \bx\}.
%\]
b) Assume that $(\cA_n)_{n\in\mathbb{N}}$ is a sequence of closed subsets of $[0, 1]^K$.
Then there is a subsequence $(\cA_{n_k})_{k\in\mathbb{N}}$ that converges to some closet subset $\cA$ of $[0, 1]^K$.\\
c) If the sets $(\cA_n)_{n\in\mathbb{N}}$ in b) are convex, then $\cA$ is convex.\\
d) $h(\bcA, \bcB) \leq h(\cA, \cB)$.

\end{lemma}
\proof{Proof.}

a)-b) This is the well-known property of the Hausdorff distance, and the compactness property of the space of closed subsets of a compact set under the Hausdorff metric;  see \cite{henrikson1999completeness,munkres1975topology}.

d) Say that $\bx, \by \in \cA$.
Then $\bx = \lim_n \bx_n$ and $\by = \lim_n \by_n$ for $\bx_n \in \cA_n$ and $\by_n \in \cA_n$.  By convexity of
each $\cA_n$, $\mathbf{z}_n := \lambda \bx_n + (1 - \lambda)\by_n \in \cA_n$.  But then, $\mathbf{z}_n \to \mathbf{z} := \lambda \bx + (1 - \lambda)\by$.
It follows that $\mathbf{z} \in \cA$, so that $\cA$ is convex.

e) Let $\epsilon := h(\cA, \cB)$.  Pick $\bx \in \bcA$.  Then $\bx = \by + \bv$ for some $\by \in \cA$ and $v \succeq 0$.
There is some $\by' \in \cB$ with $\|\by - \by'\|_{\infty} \leq \epsilon$.  Then $\bx' = \min \{\by' + \bv, \mathbf{1} \} \in \bcB$,
where $\mathbf{1}$ is the vector of ones in $\mathbb{R}^K$, i.e., $(1;k=1,\cdots,K)$, and the minimization is component-wise.  We claim that $\|\bx' - \bx\|_{\infty} \leq \epsilon$.  If $\by' + \bv \in [0, 1]^K$, this is clear.  Assume $y^\prime_k +v_k> 1$.  Then,
\[
x^\prime_k = 1 < y^\prime_k + v_k \mbox{ and } x_k = y_k + v_k \leq 1.
\]
Thus,
\[
0 \leq x^\prime_k - x_k < y^\prime_k + v_k - y_k - v_k = y^\prime_k - y_k.
\]
Hence, $|x^\prime_k - x_k| \leq |y^\prime_k - y_k|$ for any $k$. Thus, one has $\|\bx' - \bx\|_{\infty} \leq \|\by' - \by\|_{\infty} \leq \epsilon.$
 
\hfill\halmos\\
\endproof

Now we can prove the proposition. First, to show that $d$ is a metric on $\cF$, we just need to show that if $h(up(\cV),up(\cU))=0$ then $\cV= \cU$. The other properties (e.g., triangle inequality etc.) follow from the corresponding properties for the Hausdorff metric. Note that if $h(up(\cV),up(\cU))=0$, then $up(\cV)=up(\cU)$. Suppose that there is some $\bu\in\cU$ such that $\bu\notin \cV$. But since $\bu\in up(\cV)$, we have $\bu = \bv +\by$ for some $\bv\in\cV$ and some $\by\succeq \mathbf{0}$. But since $h(up(\cV),up(\cU))=0$, by the definition of the Hausdorff distance, for each $\epsilon>0$, there is a point $\bu_\epsilon$ in $up(\cU)$ such that $\bu_\epsilon\preceq \bv +\epsilon\mathbf{1}$. Consider a sequence $(\epsilon_n)_{n\in\mathbb{N}}$ such that $\epsilon_n \rightarrow 0$, and consider the corresponding sequence  $(\bu_{\epsilon_n})_{n\in\mathbb{N}}$. Now since $up(\cU)$ is compact, $(\bu_{\epsilon_n})_{n\in\mathbb{N}}$ has a convergent subsequence that converges to some $\bu^*\in up(\cU)$ such that $\bu^*\preceq \bv \preceq \bv +\by =\bu$, which contradicts the fact that $\bu \in \cU$. Thus $\cU=\cV$.

%Next, we prove statement (b). Under the given assumptions, $(\bcVn)_{n\in\mathbb{N}}$ is Cauchy in
%the Hausdorff metric, so that, by Lemma \ref{lma:hm}, 
%there is a unique closed convex set such that
%$h(\bcVn, \cA) \to 0$. But since $h(\bcVn,\bcA)\leq h(\bcVn,\cA)$ (from Lemma \ref{lma:hm}), we have that \\
%$h(\bcVn, \bcA) \to 0$ and hence $\bcA=\cA$. Thus the Pareto frontier $\cV$ of $\cA$
%is then such that $d(\cV_n, \cV) \to 0$. To show uniqueness of $\cV$, assume that there is some $\cU \in \cF$
%such that $d(\cV_n, \cU) \to 0$.  Then, the closed convex set $\bcU$
%is such that $h(\bcVn, \bcU) \to 0$.  By Lemma \ref{lma:hm},
%this implies that $\bcU = \bcV$, so that $\cU = \cV$ (from part (a) of the proposition) . 

Next, we prove statement (b). From statement (b) and (c) of Lemma \ref{lma:hm}, the subsequence $(up(\cV_{n_k}))_{k\in\mathbb{N}}$ converges to some convex set $\cA$. But since $h(up(\cV_{n_k}),\bcA)\leq h(up(\cV_{n_k}),\cA)$, we have $\bcA=\cA$. And thus the subsequence $(\cV_{n_k})_{k\in\mathbb{N}}$ converges to the Pareto frontier of $\cA$, which is in $\cF$. 

%Statement (c) directly follows from statements (a) and (b) since any compact metric space is complete.
Observe that it becomes clear from the above arguments that  $d$ induces these properties not just on $\cF$, but also on the more general space of Pareto frontiers in $[0,1]^K$ whose upset is closed.
 \hfill\halmos\\
\endproof

\proof{Proof of Proposition~\ref{prop:hausdorffequi}.}
Suppose that $\max(e(\cU,\cV),e(\cV,\cU))\leq \epsilon$. Consider a point $\bx\in \bcU$ such that $\bx=\by+\bv$ where $\by\in \cU$ and $\bv\succeq 0$. Suppose that there is no $\bx'\in \bcV$ such that $\|\bx-\bx'\|_\infty\leq \epsilon$, i.e., for any $\bx' \in \bcV$, $\|\bx-\bx'\|_\infty> \epsilon$. This means that $\bcV$ is a subset of the region $\{\bx': x'_k> x_k +\epsilon\textrm{ for some } k\}$ (this is the region $\cS$ shown in the Figure \ref{fig:metrics}). But since $\by=\bx-\bv$, we have $\by\preceq \bx$ ($\by$ is in region $\cS'$ shown in the Figure \ref{fig:metrics}). But then for any $\mathbf{w}\in \cS'$, $\|\by-\mathbf{w}\|_\infty> \epsilon$. This contradicts the fact that for $\by$ there is some $\by'\in \cV$, such that $\by+\epsilon\bone\succeq \by'$. Thus $d(\cU,\cV)\leq \epsilon$. 
\begin{figure}[htb]
\begin{center}
\includegraphics[width=2in,angle=0]{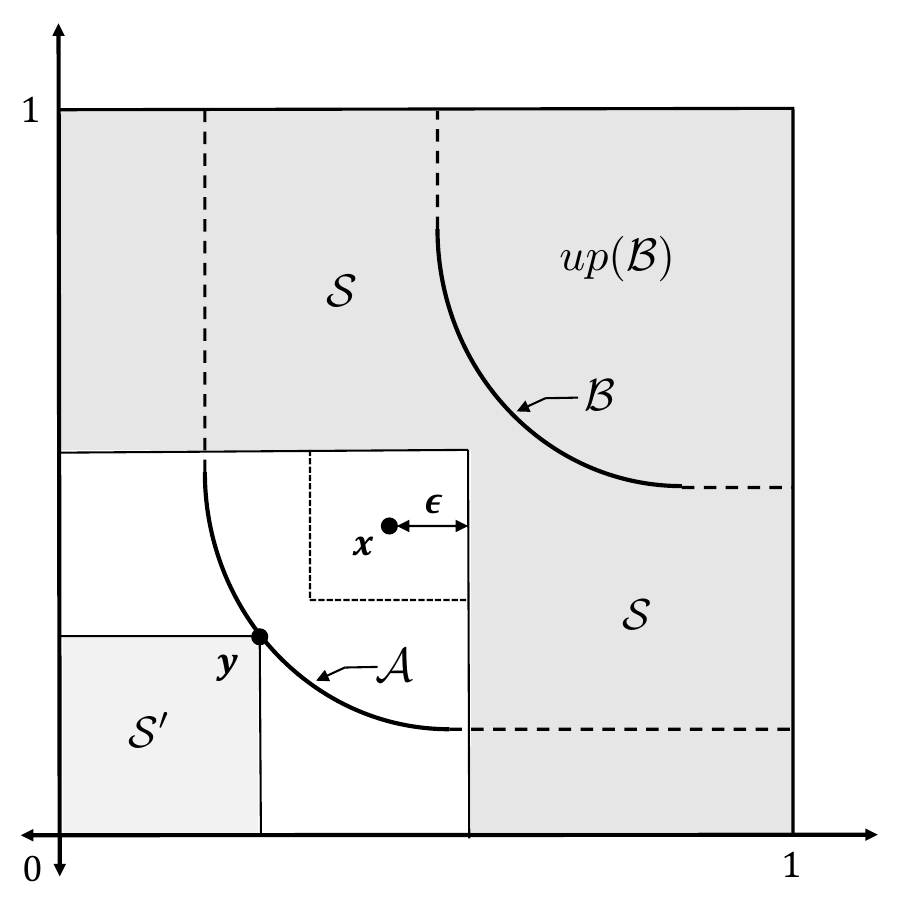}
\caption{Construction in $[0,1]^2$ for the proof of Proposition~\ref{prop:hausdorffequi}.}
\label{fig:metrics}
\end{center}
\end{figure} 
Now suppose that $d(\cU,\cV) \leq \epsilon$. Then for any $\bx\in \cU$, there is a $\bx'\in \bcV$ such that $\|\bx-\bx'\|_\infty\leq \epsilon$ where $\bx'=\by+\bv$ for $\by\in \cV$ and $\bv\succeq 0$. Thus $\bx+\epsilon\bone\succeq \bx'=\by+\bv$. The roles of $\cU$ and $\cV$ can be reversed. Thus $\max(e(\cU,\cV),e(\cV,\cU))\leq \epsilon$. Observe that this proof uses the fact that the $\sup$ and $\inf$ in the definitions of $h$ and $e$ can be replaced by $\max$ and $\min$ respectively, which is valid for the space $\cF$ as discussed in footnotes~\ref{fn:hausdorff} and~\ref{fn:e}.\hfill\halmos\\
\endproof

\proof{Proof of Lemma~\ref{lma:box}:}
In order to prove this lemma, we need a few intermediate results. We define the following notion of convexity of Pareto frontiers.

\begin{definition}\label{imp}
A Pareto frontier $\mathcal{V}$ is p-convex if for any $\bv,\,\bu \in \mathcal{V}$ and for each $\lambda\in [0,1]$, there exists a point $\br \in \mathcal{V}$ such that $\br\preceq\lambda\bv+(1-\lambda)\bu$.
\end{definition}

We then show the following equivalence.
\begin{lemma}\label{lma:equi}
For a Pareto frontier $\mathcal{V}\subset [0,1]^K$, the following statements are equivalent:
\begin{itemize}
\item[1.] $\mathcal{V}$ is in $\cF$.

\item[2.] $\mathcal{V}\subseteq [0,1]^K$ is p-convex and $up(\cV)$ is closed.
\end{itemize}
\end{lemma}
\proof{Proof.}
To show that 1 implies 2, we just need to show that $\cV$ is p-convex. To see this, suppose that $\bu$ and $\bv$ are two points in $\mathcal{V}$. Since they also belong to $up(\cV)$, which is convex, for each $\lambda\in[0,1]$, $\lambda\bu+(1-\lambda)\bv\in up(\cV)$ and thus there is some $\br \in\mathcal{V}$ such that $\br \preceq\lambda\bu+(1-\lambda)\bv$. Thus $\mathcal{V}$ is p-convex. 

To show that 2 implies 1, we just need to show that $up(\cV)$ is convex if $\cV$ is p-convex. To see this, suppose that $\bu + \bx$ and $\bv +\by$ are two points in $\up(\cV)$ where $\bu,\,\bv\in \cV$ and $\bx,\,\by\succeq 0$. By p-convexity of $\cV$, for each $\lambda\in[0,1]$, there is a $\br\in \cV$ such that $\br \preceq \lambda\bu+(1-\lambda)\bv$ and thus $\br \preceq \lambda(\bu + \bx) + (1-\lambda) (\bv + \by)$. Thus $up(\cV)$ is convex.

\hfill\halmos\\
\endproof

We can now prove Lemma~\ref{lma:box}.
%Consider the following set 
%$$\Psi(\mathcal{S})=\bigg\{\bigg(\max_{b^j}\bigg\{\sum_{a\in A}\alpha_i\big[r_1(a^i,b^j)+\beta R_1(a^i,b^j)\big]\bigg\},\max_{b^j}\bigg\{\sum_{a\in A}\alpha_i\big[r_2(a^i,b^j)+\beta R_2(a^i,b^j)\big]\bigg\}\bigg)$$
%\begin{equation}
%: (R_1(a^i,b^j),R_2(a^i,b^j))\in \mathcal{S},\bm{\alpha}\in\\mathcal{S}igma_{A^1}\bigg\}.
%\end{equation}
%Note that $\Lambda(\Psi(\mathcal{S}))=\Lambda(\Psi(\mathcal{V}))$. We will show that if $\mathcal{S}$ is closed then $\Psi(\mathcal{S})$ is closed as well.
Recall that,
$$\Psi(\cV)=\bigg\{\bigg(\max_{b\in B}\bigg\{\sum_{a\in A}\alpha_a\big[r_k(a,b)+\beta R_k(a,b)\big]\bigg\}; k=1,\cdots, K\bigg)$$
\begin{equation*}
: \bm{\alpha}\in\Delta(A),\,\,\bR(a,b)\in \cV\,\forall\, a\in A,\,b\in B\bigg\}.
\end{equation*}
First, note that $\Lambda(\Psi(\cV)) = \Lambda(\Psi(up(\cV)))$. Now one can see that $\Psi(up(\cV))$ is the image of a continuous function from the product space $up(\cV)^{m\times n}\times \Delta(A)$ to a point in $\mathbb{R}^K$, which is a Hausdorff space. Since $up(\cV)$ is closed and bounded, it is compact. Also the simplex $\Delta(A)$ is compact. Thus the product space $up(\cV)^{m\times n}\times \Delta(A)$ is compact. Hence by the closed map lemma, $f$ is a closed map and hence $\Psi(up(\cV))$ is closed. Hence $up(\Lambda(\Psi(\cV)))$ is closed.

Next, recall that any point $\bu$ in $\Lambda(\Psi(\cV))$ is of the form: 
$$\bu=\bigg(\max_{b\in B}\bigg\{\sum_{a\in A}\alpha_a\big[r_k(a,b)+\beta R_k(a,b)\big]\bigg\}; k=1,\cdots,K\bigg)$$
for some $\bm{\alpha}\in \Delta(A)$ and $\bR(a,b)\in \cV$. But since $\cV$ is p-convex from Lemma~\ref{lma:equi}, for each $b\in B$, there exists some $\bQ(b)\in \cV$ such that $\bQ(b)\preceq \sum_{a=1}^{m}\alpha_a\bR(a,b)$. Hence statement $2$ follows.

Now let $$\bu=\bigg(\max_{b\in B}\bigg\{\sum_{a\in A}\alpha_ar_k(a,b)+\beta Q_k(b)\bigg\}; k=1,\cdots,K\bigg)$$ and $$\bv=\bigg(\max_{b\in B}\bigg\{\sum_{a\in A}\eta_ar_k(a,b)+\beta R_k(b)\bigg\}; k=1,\cdots,K\bigg)$$ be two points in $\Lambda(\Psi(\cV))$, where $\bm{\alpha},\,\bm{\eta} \in\Delta(A)$ and $\bQ(b),\,\bR(b)\in \cV$ for all $b\in B$. For a fixed $\lambda\in[0,1]$, let $\bz=\bm{\alpha}\lambda +\bm{\eta}(1-\lambda)$. Then 
\begin{align*}
&\lambda\bu+(1-\lambda)\bv\\
&~=\bigg(\lambda\max_{b\in B}\bigg\{\sum_{a\in A}\alpha_ar_k(a,b)+\beta Q_k(b)\bigg\}+(1-\lambda)\max_{b\in B}\bigg\{\sum_{a\in A}\eta_ar_k(a,b)+\beta R_k(b)\bigg\}; k=1,\cdots,K\bigg)\\
&~~\succeq \bigg(\max_{b\in B}\bigg\{ \sum_{a\in A}z_a r_k(a,b)+\beta [\lambda Q_k(b)+(1-\lambda) R_k(b)]\bigg\}; k=1,\cdots,K\bigg)\\
&~~\succeq \bigg(\max_{b\in B}\bigg\{ \sum_{a\in A}z_a r_k(a,b)+\beta L_k(b)\bigg\}; k=1,\cdots,K\bigg).
\end{align*}
The first inequality holds since $\max$ is a convex function and the second follows since $\cV$ is p-convex, and hence $\mathbf{L}(b)\in \cV$ that satisfy the given relation exist. Thus $\Lambda(\Psi(\cV))$ is p-convex. Combined with the fact that $up(\Lambda(\Psi(\cV)))$ is closed, this implies that $\Lambda(\Psi(\cV))\in\cF$ using Lemma~\ref{lma:equi}. \hfill\halmos\\
\endproof

%\subsection{Proof of Lemma 3.2:  Dynamic programming operator $\Phi$ is a contraction}

\proof{Proof of Lemma~\ref{lma:contraction}.}
Suppose $e(\mathcal{U},\mathcal{V})=\epsilon$. Let 
$$\bigg(\max_{b\in B}\bigg\{\sum_{a\in A}\alpha_ar_k(a,b)+\beta R_k(b)\bigg\};k=1,\cdots,K\bigg)$$ 
be some point in $\Phi(\mathcal{V})$, where $\bm{\alpha}\in\Delta(A)$. Then for each $b$, we can choose $\bQ(b)\in \mathcal{U}$ such that $\bQ(b)\preceq \bR(b) +\epsilon\mathbf{1} $. We then have 
\begin{eqnarray*}
\max_{b\in B}\bigg\{\sum_{a\in A}\alpha_ar_k(a,b)+\beta Q_k(b)\bigg\}&=&\max_{b\in B}\bigg\{\sum_{a\in A}\alpha_ar_k(a,b)+\beta R_k(b)+\beta(Q_k(b)-R_k(b))\bigg\}\\
&\leq& \max_{b\in B}\bigg\{\sum_{a\in A}\alpha_ar_k(a,b)+\beta R_k(b)+\beta\epsilon\bigg\}\\
&=&\max_{b\in B}\bigg\{\sum_{a\in A}\alpha_ar_k(a,b)+\beta R_k(b)\bigg\}+\beta \epsilon.
\end{eqnarray*}
Thus 
\begin{eqnarray*}
&&\bigg(\max_{b\in B}\bigg\{\sum_{a\in A}\alpha_ar_k(a,b)+\beta Q_k(b)\bigg\};k=1,\cdots,K\bigg)\\
&&~~\preceq\bigg(\max_{b\in B}\bigg\{\sum_{a\in A}\alpha_ar_k(a,b)+\beta R_k(b)\bigg\};k=1,\cdots,K\bigg)+\beta \epsilon\mathbf{1}.
\end{eqnarray*}
But since 
$$\bigg(\max_{b\in B}\bigg\{\sum_{a\in A}\alpha_ar_k(a,b)+\beta Q_k(b)\bigg\};k=1,\cdots,K\bigg)\in \Psi(\mathcal{U}),$$
 and since $\Phi(\mathcal{U})=\Lambda (\Psi(\mathcal{U}))$, there exists some $\mathbf{L}\in \Phi(\mathcal{U})$ such that 

$$\mathbf{L} \preceq \bigg(\max_{b\in B}\bigg\{\sum_{a\in A}\alpha_ar_k(a,b)+\beta Q_k(b)\bigg\};k=1,\cdots,K\bigg).$$
Thus
$$\mathbf{L}\preceq  \bigg(\max_{b\in B}\bigg\{\sum_{a\in A}\alpha_ar_k(a,b)+\beta R_k(b)\bigg\};k=1,\cdots,K\bigg)+\beta \epsilon\mathbf{1}.$$
Thus
\begin{equation}\label{contract}
e(\Phi(\mathcal{U}),\Phi(\mathcal{V}))\leq \beta\epsilon= \beta e(\mathcal{U},\mathcal{V}).
\end{equation}
\hfill\halmos\\
\endproof

%\proof{Proof of Theorem~\ref{thm:main1}.}
%Since $\Phi$ is a contraction in the complete the metric space $\cF$, the statement follows from the Banach fixed point theorem. \hfill\halmos\\
% the sequence $\{\cA_n\}$  for any $\cV$ converges to is Cauchy in $\cF$. 
%Hence by Lemma~\ref{lma:hm}, $\{\cA_n\}$ converges to a Pareto frontier $\mathcal{V^*} \in \cF$. 
%The continuity of the operator further implies that $\mathcal{V^*}=\Phi(\cV^*)$.
%To show uniqueness, observe that if there are two fixed points $\cU $ and $\cV$, then we have $d(\cU ,\cV)=d(\Phi(\cU ,\Phi(\cV))\leq \beta d(\cU ,\cV)$, which implies that $d(\cU ,\cV)=0$ and hence $\cU =\cV$.
%\endproof

 \proof{Proof of Theorem~\ref{thm:main2}.}
In $\mathbb{G}^{\infty}$, fix $T \geq 1$ and consider a truncated game where Alice can guarantee the cumulative losses in $\beta^{T+1} \mathcal{V}^*$ after time $T+1$. Then the minimal losses that she can guarantee after time $T$ is the set: 
\[
\Lambda\bigg(\bigg\{\big(\max_{b\in B}\beta^{T}\sum_{a\in A}\alpha_a r_k(a, b) + \beta^{T+1} Q_k( b);k=1,\cdots,K\big) 
\mid \bm{\alpha} \in \Delta(A),\,\, \bQ(b) \in \mathcal{V}^*\, \forall\, b \in B\bigg\}\bigg).
\]
This set is $\beta^T \mathcal{V}^*$.  By induction, this implies that the set of minimal losses that she can guarantee after time $0$ is $\mathcal{V}^*$.

The losses of the truncated game and of the original game differ only after time $T+1$.  Since the losses at each step
are bounded by $(1-\beta)$, the cumulative losses after time $T+1$ are bounded by $\frac{\beta^{T+1}(1-\beta)}{1-\beta}=\beta^{T+1}$.  Consequently, the minimal losses of the original game must be in the set
\[
\bigg\{ \bu\in [0,1]^K: u_k\in [x_k- \beta^{T+1}, x_k+ \beta^{T+1}] \textrm{ for all }k, \,\, x \in \cV^* \bigg\}.
\]
Since $T \geq 1$ is arbitrary, the minimal losses that Alice can guarantee in the original game must be in $\cV^*$. \hfill\halmos\\
\endproof

\proof{Proof of Theorem~\ref{thm:optstrategy}.}
Assume that Alice can guarantee every pair $\beta^{T+1} \bu$ of cumulative losses with $\bu \in \cV^*$ after time $T+1$ by choosing some continuation strategy in $\Pi_A$. Let $\bx = \bF(\bp, \cV^*)$. We claim that after time $T$, Alice can guarantee a loss of no more than $\beta^T\bx$ on each component by first choosing $a_T = a$ with probability $\alpha_a(\bp)$ and then if Bob
chooses $b \in B$, choosing a continuation strategy that guarantees her $\bF(\bp', \cV^*)$, where $p' = \bq(b,\bp)$. Indeed by following this strategy, her expected loss on component $k$ after time $T$ is
then
\[
\{\beta^T \sum_a \alpha_a(\bp)r_k (a, b) + \beta^{T+1} F_k(\bq(b,\bp),\cV^*) \} \leq \beta^T F_k(\bp,\cV^*) = \beta^T x_k.
\]
Thus, this strategy for Alice guarantees that her loss after time $T$ is no more than $\beta^T \cV^*$.  
Hence by induction, following the indicated strategy (in the statement of the theorem) for the first $T$ steps and then using the continuation strategy from time $T+1$ onwards, guarantees that her loss is not more than $\bF(\bp_1,\cV^*)$ after time $0$. Now, even if Alice plays arbitrarily after time $T+1$ after following the indicated strategy for the first $T$ steps, she still guarantees that her loss is (componentwise) no more than $\bF(\bp_1,\cV^*)+\beta^{T+1}(1;k=1,\cdots,K)^T$. Since this is true for arbitrarily large values of $T$, playing the given strategy indefinitely guarantees that her loss is no more than $\bF(\bp_1,\cV^*)$. \hfill\halmos\\
\endproof

\proof{Proof of Proposition~\ref{lma:approx}.}
Any point $\mathbf{e}$ in $\Gamma_{N}(\cV)$ is of the form $\sum_{k=1}^M\lambda_k\bv_k$ where $\bv_k\in up(\cV)$  and $\sum_{k=1}^M\lambda_k=1$, and $M\leq K$. But then by the definition of an upset, we have $\bv'_k\in\cV$ for each $k$ such that $\bv'_k\preceq \bv_k$ and hence $\sum_{k=1}^M\lambda_k\bv'_k\preceq \sum_{k=1}^M\lambda_k\bv_k$. By the p-convexity of $\mathcal{V}$, there is some $\br \in \cV$, such that $\br\preceq \sum_{k=1}^M\lambda_k\bv'_k$, and hence $\br\preceq \mathbf{e}$. Thus $e(\Gamma_{N}(\cV),\cV)=0$.

Next, we will show that for any $\bu \in \cV$, there exists $\mathbf{e}\in\Gamma_{N}(\cV)$ such that $\mathbf{e}\preceq \bu +(1/N)\mathbf{1}$.
%$$\min \bigg\{||\bu- \bv||_{\infty}: \bv\in \Gamma_{N}(\cV)\bigg\}\leq \frac{1}{N}.$$
For the rest of the proof, all the distances refer to distances in the $\mathcal{L}^\infty$ norm. Consider a line $\mathbf{x}=t\mathbf{1}+\bp$, and suppose that the shortest distance between $\bu$ and any point on this line is $a>0$, i.e.,  $\min \big\{||\bu- \mathbf{x}||_{\infty}: \mathbf{x}=t\mathbf{1}+\bp\big\} = a$. Let $\mathbf{x}^*=t^*\mathbf{1}+\bp$ be the point on the line that is closest to $\bu$. If $a^+ \triangleq \max\{(x^*_k-u_k)^+:k=1,\cdots,K\}$ and $a^- \triangleq \max\{-(x^*_k-u_k)^-:k=1,\cdots,K\}$, then $a= \max\{a^+,a^-\}$. Consider any point $\bv$ that is the smallest point of intersection of $\mathbf{x}=t\mathbf{1}+\bp$ and the set $up(\cV)$. Then this point must lie in the set $\{t\mathbf{1}+\bp: t\in [t^*-a^+, t^* +a^-]\}$, because 
a) if $\bv= t'\mathbf{1}+\bp$ for some $t'< t^*-a^+$, then it means that $\bu$ dominates $\bv$ which contradicts the fact that $\bv \in up(\cV)$, and b) if $\bv= t'\mathbf{1}+\bp$  for some $t'> t^*+a^-$ then $\bv$ will \emph{strictly} dominate $\bu$ on each dimension, but then the point $\bv'= (t^*+a^-) \mathbf{1}+\bp$ is strictly smaller than $\bv$ and lies in $up(\cV)$ and on the line $\bv= t'\mathbf{1}+\bp$, which contradicts the definition of $\bv$. Thus $||\bu- \bv||_{\infty} \leq a^+ +a^- \leq 2a$. Thus we have shown that if the shortest distance between $\bu$ and some line $\mathbf{x}=t\mathbf{1}+\bp$ is $a$, then the distance between $\bu$ and the smallest point of intersection of $\mathbf{x}=t\mathbf{1}+\bp$ and the set $up(\cV)$ is no more than $2a$. 

Now we will show that the for any $\bu\in\cV$, there is always a line $\mathbf{x}=t\mathbf{1}+\bp$ such that the shortest distance between $\bu$ and the line is no more than $1/(2N)$. Let $u_{\min} = \min_k\{u_k\}$. Then $\bu = u_{\min}\mathbf{1} + (\bu - u_{\min}\mathbf{1})$. Now the vector $(\bu - u_{\min}\mathbf{1})$ has value $0$ on one dimension, and on every other dimension it has value in $[0,1]$ (since $\bu\in[0,1]^K$), and so it can be approximated by some $p_k\in \{0,1/N,\cdots,(N-1)/N,1\}$ where the approximation error on any dimension is at most $1/(2N)$. Thus there is a point 
$\mathbf{e}'  = u_{\min}\mathbf{1} + \bp$ where $\bp\in\mathcal{P}_N$ such that $\|\bu - \mathbf{e}'\| \leq 1/(2N)$. Thus there us always a line $\mathbf{x}=t\mathbf{1}+\bp$ such that the shortest distance between $\bu$ and the line is no more than $1/2N$.

Together, we finally have that for any $\bu\in\cV$ there is some point $\mathbf{e}''$, which is the smallest point of intersection of some line $\mathbf{x}=t\mathbf{1}+\bp$ and the set $up(\cV)$, such that $\|\bu-\mathbf{e}''\|_{\infty}\leq 2\times (1/2N) = 1/N$, and thus $\mathbf{e}'' \preceq \bu + (1/N)\mathbf{1}$. Since there is always some point $\mathbf{e}\in\Gamma_{N}(\cV)$ such that $ \mathbf{e}\preceq\mathbf{e}''$ (recall the definition~\eqref{eq:defapprox} of $\Gamma_{N}(\cV)$ as the Pareto frontier of the set $\ch\big(\big\{\bF(\bp,\cV): \bp\in \mathcal{P}_N \big\}\big)$), we have $\mathbf{e}\preceq \bu + (1/N)\mathbf{1}$. Thus $e(\cV,\Gamma_{N}(\cV))\leq 1/N$.
\hfill\halmos\\
\endproof

\proof{Proof of Proposition~\ref{prop:approxfrontier}.}
We have $\mathcal{G}_n=\Gamma_N\circ\Phi(\mathcal{G}_{n-1}))$. Consider another sequence of Pareto frontiers 
\begin{equation}
\bigg(\mathcal{A}_n=\Phi^n(\mathcal{G}_0)\bigg)_{n\in\mathbb{N}}.
\end{equation}
Then we have 
\begin{eqnarray}
d(\mathcal{A}_n,\mathcal{G}_n)&=&d(\Phi(\mathcal{A}_{n-1}),\Gamma_{N}(\Phi(\mathcal{G}_{n-1})))\nonumber\\
&\stackrel{(a)}\leq& d(\Phi(\mathcal{A}_{n-1}),\Phi(\mathcal{G}_{n-1}))+d(\Phi(\mathcal{G}_{n-1}),\Gamma_{N}(\Phi(\mathcal{G}_{n-1})))\nonumber\\
&\stackrel{(b)}\leq& \beta d(\mathcal{A}_{n-1},\mathcal{G}_{n-1}) +\frac{1}{N},
\end{eqnarray}
 where inequality (a) is the triangle inequality and (b) follows from (\ref{contract}) and Lemma \ref{lma:approx}. Coupled with the fact that $d(\mathcal{A}_0,\mathcal{G}_0)=0$, we have that 
 \begin{eqnarray}
 d(\mathcal{A}_n,\mathcal{G}_n)&\leq&\frac{1}{N}\bigg(1+\beta+\beta^2+\cdots\beta^{n-1}\bigg)\nonumber\\
&=&\frac{1}{N}\bigg(\frac{1-\beta^n}{1-\beta}\bigg).
 \end{eqnarray}
 
Since $\Phi$ is a contraction, the sequence $\{\mathcal{A}_n\}$ converges to some Pareto frontier $\mathcal{V}^*$. 
Suppose that we stop the generation of the sequences $\{\mathcal{A}_n\}$ and $\{\mathcal{G}_n\}$ at some $n$. Now since $\mathcal{A}_0=\mathcal{G}_0=\{\mathbf{0}\}$, and since the stage losses $r_k(a,b)\in [0,1-\beta]$, we have that $d(\mathcal{A}_1, \mathcal{A}_{0})\leq 1-\beta$. From the contraction property, this implies that $d(\mathcal{A}_{n+1}, \mathcal{A}_{n})\leq \beta^n(1-\beta)$. Thus $d(\mathcal{V}^*,\mathcal{A}_n)\leq \frac{\beta^n (1-\beta)}{1-\beta}=\beta^n$, and thus by triangle inequality we have
\begin{equation}
d(\mathcal{V}^*,\mathcal{G}_n)\leq \frac{1}{N}\bigg(\frac{1-\beta^n}{1-\beta}\bigg)+\beta^n.
\end{equation}
Finally, to show that $e(\mathcal{G}_n,\cV^*)\leq \beta^n$, observe that

\begin{eqnarray}
e(\mathcal{G}_n, \mathcal{A}_n)&=&e(\Gamma_{N}(\Phi(\mathcal{G}_{n-1})),\Phi(\mathcal{A}_{n-1}))\nonumber\\
&\stackrel{(a)}\leq& e(\Gamma_{N}(\Phi(\mathcal{G}_{n-1})),\Phi(\mathcal{G}_{n-1}))+ e(\Phi(\mathcal{G}_{n-1}),\Phi(\mathcal{A}_{n-1}))\nonumber\\
&\stackrel{(b)}\leq& 0 + \beta e(\mathcal{G}_{n-1},\mathcal{A}_{n-1}).
\end{eqnarray}
Since $\mathcal{A}_0=\mathcal{G}_0=\{\mathbf{0}\}$, this implies that $e(\mathcal{G}_n, \mathcal{A}_n)=0$ for all $n$. Here, (a) holds since if for three frontiers $\cU$, $\cV$ and $\mathcal{L}$, $\cU$ $\epsilon_1$-dominates $\cV$ and $\cV$ $\epsilon_2$-dominates $\mathcal{Z}$, then $\cU$ $(\epsilon_1+\epsilon_2)$-dominates $\mathcal{Z}$. (b) follows from the contraction property of $\Phi$ under $e$. Further, $e(\mathcal{A}_n,\mathcal{V}^*)\leq d(\mathcal{A}_n,\mathcal{V}^*)\leq\beta^n$ from above. Thus we have $e(\mathcal{G}_n,\cV^*)\leq e(\mathcal{G}_n, \mathcal{A}_n) + e(\mathcal{A}_n,\mathcal{V}^*)\leq\beta^n$.

\hfill\halmos\\
\endproof

\proof{Proof of Proposition~\ref{prop:approx2}.}
In order to prove this result, we need a few intermediate definitions and results.
First, we need to characterize the losses guaranteed by any $H(K,N)-$mode stationary strategy. Such a strategy $\pi$ defines the following operator on any function $\bF:\cP_N\rightarrow\mathbb{R}^K$ ($\cP_N$ is defined in \eqref{set:pn}):
\begin{equation}
\Delta_N^{\pi}(\bF)(\bp)=\bigg(\max_{b\in B}\big\{\sum_{a\in A}\alpha_a(\bp)r_k(a,b)+\sum_{k'=1}^Kz_{k'}(b,\bp) \beta F_k(\bq_{k'}(b,\bp))\big\};k=1,\cdots,K\bigg),
\end{equation}
where $\bq_{k'}(b,\bp)\in \cP_N$ for all $k'$. Now for a function $\bF:\cP_N\rightarrow\mathbb{R}^K$, define the following norm:

$$\|\bF\|=\max_{\bp\in \cP_N}\|\bF(\bp)\|_{\infty}.$$
It is easy to show that $\Delta^\pi_N$ is a contraction in the norm. We omit the proof for the sake of brevity.

\begin{lemma}
\begin{equation}
\|\Delta_N^{\pi}(\bF)-\Delta_N^{\pi}(\mathbf{G})\|\leq \beta \|\bF-\mathbf{G}\|.
\end{equation}
\end{lemma}
We can then show the following result.
\begin{lemma}
Consider a $H(K,N)$-mode strategy $\pi$. Then there is a unique function 
$$\bF^{\pi}:\cP_N\rightarrow \mathbb{R}^K$$
 such that $\Delta^\pi_N(\bF^{\pi})=\bF^{\pi}$. Further, The strategy $\pi$ initiated at mode $\bp$ where 
$\bp\in\cP_N$ guarantees the vector of losses $\bF^{\pi}(\bp)$.
\end{lemma}
The first part of the result follows from the fact that the operator is a contraction and the completeness of the space of vector-valued functions with a finite domain for the given norm. The second part follows from arguments similar to those in the proof of Theorem~\ref{thm:optstrategy}. The arguments are not repeated here for the sake of brevity. Now let 
$$\mathcal{V}^{\pi_n}\triangleq\Lambda\bigg(ch(\{\bF^{\pi_n}(\bp):\bp\in\cP_N)\bigg),$$ where $\bF^{\pi_n}$ is the fixed point of the operator $\Delta_N^{\pi_n}$.

Define a sequence of functions $\bF^n:\cP_N\rightarrow \mathbb{R}^K$ where $\bF^n(\bp)=\bF(\bp,\Phi(\mathcal{G}^{n-1}))=\bF(\bp,\mathcal{G}^n)$. We then have that
\begin{eqnarray}
d(\mathcal{V}^{\pi_n},\mathcal{V}^*)&\leq& d(\mathcal{V}^{\pi_n},\mathcal{G}_n)+d(\mathcal{G}_n,\mathcal{V}^*)\nonumber\\
&\leq& d(\mathcal{V}^{\pi_n},\mathcal{G}_n)+\frac{1}{N}\bigg(\frac{1-\beta^n}{1-\beta}\bigg)+\beta^n.
\end{eqnarray}
From the definition of $d$, it is clear that $d(\mathcal{V}^{\pi_n},\mathcal{G}_n)\leq \|\bF^{\pi_n}-\bF^n\|$. Next we have
\begin{eqnarray}
 \|\bF^{\pi_n}-\bF^n\|&\leq&  \|\bF^{\pi_n}-\Delta_N^{\pi_n}(\bF^n)\|+\|\Delta_N^{\pi_n}(\bF^n)-\bF^n\|\nonumber\\
 &\stackrel{(a)}=&\|\Delta_N^{\pi_n}(\bF^{\pi_n})-\Delta_N^{\pi_n}(\bF^n)\|+\|\bF^{n+1}-\bF^n\|\nonumber\\
&\stackrel{(b)}\leq&\beta \|\bF^{\pi_n}-\bF^n\|+\|\bF^{n+1}-\bF^n\|.
\end{eqnarray}
Here $(a)$ holds because $\Delta_N^{\pi_n}(\bF^n)=\bF^{n+1}$ by the definition of the strategy $\pi_n$, and because $\bF^{\pi_n}$ is a fixed point of the operator $\Delta_N^{\pi_n}$. $(b)$ holds because $\Delta_N^{\pi_n}$ is a contraction. Thus we have
\begin{equation}
d(\mathcal{V}^{\pi_n},\mathcal{G}_n)\leq \|\bF^{\pi_n}-\bF^n\| \leq \frac{\|\bF^{n+1}-\bF^n\|}{1-\beta}.
\end{equation}
And finally we have:
\begin{equation}\label{inter}
d(\mathcal{V}^{\pi_n},\mathcal{V}^*)\leq \frac{1}{N}\bigg(\frac{1-\beta^n}{1-\beta}\bigg)+\beta^n+ \frac{\|\bF^{n+1}-\bF^n\|}{1-\beta}.
\end{equation}
To finish up, we need the following result:

\begin{lemma}
$$\|\bF^{n+1}-\bF^{n}\|\leq d(\mathcal{G}_{n+1},\mathcal{G}_{n}).$$
\end{lemma}
\proof{Proof.}
Let $\bu=\bF^{n+1}(\bp)$ and $\bv=\bF^{n}(\bp)$ for some $\bp$. Now $\bu$ is the point of intersection of $\mathcal{G}_{n+1}$ and the line $\mathbf{x}=t\mathbf{1}+\bp$. $\bv$ is the point of intersection of the frontier $\mathcal{G}_{n}$ and the line $\mathbf{x}=t\mathbf{1}+\bp,$. 
Now suppose that $\|\bu-\bv\|_{\infty}>d(\mathcal{G}_{n+1},\mathcal{G}_{n})$. Then either for $\bu$, there is no $\br\in \mathcal{G}_{n}$ such that $\br\preceq \bu+\mathbf{1}d(\mathcal{G}_{n+1},\mathcal{G}_{n})$ or for $\bv$, there is no $\br\in \mathcal{G}_{n+1}$ such that  $\br\preceq \bv+\mathbf{1}d(\mathcal{G}_{n+1},\mathcal{G}_{n})$. Either of the two cases contradict the definition of $d(\mathcal{G}_{n+1},\mathcal{G}_{n})$. Thus $\|\bu-\bv\|_{\infty}\leq d(\mathcal{G}_{n+1},\mathcal{G}_{n})$. \hfill \halmos
\endproof
Finally, by the triangle inequality we have 

\begin{eqnarray}
d(\mathcal{G}_{n+1},\mathcal{G}_{n})&\leq& d(\mathcal{A}_{n+1},\mathcal{A}_n) + d(\mathcal{G}_{n+1},\mathcal{A}_{n+1})+d(\mathcal{G}_{n},\mathcal{A}_{n})\nonumber\\
&\leq&(1-\beta)\beta^n+\frac{1}{N}\bigg(\frac{1-\beta^{n+1}}{1-\beta}\bigg)+\frac{1}{N}\bigg(\frac{1-\beta^n}{1-\beta}\bigg).
\end{eqnarray}
Combining with (\ref{inter}) we have the result.
\hfill\halmos\\
\endproof

\section{Some remarks on Pareto frontiers of closed and convex sets.}\label{sec:convclo}
The Pareto frontier of a closed set is not necessarily closed. Figure~\ref{fig:upset} is one example -- the upset is closed, but its Pareto frontier is open. But we can show that the Pareto frontier of a closed and convex set in $\mathbb{R}^2$ is closed.\footnote{Of course, the Pareto frontier of a closed set may be empty, e.g., $\{(x,y)\in\mathbb{R}^2: x=y\}$, in which case it is trivially closed.}
\begin{figure}[H]
\begin{center}
\includegraphics[width=2in,angle=0]{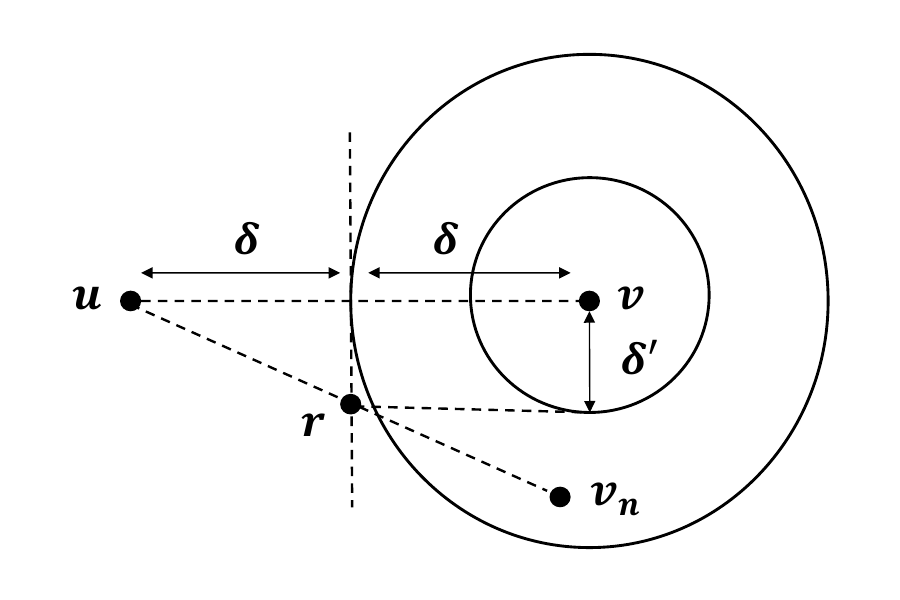}
\caption{Construction in the proof of Proposition $\ref{lma:closed}$.}
\label{fig:convclo}
\end{center}
\end{figure} 
\begin{proposition}\label{lma:closed}
Let $\mathcal{V}$ be the lower Pareto frontier of a closed and convex set $\cS$ in $\mathbb{R}^2$. Then $\mathcal{V}$ is closed. 
\end{proposition}
\proof{Proof.}
Suppose that $\{\bv_n\}$ is a sequence of points in $\mathcal{V}$ that converge to some point $\bv$. Then since $\cS$ is closed, $\bv\in \cS$. We will show that $\bv\in \cV$. Suppose not. Then there is some $\bu\in \mathcal{V}$ such that $\bu\preceq \bv$. Suppose first that 
$u_1<v_1$ and $u_2< v_2$. Then let $\epsilon =(\min(v_1-u_1,v_2-u_2)/2$ and consider the $\mathcal{L}^2$ ball of radius $\epsilon$ around $\bv$, i.e. 
$$B_{\bv}(\epsilon)=\{\by\in \mathbb{R}^2:\|\by-\bv\|_2 \leq \epsilon\}.$$
Then for any point $\by$ in $B_{\bv}(\epsilon)$, we have that $\bu\preceq \by$. But since $\{\bv^n\}$ converges to $\bv$, there exists some point in the sequence that is in $B_{\bv}(\epsilon)$, and $\bu$ is dominated by this point, which is a contradiction.
Hence either $u_1=v_1$ or $u_2=v_2$. Suppose w.l.o.g. that $u_1<v_1$ and $u_2=v_2$. See Figure~\ref{fig:convclo}. Let $\delta = (v_1-u_1)/2$ and consider the ball of radius $\delta$ centered at $\bv$, i.e. $B_{\bv}(\delta)$. Let $\bv^n$ be a point in the sequence such that $\bv^n\in B_{\bv}(\delta)$. Now $v_{n,1}>u_1$ and hence it must be that $v_{n,2}<u_2$. Now for some $\lambda\in (0,1)$, consider a point $\br=\lambda\bu+(1-\lambda)\bv_n$ such that $r_1=u_1+\delta$. It is possible to pick such a point since a) $v_1=u_1+2\delta$ and b) $|v_{n,1}-v_1|\leq \delta$, which together imply that $v_{n,1}\geq u_1+\delta$ (please see the figure).
Now $\br\in S$ since $S$ is convex. Next $r_1=v_1-\delta <v_1$, and also $r_2<u_2=v_2$ since $\lambda>0$ and $v_{n,2}<u_2$. Let $\delta'=v_2-r_2$. Then consider the ball $B_{\bv}(\delta')$ centered at $\bv$. Clearly $\br\preceq\by$ for any $\by\in B_{\bv}(\delta')$. But since $\{\bv_n\}$ converges to $\bv$, there exists some point in the sequence that is in $B_{\bv}(\delta')$, and $\br$ is dominated by this point, which is again a contradiction.  
Thus $\bv\in \mathcal{V}$.\hfill\halmos\\
\endproof

Interestingly, this result doesn't hold for closed and convex sets in $\mathbb{R}^K$ for $K>2$. A counterexample can be found in Kruskal \cite{kruskal1969two}. A variant of this counterexample is depicted in Figure~\ref{fig:counter} for completeness. The closed and convex set $\cS$ is a solid 3-dimensional cone with apex $(0,0,1)$ and base being the semicircular disc defined by the set $\{(x,y,z)\in\mathbb{R}^3: x= 5,\, (y-1)^2 + (z-1)^2 \leq 1,\, z\leq 1\}$. The sequence $(\bv_n)_{n\in \mathbb{N}}$ lies on the Pareto frontier of this set as shown in the figure, but it converges to the point $(5,0,1)$, which dominates the point $(0,0,1)$.

\begin{figure}[H]
\begin{center}
\includegraphics[width=4in,height=3in,angle=0]{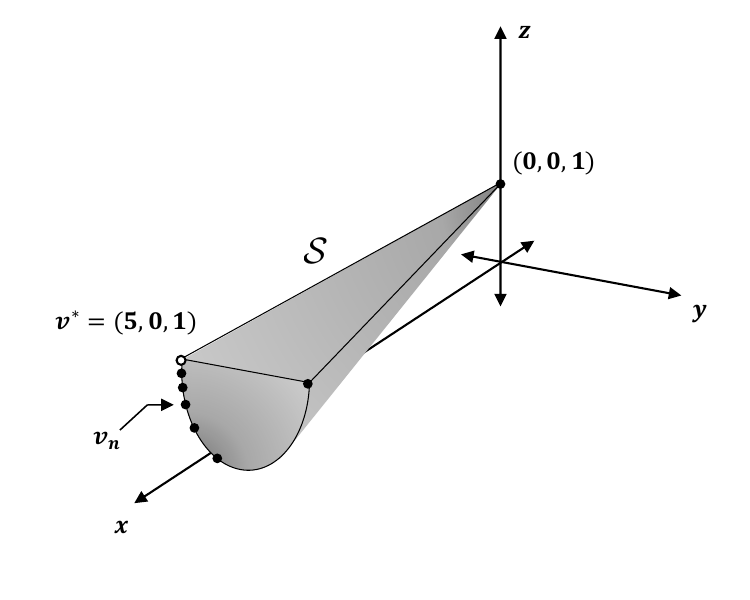}
\caption{An example of a compact and convex set in $\mathbb{R}^3$ whose Pareto frontier is not closed. The sequence $(\bv_n)$ on the Pareto frontier converges to the point $(5,0,1)$, which dominates $(0,0,1)$.}
\label{fig:counter}
\end{center}
\end{figure} 
\section{Solving Linear Program~\eqref{operator}.}\label{apx:linprog}
When $\cV$ is the lower Pareto frontier of a convex polytope, \eqref{operator} is a linear program. In this program, $\bx$ is a dependent vector and can be eliminated. The only question remains is that of addressing the constraint $\mathbf{Q}(b)\in\cV$. The vertices of $\cV$ are a subset of $\{\bF(\bp,\Phi(\cV)): \bp\in\mathcal{P}_N\}$. Thus $\mathbf{Q}(b)$ can be chosen as a convex combination of points in $\{\bF(\bp,\Phi(\cV)): \bp\in\mathcal{P}_N\}$. This introduces $H(K,N)$ variables for each $b\in B$ along with the constraint that these variables sum to $1$, thus contributing $mH(K,N)$ variables and $m$ constraints. Along with the variables $\mathbf{\alpha}$ and $t$, this makes $mH(K,N)+l+1$ variables in total. And along with the $K$ domination constraints for each $b\in B$ and the constraint that $\sum_{a}\alpha_a = 1$, this makes $Km + m +1$ constraints in total (ignoring non-negativity constraints on all variables except $t$). 

Of the $H(K,N)$ variables associated with each $b\in B$ that determine the point $\mathbf{Q}(b)$, we know that at most $K$ will be non-zero. This sparsity constraint can potentially be utilized to speed up the computation, although we didn't attempt to do so in our computations.

\section{Evaluation of finite-mode stationary policies} \label{apx:pol-eval}
In this section, we show how the upper bounds on the losses guaranteed by a finite-mode stationary policy can be efficiently computed as the solution to a linear program. Consider an $M$-mode stationary policy $\pi$, with a set of modes $\mathcal{M} = \{1,\cdots,M\}$, where each mode $i\in\mathcal{M}$ is associated with a probability distribution $\bm{\alpha}_i\in\Delta(A)$ over immediate actions and a transition rule $(\mathbf{z}_i(b)\in\Delta(\mathcal{M}); \, b\in B)$. Let $\bv_i$ be the vector of smallest upper bounds on the total discounted losses guaranteed by this policy starting from mode $i$. Then $\{\bv_i;\, i\in\mathcal{M}\}$ can be computed as the solution of the following linear optimization problem.
\begin{subequations}
\begin{align}
 \min_{(\bv_i)_{i\in\mathcal{M}}} \sum_{i \in \mathcal{M}} \mathbf{1}^T\bv_i
  \end{align}
\begin{equation}\textrm{s.t. } \bv_i\succeq \sum_{a\in A}\alpha_{i,a}\br(a,b) + \beta\sum_{j\in \mathcal{M}}z_{i,j}(b) \bv_j,\textrm{ for all }b\in B\textrm{ and } i\in\mathcal{M}.\label{bellapprox}\end{equation}
\end{subequations}
It is clear that if $\bv_i$ is the vector of smallest upper bounds on the losses corresponding to mode $i$, then the set $(\bv_i)$ satisfies \eqref{bellapprox}, which captures the Bellman one-step optimality conditions. In other words, $(\bv_i)$ is feasible in the above program. On the other hand, due to the same inequalities in \eqref{bellapprox}, any feasible set $(\bv_i)$ can be approximately guaranteed by the given stationary policy in a long but finite discounted repeated game, where the approximation error goes to $0$ as the length of the game approaches infinity. Hence, the upper bounds $(\bv_i)$ can be guaranteed in the infinitely repeated game (this argument is analogous to the one used in the proof of Theorem~\ref{thm:main2}). We thus conclude that the solution to the linear program above yields the smallest lower bounds on the losses corresponding to the different modes. Note that in the objective function, the weights for the different components of $\bv_i$ for the different modes $i$ can be chosen to be any positive numbers. Finally, note that this linear program decouples across dimensions and hence, can be solved for each dimension $k$ to obtain $\{v_{k,i};\, i\in\mathcal{M}\}$. 
%Each of these linear programs in a single dimension is identical to the linear program for computing the value of a zero-sum, single-controller stochastic game, in which the state transitions are controlled by the actions of a single player (in our case the mode transitions are dependent only on the adversary's actions) \cite{parthasarathy1981orderfield,raghavan1991algorithms}.
\section{Benchmark algorithms}
\subsection{Hedge}\label{apx:hedge}
In this algorithm, if $L_t(i)$ is the cumulative loss of expert $i$ till time $t$, then the probability of choosing expert $i$ at time $t+1$ is
$$p_i(t+1)\propto \exp(-\eta L_t(i)),$$
where $\eta$ is a parameter. In the undiscounted problem, choosing $\eta =\sqrt{8\log K/T}$ when the time horizon $T$ is known attains an upper bound of $\sqrt{T\log{K}/2}$ on the expected cumulative regret (Thm. 2.2, \cite{Cesa-Bianchi06a}). In a certain sense, this is shown to be asymptotically optimal in $K$ and $T$ for general loss function taking values in $[0,1]$ (Thm. 3.7, \cite{Cesa-Bianchi06a}). In our implementation, we use discounted cumulative losses in this algorithm, and choose $\eta = \sqrt{8\log K(1-\beta^2)}$. This resulting algorithm achieves an upper bound of $\sqrt{\log{K}/2(1-\beta^2)}=\sqrt{\log{K}/(2(1-\beta)(1+\beta))}$ on the expected discounted regret in the infinitely repeated game (see proof of Thm. 2.2, and Thm. 2.8 in \cite{Cesa-Bianchi06a}). 
\subsection{GPS}\label{apx:gps}
The GPS algorithm for $K=2$ experts is defined as follows \cite{gravin2014towards}. Let $\xi=(1-\sqrt{1-\beta^2})/\beta$.Let $d$ be the difference in the cumulative (undiscounted) losses of the leading expert (the one with the lower cumulative loss) and the lagging expert (the one with the higher loss). Then at every stage, the algorithm chooses the leading expert with probability $1-(1/2)\xi^d$ and the lagging expert with probability $(1/2)\xi^d$.

\section{Remarks on the model of expert selection considered in GPS \cite{gravin2014towards}.}\label{apx:gps2}
In the expert selection game considered in \cite{gravin2014towards}, at each stage, the game ends with a probability $1-\beta$ and continues with probability $\beta$. This is essentially the same as our model of discounted losses, where the discount factor is interpreted as the probability of continuation at each stage. But the difference between that formulation and our formulation is in the definition of regret. In their formulation, the loss of the decision-maker is compared to the expected loss of the best expert in the \emph{realized} time horizon (where the expectation is over the randomness in the time horizon), i.e., to $\textup{E}_T[\min_{i=1,2}\sum_{t=1}^Tl_t(i)]$, where $l_t(i)$ is the loss of expert $i$ at time $t$. On the other hand, in our formulation, the loss of the decision-maker is compared to the expert with the lowest expected loss, i.e., to $\min_{i=1,2}\textup{E}_T[\sum_{t=1}^Tl_t(i)]=\min_{i=1,2}\sum_{t=1}^\infty\beta^{t-1}l_t(i)$. Naturally, the optimal regret in their formulation is at least as high as the optimal regret in our formulation, and in fact it turns out to be strictly higher. For example, for $K=2$ and $\beta=0.9$, the optimal expected total regret in their formulation is $\approx 1.147$,\footnote{The optimal regret for \cite{gravin2014towards}'s formulation is $\frac{1}{2\sqrt{1-\beta^2}}$ (or an average discounted regret of $\frac{1-\beta}{2\sqrt{1-\beta^2}}=(1/2)\times\sqrt{(1-\beta)/(1+\beta)}$). The expression in \cite{gravin2014towards} is off by a factor of $\beta$ compared to this expression since they discount the first period by $\beta$ (i.e., the game could end before the first stage begins), whereas we discount it by $1$. \label{fn:gps}} while in our formulation, the optimal regret in this case is at most $\approx 0.9338$ (see Figure~\ref{fig:cumreg}). Further, it is clear that their optimal strategy gives the same guarantee for our definition of regret as the optimal regret in their definition, i.e., for instance, for $\beta=0.9$, their strategy will guarantee an expected regret of at most $\approx1.147$ according to our definition. Note that in the context where discounting captures the temporal change in the present value of money, our formulation of regret is arguably more natural. %The GPS algorithms for $K=2$ and $3$ are described in Appendix~\ref{apx:gps}.\\

%\section{A Blackwell-type minimax theorem doesn't hold for finite games.}\label{sec:minimax-counter}
%Here we give an example showing that a Blackwell-type minimax result doesn't hold for expected losses in finite games, even if the losses are scalar. Consider the matrix game shown in Figure~\ref{fig:minimax-scalar}. And consider the convex set consisting of a single point 0.4 on the real line. Then observe that for any mixed strategy $(\alpha, 1-\alpha)$ of player 1, the expected loss if player 2 chooses action 1 is $(1-\alpha)$ and if she chooses action 2, it is $0.25\alpha +0.5(1-\alpha)= 0.5-.25\alpha$. Thus no strategy of player 1 can guarantee that the expected loss is exactly 0.4 irrespective of the actions of player 2. But neither does player 2 have any strategy that guarantees that the expected loss in the game is \emph{not} 0.4. To see this, observe that for any strategy $(\gamma, 1-\gamma)$ of player 2, player $1$ can choose a randomization over her two actions that will result in an expected loss of exactly 0.4. 

%\begin{figure}[H]
%\begin{center}
%\includegraphics[width=1.5in,angle=0]{minimax-scalar.pdf}
%\caption{A game with scalar losses.}
%\label{fig:minimax-scalar}
%\end{center}
%\end{figure} 

\section{Hedge and GPS are suboptimal for the expert selection problem with $K=2$ and $\beta =0.8$.}\label{apx:badadv}
\begin{figure}[H]
\centering
\begin{minipage}[H]{.45\textwidth}
\vspace{0in}
\begin{table}[H]\caption{Upper bounds on the average discounted regret under different policies for $\beta = 0.8$}\label{tbl:ub1}
\vspace{0.1in}
\centering
{\small
\begin{tabular}{cc}\toprule
Policy                       & Regret upper bound \\
\midrule
Hedge                        & 0.1962             \\
GPS                          & 0.1666             \\
203-mode & 0.1357             \\
21-mode  &  0.1374     \\
\bottomrule    
\end{tabular}}
\end{table}\end{minipage}
\hspace{1cm}
\begin{minipage}[H]{.45\textwidth}
\vspace{0in}
\centering
\includegraphics[width=2.5in]{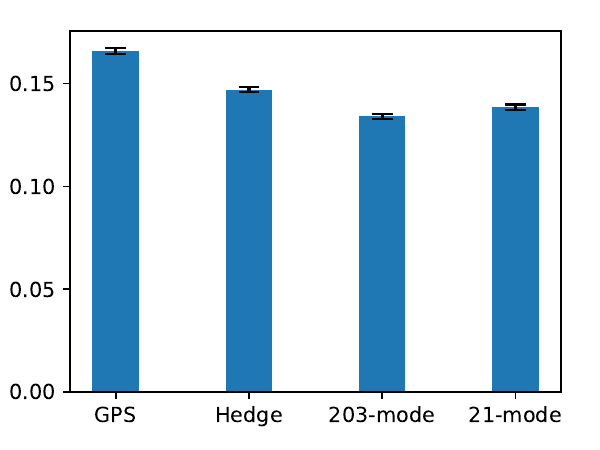}
\caption{Estimates of the expected average discounted regret of different algorithms for $K=2$ and $\beta= 0.8$ against Adversary A, along with associated error bars.}
\label{fig:runs8K2}
\end{minipage}
\end{figure}

%\begin{figure}[h]
%\centering
%\begin{minipage}[h]{.45\textwidth}
%\begin{table}[H]\caption{$\beta = 0.8$}\label{tbl:ub1}
%\centering
%{\small
%\begin{tabular}{|c|c|}\hline
%Policy                       & Regret upper bound \\
%\hline\hline
%Hedge                        & 0.1962             \\
%GPS                          & 0.1666             \\
%203-mode & 0.1357             \\
%21-mode  &  0.1374     \\
%\hline     
%\end{tabular}}
%\end{table}\end{minipage}
%\hspace{1cm}
%\begin{minipage}[t]{.5\textwidth}
%\centering
%\includegraphics[width=2.5in]{beta-8-K-2-new.pdf}
%\caption{Estimates of the expected average discounted regret of different algorithms for $K=2$ and $\beta= 0.8$ against Adversary A, along with associated error bars.}
%\label{fig:runs8K2}
%\end{minipage}
%\end{figure}
Consider the experts problem with $K=2$ experts and a discount factor $\beta = 0.8$. Consider the following strategy for the adversary. 

\noindent{\bf Adversary A:} In this strategy, the probability that expert $1$ incurs a loss (and expert $2$ doesn't) at time $t$ is $0.9^{1/t}$ if $t$ is odd, and $0.9^t$ if $t$ is even. This adversary never gives equal losses to both experts.

We compare the performance of Hedge and GPS against this adversary to that of two approximately optimal policies that we design. The first policy is a $203-$mode ($H(2,101)$) stationary policy and the second is a $21-$mode ($H(2,10)$) stationary policy ($n=28$ in both cases). Table~\ref{tbl:ub1} shows the theoretical upper bounds on the regret guaranteed by these algorithms.

Figure \ref{fig:runs8K2} compares the expected average discounted regret incurred by the four algorithms. The expected regret is estimated in each case by averaging over 10000 runs, where each run is a game with time horizon $T=100$. The associated error bars %($\pm1.96\,\times$ standard error) 
are shown in the graph. Note that our strategies significantly outperform Hedge and GPS.
%Observe that against both adversaries, the regret incurred by our strategies does not significantly exceed the upper bounds of $\approx 0.6886$ for $\beta = 0.8$ and $\approx 0.9338$ for $\beta = 0.9$. 
Importantly, the regrets of Hedge and GPS significantly exceed the upper bounds on the regret guaranteed by our algorithms. This eliminates the possibility of these algorithms being optimal for our problem with high probability. 
%As expected, the average regret incurred by the GPS algorithm does not significantly exceed the values of $\approx0.8333$ and $\approx 1.147$ respectively, which are the optimal values of regret as defined by GPS. 
%Similarly, considering the performance against Adversary B, Hedge appears to be sub-optimal by a large margin for $\beta= 0.8$. But its performance comes close to the upper bound of $\approx 0.9338$ for $\beta = 0.9$. This seems aligned with the possibility of Hedge being asymptotically optimal for our problem as $\beta\rightarrow 1$. 
%Finally, it is also interesting to note that there is no significant difference in the performance of the $203-$mode and $21-$mode strategies for $\beta = 0.8$, or  in the performance of the $403-$mode and $41-$mode strategies for $\beta = 0.9$. This suggests the possibility of obtaining better performance bounds for less complex strategies, potentially using some other approximation schemes. We explore this possibility in the next section. 

\section{Exact characterization of $\mathcal{V}^*$ in the expert selection problem with $K=2$ and $\beta=0.5$.}\label{sec:exact}
In some simple examples, we can exactly determine the set $\mathcal{V}^*$ by ``solving'' the fixed point relation given by the dynamic programming operator. We demonstrate this by determining the optimal Pareto frontier in the game of combining expert advice (Figure~\ref{fig:ssregrets}) from 2 experts for $\beta = 0.5$. Note that the points $(0,1/(1-\beta)) = (0,2)$ and $(1/(1-\beta),0)= (2,0)$ lie on $\cV^*$ (achieved by choosing Expert 1 always or Expert 2 always, respectively). We can thus represent $\cV^*$ by a convex and decreasing function $f(x)$ defined on $x\in[0,2]$ such that $f(0) = 2$ and $f(2) = 0$, so that $\cV^* = \{(x,f(x)):x\in [0,2]\}$. $\beta \cV^*$ for $\beta=0.5$ is thus the set $\{(\beta x,\beta f(x)):x\in [0,2]\} = \{(x,0.5 f(x/0.5):x\in [0,1]\}$, which thus can be represented by the convex, decreasing function $\bar{f}(x) = f(2x)/2$ defined on $x\in [0,1]$, where $\bar{f}(0) =1$ and $\bar{f}(1) = 0$.

\begin{figure}[h]
\centering
\begin{minipage}[t]{.4\textwidth}
\centering
\includegraphics[width=2.5in,height=2.5in]{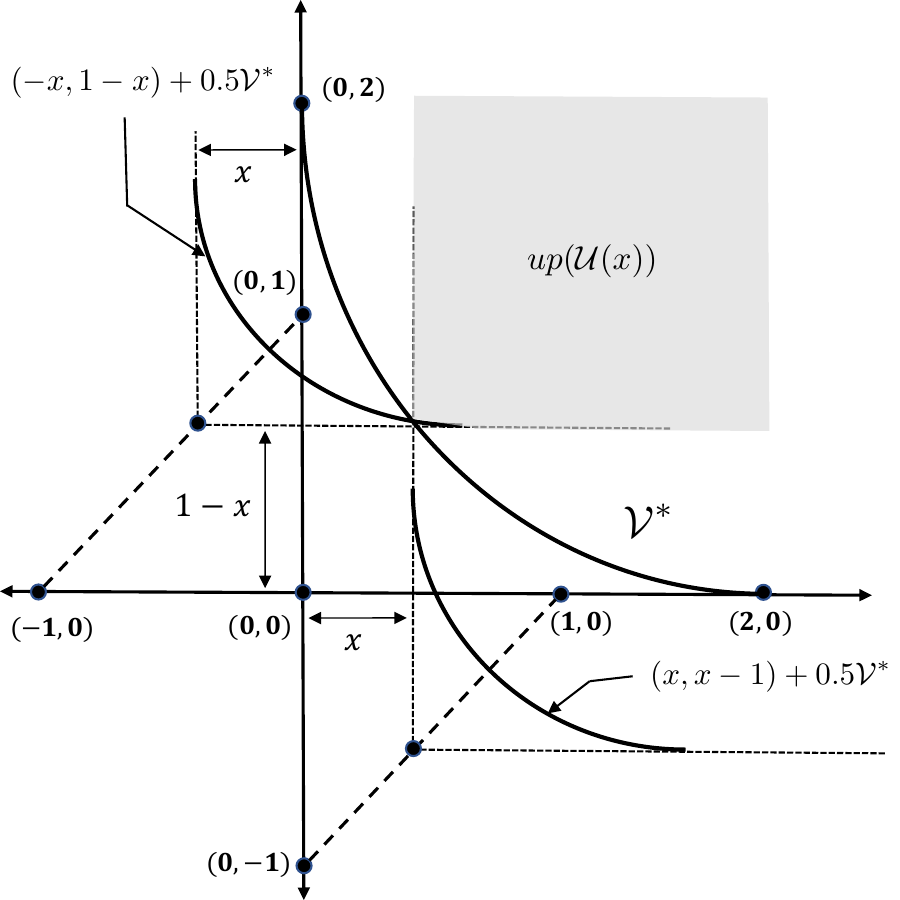}
\caption{Construction of $up(\mathcal{U}(x))$.}\label{fig:midbeta}
\end{minipage}\hfill
\begin{minipage}[t]{.5\textwidth}
\centering
\includegraphics[width=2.5in,height=2.5in,angle=0]{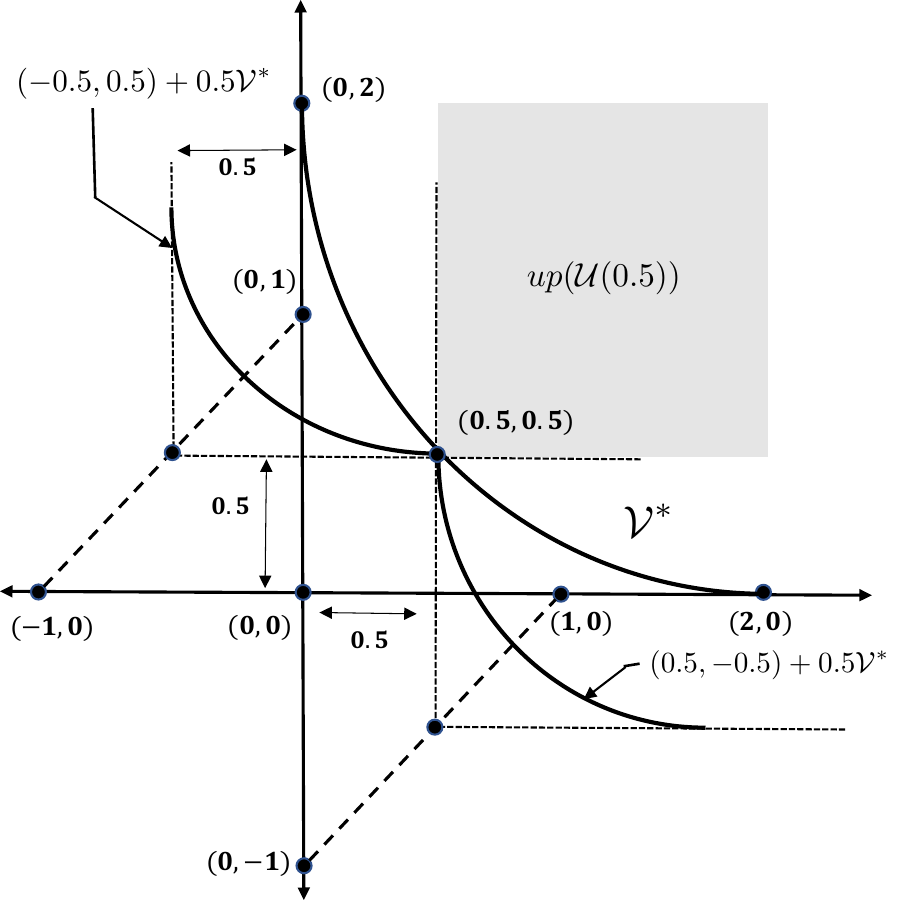}
\caption{Construction of $up(\mathcal{U}(0.5))$.}
\label{fig:midbeta-mid}
\end{minipage}
\end{figure}

Now for a fixed randomization over Alice's actions, $(1-x,x)$, by choosing different points in $\mathcal{V}^*$ from the next stage onwards, one obtains the set of guarantees
\begin{align}
&\mathcal{U}(x) = \bigg\{\big(\max (-x + 0.5\, Q_1(1),\,x +0.5 \,Q_1(2)),\nonumber\\
&~~ \max(1-x + 0.5\, Q_2(1),\, x-1+ 0.5 \,Q_2(2))\big): \mathbf{Q}(1),\,\mathbf{Q}(2)\in \cV^*\bigg\}.
\end{align}
If we denote the set $(-x,1-x) +0.5\cV^*$ (which is obtained by mapping each element $\mathbf{u}$ of $\cV^*$ to $(-x,1-x)+0.5\mathbf{u})$), by $\mathcal{U}_1(x)$, and the set $(x,x-1) +0.5\cV^*$ by $\mathcal{U}_2(x)$, it is straightforward to see that 
$$up(\mathcal{U}(x)) = up(\mathcal{U}_1(x))\cap up(\mathcal{U}_2(x)),$$
where $up(.)$ is the upset of the set in $[0,2]^2$. This is depicted in Figure~\ref{fig:midbeta}.
The fixed point relation says that
$$\cV^* = \Lambda \bigg(\cup_{x\in[0,1]}up(\mathcal{U}(x))\bigg).$$
From the figure, one can see that $\cV^*$ is the curve traced by the lower left corner point of $up(\mathcal{U}(x))$ as $x$ varies between $0$ and $1$. Since we already know that the two extreme points on $\cV^*$ are $(0,2)$ and $(2,0)$, for $x= 0.5$, we know that the lower left corner point of $up(\mathcal{U}(0.5))$ is $(0.5, 0.5)$, and hence is contained in $\mathcal{V}^*$, as shown in Figure~\ref{fig:midbeta-mid}. Since we know that $\cV^*$ is symmetric around the line $x=y$, we know that $f(x) = f^{-1}(x)$, and thus it is sufficient to determine $f(x)$ in the range $x\in[0,0.5]$. In this range, $f$ satisfies the following fixed point relation (again, see Figure~\ref{fig:midbeta}):
\begin{align}
f(x)&= \bar{f}(2x) +1-x\nonumber\\
&=f(4x)/2 +1-x.
\end{align}
Taking the derivative twice on both sides, we obtain: 
$$f''(x) = 8f''(4x).$$
This gives us $f''(x) = ax^{-\frac{3}{2}}$ for any $a\in\mathbb{R}$. Integrating, we obtain $f(x) = a\sqrt{x} +x +2$. Since we want $f(0.5) = 0.5$, we obtain $a=-2\sqrt{2}$. Thus we have $f(x) = -2\sqrt{2x} +x+2$. Note that $f(2) = 0$, and it turns out that $f(x)$ restricted to the domain $x\in[0,2]$ is such that $f(x) = f^{-1}(x)$. Thus $f(x)$ is the function we are looking for and $\cV^*= \{(x,-2\sqrt{2x} +x+2):x\in[0,2]\}$. 

We can compare this exact characterization with the approximate frontier that we computed using our approximation procedure for $\beta=0.5$ (approximation error less than $0.06$). Both these frontiers are plotted in Figure~\ref{fig:compare}. As we can observe, the two frontiers are close to identical.

\begin{figure}[htb]
\begin{center}
\includegraphics[width=3in,angle=0]{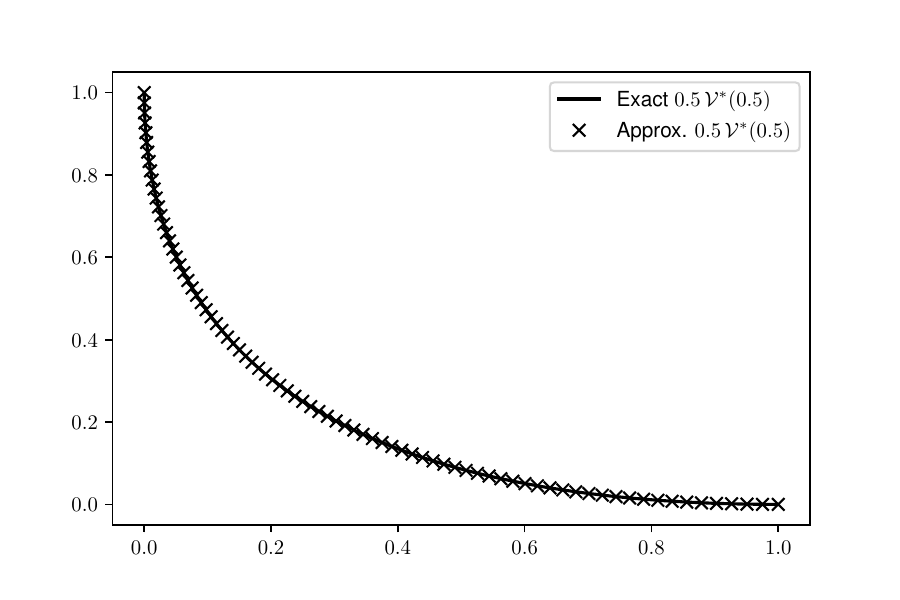}
\caption{Comparison of the approximation of $0.5\cV^*(0.5)$ and the exact characterization $0.5\cV^*(0.5)= \{(x,-2\sqrt{x} +x+1):x\in[0,1]\}$.}
\label{fig:compare}
\end{center}
\end{figure} 

{\bf Optimal Policy:} To attain the point $(x,f(x))$ for $x\in[0,0.5]$, the optimal strategy of Alice chooses a randomization $(1-x,x)$; then if the adversary chooses action $1$, the next point she chooses to attain is $(4x,f(4x))$, whereas if he chooses action $2$ then the next point she chooses to attain is $(0,f(0))$. To attain the point $(f(x),x)$ for $x\in[0,0.5]$, the optimal strategy of Alice chooses a randomization $(x,1-x)$; then if the adversary chooses action $2$, the next point she chooses to attain is $(f(4x),4x)$, whereas if he chooses action $1$, then the next point she chooses to attain is $(f(0),0)$.

\section{Zero-sum repeated games with scalar losses: a review of results}
The scalar counterpart of the vector-valued repeated game we study in the paper is a relatively much simpler object of study. Preserving the notation in the paper, suppose that in the single stage game $\mathbb{G}$, a pair of actions $a\in A$ for Alice and $b\in B$ for Bob leads Alice to incur a scalar loss $r(a,b)$. The value of game $\mathbb{G}$ is then defined to be,

\begin{align}
v^* =\min_{\bm{\alpha}\in\Delta(A)}\max_{b\in B}\sum_{a\in A}\alpha_a r(a,b) \overset{(a)}{=} \max_{\bm{\nu}\in\Delta(B)}\min_{a\in A}\sum_{b\in B}\nu_b r(a,b), 
\end{align}
where the equality $(a)$ follows from the Von Neumann minmax theorem \cite{neumann1928theorie}. It is well-known that both the above optimization problems can be solved as linear programs. For example, the minmax optimization problem for Alice can be solved as the following linear program.
\begin{subequations}
\begin{align}
\min v
\end{align}
\begin{equation}
v\geq \sum_{a\in A}\alpha_a r(a,b),\,\textrm{ for all }b\in B,
\end{equation}
\begin{equation}
\bm{\alpha}\in\Delta(A).
\end{equation}
\end{subequations}

Let $\bm{\alpha}^*$ be an arbitrary minmax optimal strategy for Alice and $\bm{\nu}^*$ be an arbitrary maxmin optimal strategy for Bob in $\mathbb{G}$ (since optimal strategies may not be unique). 

Now consider a repeated game $\mathbb{G}^T$, in which $\mathbb{G}$ is repeated $T$ times, with the cumulative loss of $\mathbb{G}^T$ defined to be 
$$\frac{\sum_{t=1}^T \beta^{t-1}r(a_t,b_t)}{\sum_{t=1}^T \beta^{t-1}},$$
which is the average discounted loss with a discount factor $\beta\in [0,1]$ (we obtain the simple average loss for $\beta = 1$). It is straightforward to argue that the smallest upper bound on the expected loss that Alice can guarantee in $\mathbb{G}^T$ is simply $v^*$, i.e., the value of the game $\mathbb{G}$. Or in other words, the value of the game $\mathbb{G}^T$ is $v^*$ for any $T$ and $\beta \in [0,1]$. To see this, note that by playing $\bm{\alpha}^*$ in every repetition, Alice can guarantee that the expected stage loss in every repetition is {\it at most} $v^*$. Similarly, by playing $\bm{\nu}^*$ all the time, Bob can guarantee that expected stage loss in every repetition is {\it at least} $v^*$. Thus irrespective of how one averages the daily losses, the optimal guarantee on the average loss that Alice can guarantee is $v^*$. And the strategy that achieves this guarantee is the one where Alice plays any equilibrium strategy of $\mathbb{G}$ in each repetition. Now consider the game $\mathbb{G}^\infty$, in which $\mathbb{G}$ is repeated infinitely often, with the cumulative loss defined to be $$(1-\beta)\sum_{t=1}^T \beta^{t-1}r(a_t,b_t)$$
for $\beta \in [0,1)$. The previous argument extends to this game as well and we can conclude that the optimal guarantee on the average loss that Alice can guarantee is again $v^*$.

\subsection{Simultaneous vector guarantees vs. guarantees on the combined scalar loss}\label{apx:scalar}
In the present paper, we concerned ourselves with characterizing the best {\it simultaneous guarantees} that Alice can guarantee across the vector components of the losses. A natural question is whether the frontier of such simultaneous guarantees can be characterized by solving a set of scalar repeated games, where each game is obtained from a different weighted combination of the vector losses. The example below shows that this is not the case. The key point is that optimal strategies that achieve simultaneous guarantees across the different dimensions must adapt to the evolution of the profile of losses across the different dimensions over time. If one dimension suffers excessive losses, these strategies must shift to focusing on minimizing losses on that dimension. This profile information is lost when one combines the losses into a single scalar. In other words, when we combine the dimensions, the optimal strategy in the resulting scalar repeated game only guarantees an upper bound on the combined loss, as opposed to simultaneously guaranteeing upper bounds on losses across the different dimensions. Hence, one must directly address the multi-dimensional nature of the game as we do in the paper as opposed to attempting a reduction from the scalar case.

{\bf Example.} Consider the game with vector losses shown in Figure~\ref{fig:one-shot-a}. This is the game that corresponds to the single-stage vector regrets in the expert selection problem with $K=2$ experts.  If the losses are combined across the two dimensions with weights $(\alpha, 1-\alpha)$, then the resulting game is shown in Figure~\ref{fig:one-shot-b}. In this game, the unique minmax optimal strategy for Alice is to play action $1$ with probability $\alpha$ and action $2$ with probability $1-\alpha$. Now suppose that this scalar game is repeated infinitely often, with losses in the $n^{\textup{th}}$ repetition discounted by $(1-\beta)\beta^{n-1}$. In this repeated game, as we argued above, the unique minmax optimal strategy for Alice is to play $(\alpha, 1-\alpha)$ in every repetition. Against this strategy, the total discounted vector losses corresponding to the two actions (also repeated forever) of the adversary are depicted in Figure~\ref{fig:one-shot-c}. Considering worst-case choices of the adversary, we can deduce that Alice guarantees a maximum loss of $1-\alpha$ on dimension 1 and $\alpha$ on dimension 2, i.e., this strategy achieves the vector guarantee $(1-\alpha, \alpha)$. Let $\mathcal{V}^*_s$ denote the set of vector guarantees achievable using this scalar reduction by varying $\alpha$. $\mathcal{V}^*_s$ is shown in Figure~\ref{fig:one-shot-d}: it is simply the line segment joining $(0,1)$ and $(1,0)$, irrespective of the value of $\beta$. However, note that we have shown in Section~\ref{sec:exact} above that we can achieve significantly better guarantees for $\beta = 0.5$;  see Figure~\ref{fig:compare}. This demonstrates that the optimal guarantees cannot be achieved via a scalar reduction of the vector-valued game.
\begin{figure}[]
\centering
\begin{minipage}[t]{.45\textwidth}
\centering
\includegraphics[width=1.5in]{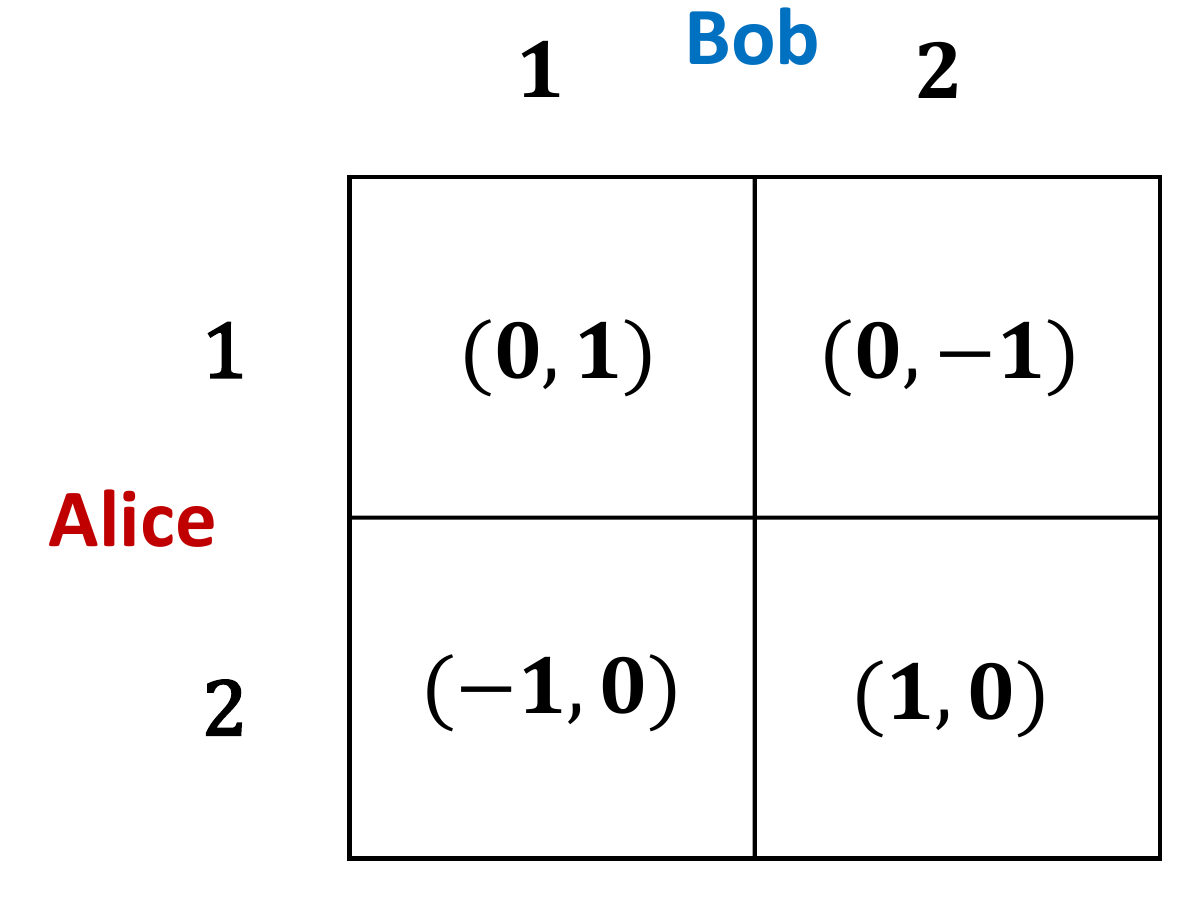}
\caption{A game with vector losses.}\label{fig:one-shot-a}
\end{minipage}
\hspace{1cm}
\begin{minipage}[t]{.45\textwidth}
\centering
\includegraphics[width=1.5in]{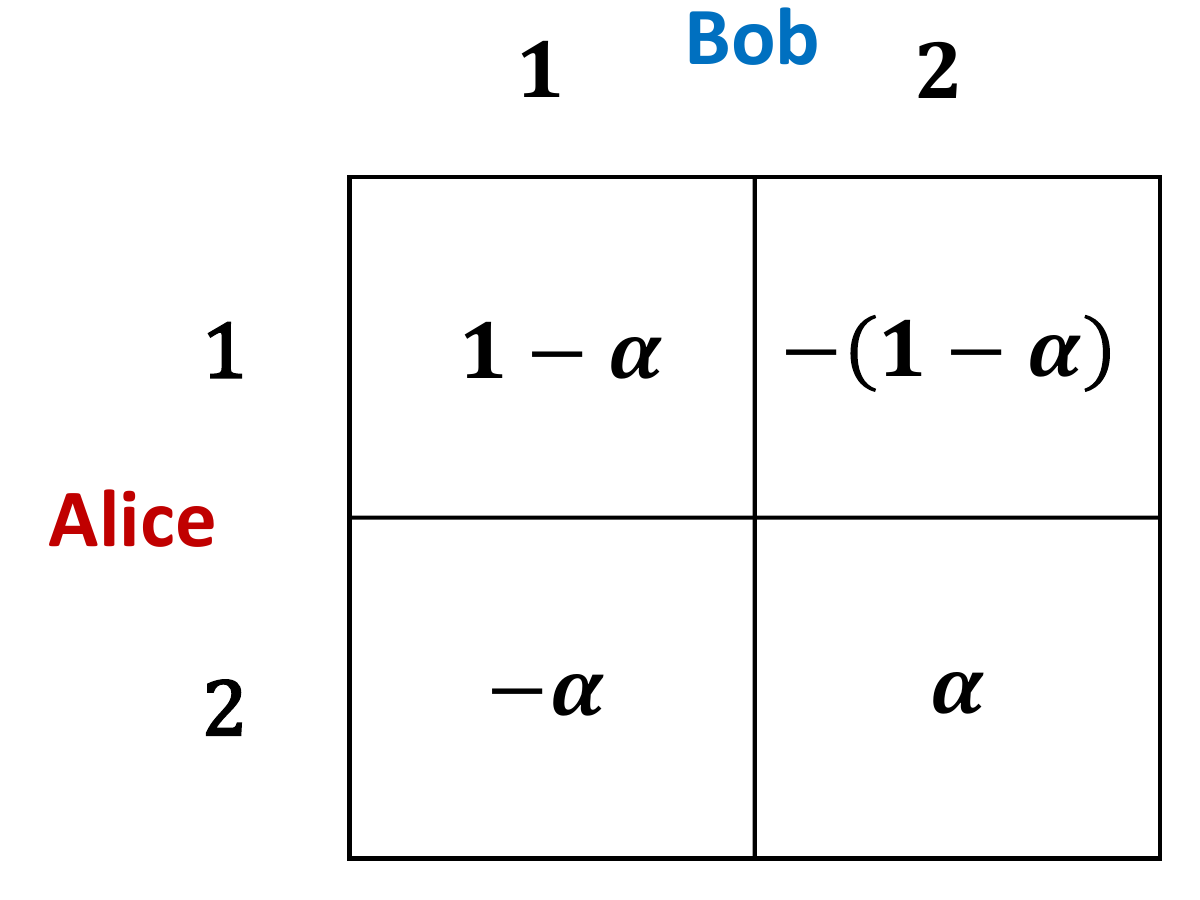}
\caption{A scalar game corresponding to weights $(\alpha, 1-\alpha)$.  }
\label{fig:one-shot-b}
\end{minipage}
\begin{minipage}[t]{.45\textwidth}
\centering
\includegraphics[width=1.5in]{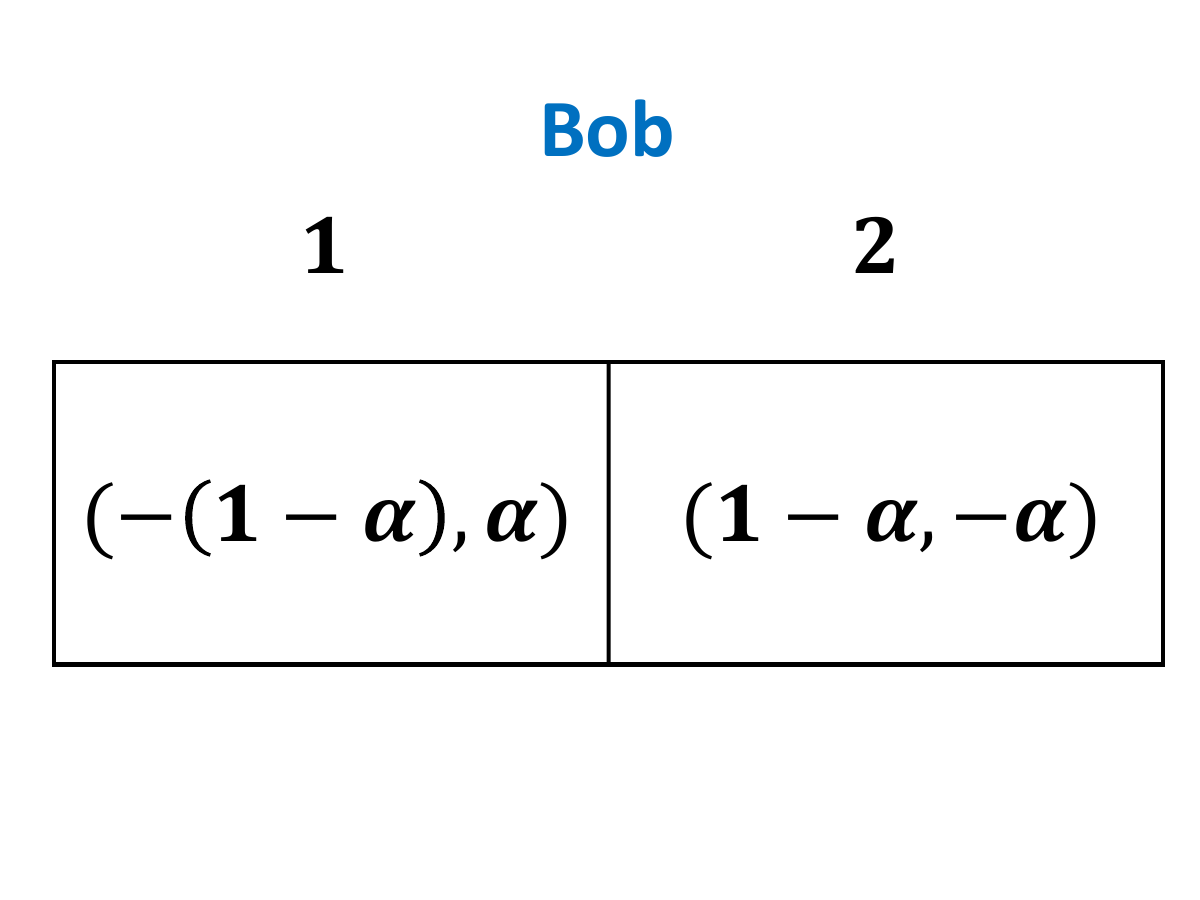}
\caption{Vector losses as a function of Bob's actions given Alice's stationary strategy $(\alpha, 1-\alpha)$.}\label{fig:one-shot-c}
\end{minipage}
\hspace{1cm}
\begin{minipage}[t]{.45\textwidth}
\centering
\includegraphics[width=2in]{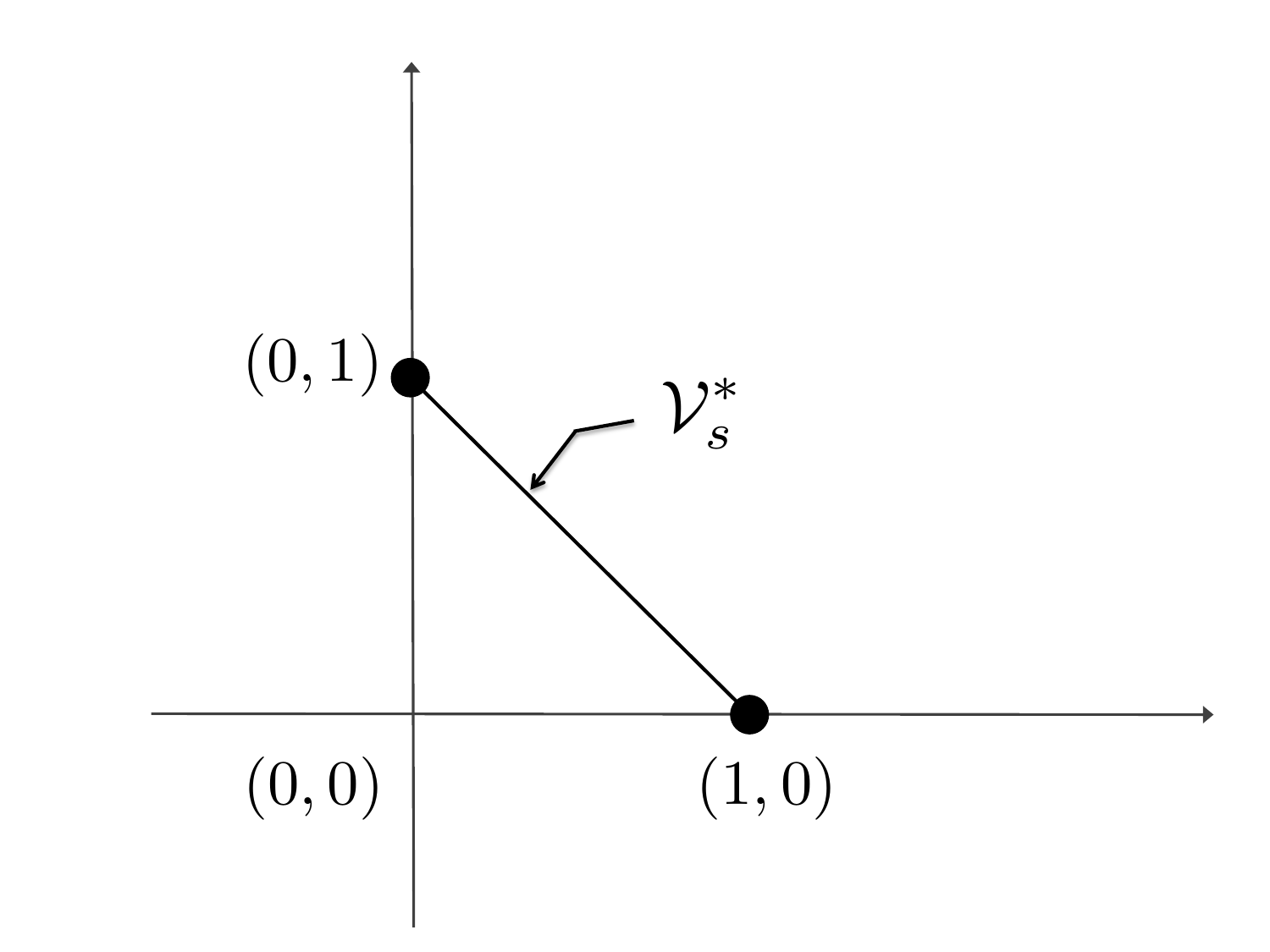}
\caption{The set of vector guarantees on losses achievable by varying $\alpha$.}\label{fig:one-shot-d}
\end{minipage}
\end{figure}

\end{APPENDIX}

%\subsection{Proof of Lemma 3.1}
%\begin{proof}
%For some $p\in (0,1)$ consider the minimization problem:
%$$\min_{x\in S} f(x)= px_1+(1-p)x_2.$$
%Since $f(x)$ is a continuous function defined on a compact set, it achieves this minimum value at some point $\mathbf{x}(p)\in S$. Hence there cannot be any point $\bx'\preceq \mathbf{x}(p)$, which means that $\mathbf{x}(p)$ is on the Pareto frontier of $S$. 
%\end{proof}

% Appendix here
% Options are (1) APPENDIX (with or without general title) or 
%             (2) APPENDICES (if it has more than one unrelated sections)
% Outcomment the appropriate case if necessary
%
% \begin{APPENDIX}{<Title of the Appendix>}
% \end{APPENDIX}
%
%   or 
%
% \begin{APPENDICES}
% \section{<Title of Section A>}
% \section{<Title of Section B>}
% etc
% \end{APPENDICES}

% Acknowledgments here
%\section*{Acknowledgments.}
% Enter the text of acknowledgments here

% References here (outcomment the appropriate case) 

% CASE 1: BiBTeX used to constantly update the references 
%   (while the paper is being written).
%\bibliographystyle{informs2014} % outcomment this and next line in Case 1
%\bibliography{<your bib file(s)>} % if more than one, comma separated

% CASE 2: BiBTeX used to generate mypaper.bbl (to be further fine tuned)
%\input{mypaper.bbl} % outcomment this line in Case 2

\end{document}